\documentclass[aps,rmp,twocolumn,superscriptaddress
]{revtex4}

\usepackage{amssymb,amsmath,epsfig,wasysym,float}
\usepackage[sort&compress]{natbib}
\usepackage{amsfonts}
\usepackage{amsmath}
\usepackage{amssymb}
\usepackage{upgreek}
\usepackage[usenames,dvipsnames]{color}

\begin{document}
\title{L\'{e}vy walks}
\author{V. Zaburdaev}
\email{vzaburd@pks.mpg.de}
\affiliation{Max Planck Institute for the Physics of Complex Systems, \mbox{N\"{o}thnitzer Str. 38, D-01187 Dresden, Germany}}

\author{S. Denisov}
\email{sergey.denisov@physik.uni-augsburg.de}
\affiliation{\mbox{Institute of Physics, University of Augsburg}, \mbox{Universit\"{a}tstr.~1, D-86159 Augsburg, Germany}} 
\affiliation{Lobachevsky State University,
\mbox{Gagarin Avenue 23, 603950 Nizhny Novgorod, Russia}}
\affiliation{Sumy State University, \mbox{Rimsky-Korsakov Street 2, 40007 Sumy, Ukraine}}
\author{J. Klafter}
\email{klafter@post.tau.ac.il}
\affiliation{\mbox{School of Chemistry, Tel Aviv University}, \mbox{69978 Tel Aviv, Israel}}

\begin{abstract}
Random walk is a fundamental concept with applications ranging from quantum physics to econometrics. 
Remarkably, one specific model of random walks appears to be 
ubiquitous across many fields as a tool to analyze  transport phenomena in which
the dispersal process is faster than dictated by Brownian diffusion. 
The L\'{e}vy walk model combines two key features, the ability to generate anomalously fast 
diffusion and a finite velocity of a random walker.  Recent results in optics, 
Hamiltonian  chaos, cold atom dynamics, bio-physics, and 
behavioral science demonstrate that this particular
type of random walks provides significant insight into 
complex transport phenomena. This review provides a self-consistent introduction to L\'{e}vy 
walks, surveys their existing applications, including latest advances, 
and outlines further perspectives. \end{abstract}

\maketitle
\tableofcontents

\section{Introduction}\label{Introduction}

In this review we want to demonstrate how a simple idea of 
the giving a finite velocity to a diffusing particle increases flexibility and diversity 
of diffusion models in describing complex transport phenomena. We  consider  
processes resulting from a motion of many identical  non-interacting particles. 
There are two key complimentary approaches to statistical description of such  motion. 
The first approach is based on the concept of random walks \cite{Weiss1994}  while the 
second is based on stochastic differential equations. Among the latter are the 
Langevin equation \cite{Coffey2012} and the concept of Brownian motion \cite{morters2010}. 
Although having  different terminologies and mathematical apparatus, 
the two approaches are closely related and  their exact equivalence can be demonstrated in some cases. 
The framework of our review is random walks and we would like to start with a 
historical overview of the  development of the concept,  with a special focus on how the idea 
of the finite velocity of walking particles was born and matured over the years.

As if to predict its interdisciplinary future, the theory of random walks was developed independently 
in the context of biology \cite{brown}, probability theory \cite{bernoulli}, finance \cite{bachelier}, 
and physics \cite{reyleigh, pearson}. The seminal works of \textcite{einstein} and 
\textcite{smoluchowski} marked the start of rigorous and quantitative approach
connecting microscopic dynamics of particles to the macroscopic process of diffusion 
[see, for example, \cite{nelson} for more historical background on Brownian motion]. 
The diffusion equation was already known for nearly a century as it was derived to describe 
the heat conduction by \textcite{fourier}. Despite the success of this equation 
in various applications, it had one particular drawback that did not escape the attention of contemporary scientists. 
According to the diffusion equation, when starting with a localized initial condition, even after an infinitesimally short
elapsed time there will be a nonzero density of diffusing particles at any arbitrary distance from the staring point. 
This implies an infinite propagation speed of some particles and thus contradicts 
our understanding of how physical objects move [see an interesting discussion on this issue in the context
of relativistic statistical physics in \textcite{Dunkel2009, Dunkel2009a}]. 
The infinite speed is also inconsistent with the original schematization of a random walk process by 
Karl Pearson \cite{pearson}: ``{\em A man starts from a point $0$ and walks $l$ yards in a straight line; he then turns through any angle whatever 
and walks another $l$ yards in a second straight line. He repeats this process n times. I require the probability that after 
these n stretches he is at a distance between $r$ and $r + dr$ from his starting point, $0$.}'' After this drawback was noted, 
two  approaches to resolve the issue have been proposed. In 1920, G. I. Taylor, concerned 
with the problem of turbulent transport, formulated a random walk model in which the motion 
of a particle between two turning events was characterized by a finite velocity \cite{taylor1922}. 
The same year the finiteness 
of  the velocity was mentioned 
by  \textcite{fuerth1920} in the context of the so-called persistent Brownian motion. 
Both these models assume that there should be no particles outside the 
ballistic cone defined by the maximal velocity of the particles. 
In 1935 Davydov proposed to use the telegraph 
equation, which contains additional second order time derivative, to address the existence of the ballistic cone \cite{davydov1934,bakunin2003}. 
As with the diffusion equation, the telegraph equation was discovered much earlier by Kirchhoff and Heaviside in the 
context of electric current transmission through a conducting line. Around 1950  it was demonstrated that the telegraph equation could be 
derived from the random walk model proposed by Taylor \cite{goldstein}. The next milestone in the development of the modern random walk theory 
was due to Montroll and Weiss who introduced the continuous time random walks model (CTRW)\cite{montroll1965,scher1975}. 
The main innovation of that model is that a particle has to wait for a random time before moving to another point. 
This model provided the framework necessary for describing anomalous diffusion with the spreading of particles  \textit{slower} than in 
the Brownian diffusion, a process that was named ``subdiffusion''.

\textcite{richardson} pointed out the possibility of the anomalous diffusion 
in turbulent flows,  where particles spread \textit{faster} than in normal diffusion, and referred to as ``superdiffusion''. To accommodate for superdiffusive transport, the random walk model was modified  to allow  
particles to perform very long excursions. To step beyond the premises of the central limit theorem (CLT), 
slow decaying functions with power law tails and diverging second moment were used as 
the distributions of the excursion lengths.  
The  scaling properties of  the corresponding particle distributions were found to be different from those 
of the standard Brownian diffusion and thus required a new mathematical apparatus. At this point, a link between the superdiffusion
 and  L\'{e}vy stable distributions \cite{levy1937,gnedenko1954} was established. 
The random walk model with walkers covering long distances instantaneously received the 
name of \textit{L\'{e}vy flight} \cite{mandelbrot1982}. In its simplest schematization, this stochastic process could
drive a particle over  very long distance in a single motion event, that is called ``flight'' (although, in fact, it is a \textit{jump}),
so that the mean squared flight length is infinite \cite{shlesinger19862}. 
Similar to the concept of Brownian diffusion, L\'{e}vy flights served well to describe different transport phenomena. 
However, L\'{e}vy flights have the same trait of infinite propagation 
speed as the diffusion equation.
In addition, the  distribution of the particles performing  L\'{e}vy flight has a divergent second
and all higher moments\footnote{It is not correct, however, to think of 
the L\'{e}vy flight as an abstract mathematical formalism. 
The mechanisms leading to the dispersion of the observable 
of interest may not be related to a physical motion of an entity in Euclidean space.
For example, it may be caused by long-range interactions \cite{Barkai2003a},  
or by a nontrivial ``crumpled'' topology of a
phase (or configuration) space of polymer systems \cite{Sokolov1997,Brockmann2003} and 
small-world networks \cite{Kozma2005}, or by
spectral characteristics of 
disordered media, amorphous materials, and glasses \cite{Klauder1962,Zumofen94};
see  review by \textcite{bouchaud1990} 
for more information and other examples where the L\'{e}vy flights are of relevance.}. This poses a significant difficulty in relating L\'{e}vy flight models 
to experimental data, especially when analyzing the scaling of the measured moments in time.
Akin to the Taylor model, the L\'{e}vy flight 
model was then equipped with a finite velocity of moving particles and therefore produced distributions which are 
confined to ballistic cones and thus have finite moments. As a contrast to the flight process with instantaneous jumps, 
the name \textit{L\'{e}vy walk } was coined by 
\textcite{shlesinger1982}. The aim of this review is to show how versatile and powerful the 
concept of L\'{e}vy walks is in describing a wide spectrum of physical and biological processes involving 
stochastic transport.

In order  to orient the reader in the existing literature on 
random walks in the context of anomalous diffusion, we would like to mention several  
monographs which can serve as a good introductory material to continuous time random walks \cite{klaftersokolov2011}, 
anomalous diffusion, L\'{e}vy flights and subdiffusion \cite{montrollshlesinger1984, havlin1987, bouchaud1990, isichenko1992, 
metzlerklafter2000, metzlerklafter2004}, and some reviews on particular applications of these formalisms \cite{balescu2005, bardou2001}. 

\subsection{L\'{e}vy stable laws}\label{stable_laws}
One of the fundamental theorems in probability theory is the central limit theorem. It states that 
the sum of independent identically distributed random variables with a finite second moment is  a 
random variable with the distribution tending to a normal distribution as the number of summands increases. 
It has a history of development spanning several hundred years, from initial considerations by Laplace and 
Poisson at the end of the $18^{\text{th}}$ century to the stage of rigorous analysis by Markov, Chebyshev, 
Lyapunov, Feller, L\'{e}vy and others in the beginning of the $20$th century, see  \textcite{fischer2010} 
for historical overview. 
Remarkably, the CLT can be cast into the dynamical problem of a particle hopping at random 
distances. The sum of all displacements (independent random variables) will then determine the final position of a particle (their sum). 
As a result, the distribution of particle's position is normal (or Gaussian) 
if the second moment of the displacement length distribution is finite.  
Normal distributions are also known to be stable distributions meaning 
that the sum (or, more generally, a linear combination with positive weights) of two independent random variables has 
the same distribution (up to a scaling factor and shift). Around 1920, Paul L\'{e}vy showed that there are other 
stable distributions which now bare the name of L\'{e}vy alpha-stable distributions. In particular, 
they have power law tails and  diverging  second moments. The generalized central limit theorem (gCLT) was then formulated to state 
that the sum of identically distributed random variables with distributions having power law tails converges to 
one of the L\'{e}vy distributions. We can now look at the total displacement 
of a particle whose individual hops are distributed as a power law. The position of the particle after many hops will be 
described, according to the gCLT, by a L\'{e}vy distribution, see e.g. \cite{uchaikin2003}. That is why such 
random walks are also known as L\'{e}vy flights. It has been found that a big variety of natural and man-made phenomena 
exhibit power law statistics \cite{bou1995,uchaikin1999,clauset2009}. While relating the empirical data 
to the theoretical models with L\'{e}vy distributions, it became clear that the model solutions could not be characterized 
by the second moment: Like every individual jump in a sequence, the distribution of particle's final positions has
an infinite variance. 

One of the most straightforward ways to resolve this inconsistency is to regularize the power-law distributions
by truncating them at large values \cite{mantegna1994}. That would make the moments of the distribution finite while still 
retaining some properties of the power-law distributions for intermediate values. However, the truncation introduces
a certain arbitrariness and, as a phenomenological procedure, it 
can not be always justified  in a particular physical (or economical, biological, etc.) context. 

Importantly, there is an alternative way to remedy the problem of divergent moments. 
A fundamental property of having a finite velocity while moving 
couples the displacement of a walker and time it takes to cover the corresponding distance and puts 
a larger time-cost to a longer displacement. 
In the simple picture of a hopping particle, that would mean that at any moment of time the position of the
particle after many hops is bounded by the ballistic cone with the fronts matching the maximal 
particle velocity multiplied by the observation time. 
In between these fronts, the long displacements of the particle would still exist, 
as necessary for the L\'{e}vy-like statistics,
but all moments of the distribution of particle's position will be finite for any given time \cite{shlesinger1986}. 

We hope that at this point we already convinced the reader that random walks
is an appropriate language to describe the stochastic transport phenomena. 
We now proceed to introduce the theoretical framework of continuous-time random walks \cite{klaftersokolov2011} 
and describe how it changes when the finite velocity of walkers is taken into account. 
Below, we mainly focus on one-dimensional
systems (some open problems concerning the generalization to higher dimensions are 
mentioned in the Outlook section).

\subsection{Continuous time random walks}\label{ctrw}
Consider a random motion of passive particles in homogeneous media. We are interested in the macroscopic
behavior of the density of particles $P(x,t)$ as a function
of space and time. 
Each particle can make instantaneous jumps  to the left or  to the right with equal
probabilities. The probability density function (PDF) of the
jump lengths $x$, $g(x)$, is chosen to be symmetric $g(x)=g(-x)$ and independent of
the starting point. Before making a jump, a particle waits for a
time $\tau$ defined by another PDF
$\psi(\tau)$, see Fig. 1(a). Both distributions are normalized:
$\int_{-\infty}^{\infty}g(x)dx=1$ and
$\int_{0}^{\infty}\psi(\tau)d\tau=1$. These two
functions determine the macroscopic properties of the transport process. 
In the standard continuous time random walk model, random variables $x$ and $\tau$ are independent from each other.
We can define the survival probability $\Psi(t)$, that is the probability for a particle not to
jump away until time $t$, as
\begin{equation}
\Psi(t)=1-\int\limits_{0}^{t}\psi(\tau)d\tau.
\label{survival}
\end{equation} 
The first transport equation governs the outgoing flow of particles $Q(x,t)$, which defines
how many particles leave the point $x$ per unit of time. The equation
connects the flux at the current point in space and time to the flux from all neighboring points in the past \cite{klafter1980}:
\begin{equation}Q(x,t)=\int\limits_{-\infty}^{+\infty}g(y)\int\limits_{0}^{t}\psi(\tau)Q(x-y,t-\tau)dy
d\tau
+P_{0}(x)\psi(t).\label{Qctrw}
\end{equation}
It is time for a particle to leave from the point $x$ 
if its waiting time $\tau$ has elapsed, 
which is taken care of via the multiplication by $\psi(\tau)d\tau$. 
The particle could arrive to the point $x$ time $\tau$ ago from some other 
point $x-y$ by making a jump of length $y$ with probability density $g(y)$. 
We next integrate over all possible waiting times and jump distances. 
The last term on the right hand side assumes that at the moment of time $t=0$ 
particles had an initial distribution $P_{0}(x)=P(x,t=0)$. 
Particles  gradually leave their initial spots 
according to the waiting time distribution. 
We also assume that all particles were introduced to the system 
at $t=0$ and the probability density of making the first jump is given by $\psi(\tau)$. 
The situation is different if the particles have some pre-history. 
In that case, the probability distribution of the first waiting time is in general different from $\psi(\tau)$ \cite{haus1987}, 
see Section \ref{memory_effects} for more detailed discussion.

The next step is to connect the outgoing flux in the past to the current density of particles at a given point in space and time, 
\begin{equation}P(x,t)=\int\limits_{-\infty}^{+\infty}g(y)\int_{0}^{t}\Psi(\tau)Q(x-y,t-\tau)dy
d\tau
+P_{0}(x)\Psi(t).\label{Pctrw}
\end{equation}
The density is a sum of outgoing particles from all other points
at different times, weighted by the jump length
probability, provided the particles survived for a time $\tau$ after their arrival to $x$
at $t-\tau$. The last term on the right hand side of Eq.
(\ref{Pctrw}) accounts for the particles which stay in their starting points until the observation time $t$.

The above set of equations specifies the standard continuous-time random walk (CTRW) process with an
arbitrary initial condition. These integral equations can be 
solved by using the combination of Fourier (with respect to space) and 
Laplace (with respect to time) integral transforms \cite{fourierlaplace}. 
We exploit the fundamental property of these transforms which turns  convolution integrals into  products 
in the Fourier-Laplace space. 
We use $k$ and $s$ to denote coordinates in Fourier and Laplace space, respectively. 
By explicitly providing the argument of a function, we will distinguish 
between the normal or transformed space, for example $\psi(\tau)\rightarrow\psi(s)$ and $g(x)\rightarrow g(k)$. 
As a result, the solution for the density of particles is given by 
the Montroll-Weiss equation \cite{montroll1965, klafter1987}:
\begin{equation}
P(k,s)=\frac{\Psi(s)P_{0}(k)}{1-\psi(s)g(k)}, \label{Pksctrw}
\end{equation}
where $\Psi(s)=[1-\psi(s)]/s$. This solution allows us to reduce the pair of original equations (\ref{Qctrw})-(\ref{Pctrw}) to a single equation for the density $P(x,t)$:
\begin{equation}P(x,t)=\int\limits_{-\infty}^{+\infty}g(y)\int\limits_{0}^{t}\psi(\tau)P(x-y,t-\tau)dy
d\tau +\Psi(t)P_0(x).\label{Psingle}
\end{equation}
In our derivation, we intentionally used an intermediate step of introducing the flow of particles $Q(x,t)$. For a more general initial condition with a non-trivial distribution of particles over their lifetimes, it provides the proper way to obtain the corresponding transport equation for the density of particles \cite{zaburdaev20082}. In addition, a very similar set of equations will be used for the models that incorporate velocity of particles.

Below we will frequently use the notion of the \textit{propagator} (and sometimes, depending on the context, Green's function).
It is the solution of the transport equation for the
delta-like initial distribution\footnote{
A biology-oriented definition of the propagator was nicely put by 
Ronald Ross, the Nobel laureate for medicine in 1902 \cite{Ross1905}:
``\textit{...suppose a box containing a million 
gnats were to be opened in the centre of a large plain, and that the 
insects were allowed to wander freely in all directions, how many 
of them would be found after death at a given distance from the 
place where the box was opened?}''}  $P_{0}(x)=\delta(x)$.
From Eq. (\ref{Pksctrw}) the propagator can be identified as
\begin{equation}
G(k,s)=\frac{\Psi(s)}{1-\psi(s)g(k)}. \label{Gksctrw}
\end{equation}
Then, for any arbitrary initial distribution, the solution
will be given by the convolution  of an initial profile
with the propagator,
\begin{equation}\label{nG}
    P(x,t)=\int\limits_{-\infty}^{\infty}G(x-y)P_{0}(y)dy.
\end{equation}

The above Eqs. (\ref{Pksctrw}) and (\ref{Gksctrw}) give a formal 
solution of the transport equation. When given functions $g(y)$,
$\psi(\tau)$, and $P_{0}(x)$, one should  find their Fourier and
Laplace transforms, insert them into  Eq. (\ref{Pksctrw}), and calculate their
inverse transform. Unfortunately, in general it is almost
impossible to find this inverse transform
analytically. 
Instead an asymptotic analysis for large time and space scales can be employed. 
To proceed, we
must specify the probability densities $g(y)$ and $\psi(\tau)$.
Motivated by applications and also mathematical convenience, we choose a power law form of these PDFs. By varying the exponent of their power law tails, 
different regimes of diffusion can be accessed. 
Assume the following particular forms:
\begin{eqnarray}
\psi(\tau)&=&\frac{1}{\tau_0}\frac{\gamma}{(1+\tau/\tau_0)^{1+\gamma}},\quad \gamma>0;\label{psi}\\
g(x)&=&\frac{\Gamma\left[\beta+1/2\right]}{x_0\sqrt{\pi}\Gamma \left[\beta \right](1+(x/x_0)^2)^{\beta+1/2}}, \quad \beta>0.\label{gx}\end{eqnarray}
The exact details of these distributions are not qualitatively important in the asymptotic limit. 
Crucial are their power law tails which determine the behavior of their moments. 
All other details will be absorbed into constant pre-factors; yet we will keep track of those for the sake of completeness.

There are two important moments of these PDFs, the mean squared jump length,
\begin{equation}
\left<x^2\right>=\int_{-\infty}^{\infty}x^2g(x)dx,\label{second_moment}
\end{equation}
and the mean waiting time,
\begin{equation}\left<\tau\right>=\int_{0}^{\infty}\tau\psi(\tau)d\tau.\label{average_time}
\end{equation}
If these moments exist, for the chosen functions in Eqs.(\ref{psi})-(\ref{gx}), they are given by simple expressions:
\begin{equation}
\langle\tau\rangle=\frac{\tau_0}{\gamma-1}, \gamma>1;\quad \langle x^2 \rangle=\frac{x_0^2}{2(\beta-1)}, \beta>1.
\label{moments}
\end{equation}
When
both quantities are finite, the resulting transport equation reduces to the
standard diffusion equation with a diffusion coefficient
$D=\left<x^2\right>/(2\left<\tau\right>)$,
\begin{equation}
\frac{\partial P}{\partial t}=D\triangle P(x,t).
\label{diffusion_equation}
\end{equation} 
It is easy to demonstrate by assuming that the typical
spatial and temporal scales of interest are significantly
larger then $\langle x^2 \rangle $ and $\langle \tau \rangle$ respectively. It is then possible to expand the expression under the integral in Eq. (\ref{Psingle}) in a Taylor series with respect to $y$ and $\tau$ yielding the diffusion equation above; its propagator is the well known Gaussian distribution:
\begin{equation}
P(x,t)=\frac{1}{\sqrt{4\pi Dt}}e^{-\frac{x^2}{4Dt}}
\label{diffusion_solution}.
\end{equation}
The second moment of this distribution is the mean squared displacement (MSD) which scales linearly with time:
\begin{equation}
\left<x^{2}(t)\right>=\int\limits_{-\infty}^{\infty}x^2P(x,t)dx=2Dt.
\label{diffusion_msd}
\end{equation}
Another important property of the diffusion process is the scaling of the density profile. 
As we will see for different models and regimes of random walks, in the limit of large times, 
the propagator may be represented as
\begin{equation}
G(x,t)=t^{-\alpha}F\left(\frac{x}{t^{\alpha}}\right),
\label{selfsimilarity_ctrw}
\end{equation}
where $F(\xi)$ is a scaling function (for example, Gaussian, in the case of normal diffusion)
and $\alpha$ is a model specific scaling exponent (with $\alpha = 1/2$ in the case of normal diffusion).
Such a functional form suggests a scaling variable $\xi=x/t^{\alpha}$, meaning 
that a characteristic spatial scale on which the density changes, 
$\bar{x}$, scales with time as $\bar{x}\propto t^{\alpha}$.
From the  solution given by Eq. (\ref{diffusion_solution}), we see that the width of the cloud of particles grows 
as $\overline{x}\propto t^{1/2}$. 

The asymptotic limit $x,t\rightarrow \infty$ corresponds to the dual
transition $k,s\rightarrow 0$ in the Fourier-Laplace domain. That is why, instead of the full
Fourier and Laplace transforms of $g$ and $\psi$ in Eq. (\ref{Pksctrw}), their
expansion in Taylor series with respect to small $k$ and $s$ can
be used\footnote{Taylor series are normally
understood as an expansion in integer powers of the argument. In
our case the leading terms of expansion with respect to the small
$k$ and $s$ may involve fractional powers. Below we use the notion
of Taylor series in this extended sense.}. In the Fourier-Laplace coordinates, the leading terms of the expansion for the chosen power-law
functions are \cite{prudnikov1986}:
\begin{eqnarray}
g(k)\!\!&=&\!\!
1\!-\!\frac{x_0^2}{\beta-1}\frac{k^2}{4}-\frac{x_0^{2\beta}\Gamma[1-\beta]}{2^{2\beta}\Gamma[1+\beta]}|k|^{2\beta}\!\!+\!O(k^{2+2\beta})
\label{gk}\\
\psi(s)\!\!&=&\!\!1\!-\!\frac{\tau_0}{\gamma-1}s-\tau_0^\gamma\Gamma[1-\gamma]s^{\gamma}\!\!+\!O(s^{1+\gamma})\label{psis}
\end{eqnarray}
In the marginal cases, when $\gamma$ or $\beta$ have values where one of the moments starts to diverge (for example, $\gamma,\beta=1$), 
there are logarithmic correction terms appearing in this expansion. 
We will not consider these cases here (see for example \textcite{Zumofen1993, chukbar1995} for more information). 
The finite waiting time and the mean squared jump distance correspond to $\gamma,\beta > 1$. 
In this case, the first two terms in the above expansions are dominant for small $k$ and $s$. The pre-factors in front of $k^2$ and $s$ can be recognized as the mean squared jump distance and the mean waiting time from Eq.(\ref{moments}) respectively. 
By substituting them into the formal solution, Eq. (\ref{Pksctrw}), we can rewrite it as the diffusion equation in the 
Fourier-Laplace space, or simply compute the inverse transforms and obtain the Gaussian distribution Eq. (\ref{diffusion_solution}). 
For $\gamma<1$ and $\beta<1$, terms with fractional powers dominate over linear and quadratic terms in Eqs. (\ref{psis}) and (\ref{gk}), 
respectively. By substituting them in Eq. (\ref{Pksctrw}) we obtain the following
equation in the Fourier-Laplace space 
\begin{equation}
s^{\gamma}\Gamma[1-\gamma]P(k,s)=-|k|^{2\beta}K'P(k,s)
+ s^{\gamma-1}\Gamma[1-\gamma]P_{0}(k),\label{levyflight_equation_ks}
\end{equation}
where $K'=(x_0/2)^{2\beta}\Gamma[1-\beta]/\left(\tau_0^{\gamma}\Gamma[1+\beta]\right)$.
After returning to the original space-time domain, we obtain an
equation with integral nonlocal operators:
\begin{equation}
\frac{\partial}{\partial
t}\int\limits_{0}^{t}\frac{P(x,\tau)}{(t-\tau)^{\gamma}}d\tau=K
\int\limits_{-\infty}^{\infty}\frac{P(y,t)}{|x-y|^{2\beta+1}}dy
+\frac{P_{0}(x)}{t^{\gamma}}, \label{levy_flight_equation_xt}
\end{equation}
where $K=(x_0)^{2\beta}\Gamma[\beta+1/2]/\left(\tau_0^{\gamma}\sqrt{\pi}\Gamma[\beta]\right)$.
The integral nonlocal operators in the above equation are the
{\em fractional derivatives} (the integral on the right hand side is
understood as the principle value) \cite{kilbas2006,podlubny1999,samko1993,west2014}. 
The notion of fractional derivative allows us to rewrite the asymptotic
balance equation in a compact form of the fractional diffusion equation \cite{Saichev1997,metzlerklafter2000,barkai2002}:
\begin{equation}\label{FDE}
\frac{\partial^\gamma P}{\partial
t^{\gamma}}=K_{\beta,\gamma}\triangle^{\beta}P+\frac{P_0(x)}{t^{\gamma}}.
\end{equation}
This equation describes the stochastic spreading of a cloud of particles and besides the case of normal diffusion has several interesting regimes.
If the mean squared jump length is finite but the waiting times are anomalously long, 
$\left<\tau\right >= \infty$, the resulting dispersal is anomalously slow or subdiffusive.  
The typical width of the cloud scales as
$\overline{x}\propto t^{\gamma/2}$, with  $\gamma<1$. 
In the opposite case, when the average waiting time is finite but jumps are very long, $\left<x^2\right>=\infty$, 
the equation describes superdiffusion. The typical width of the distribution of particles scales as $\overline{x}\propto t^{1/2\beta}$. 
Finally, when long waiting times compete with long jumps, 
the scaling is defined by both distributions of waiting times and jump lengths, $\overline{x}\propto t^{\gamma/2\beta}$. 
The stochasticity of the transport process reveals itself in the ``forgetting'' of the initial
distribution and the tendency of the particles' density to the
universal self-similar profile of the corresponding propagator, Eq. (\ref{selfsimilarity_ctrw}), with $\alpha=\gamma/2\beta$.

Before closing this section, let us have a closer look at the superdiffusion regime. 
The jump length distribution has a diverging second moment ($\beta<1$) 
whereas the mean waiting time is finite. Therefore the leading terms in the expansion of $\psi(s)$ in Eq. (\ref{psis}) could be 
written as: $\psi(s)\simeq 1+s\left<\tau\right>$. 
The propagator in the Fourier-Laplace space is then given by
\begin{equation}
G(k,s)=\frac{1}{s+K_{\beta}|k|^{2\beta}}\label{levy_flight_ks}
\end{equation}
with $K_{\beta}=(x_0/2)^{2\beta}\Gamma[1-\beta]/\left(\left<\tau\right>\Gamma[1+\beta]\right)$.
By performing the inverse Laplace transform we obtain:
\begin{equation}
G(k,t)=\exp{\left(-K_{\beta}|k|^{2\beta}t\right)}.\label{levy_flight_tk}
\end{equation}
This expression is the Fourier transform (a characteristic function) of the symmetric L\'{e}vy 
distribution $L_{\kappa}[x,\sigma(t)]$, which describes the distribution of the sum of independent and identically distributed 
variables with power law PDFs \cite{uchaikin2003}. 
Here $\kappa=2\beta$ is  the L\'{e}vy index and $\sigma^{\kappa}=K_{\beta}t$ is the scaling parameter. 
For some particular values of $\kappa$, the L\'{e}vy distribution has an analytical expression in coordinates space, 
such as Cauchy distribution ($\kappa=1$), L\'{e}vy-Smirnov ($\kappa=1/2$) or Holtsmark distribution ($\kappa=3/2$) \cite{uchaikin2003, klaftersokolov2011}. 
The key feature of all of these distributions is the asymptotic power law 
tail $G(x,t)\sim t/|x|^{\kappa+1}$ \cite{klaftersokolov2011, chukbar1995}. 
In the scaling relation given by Eq.(\ref{selfsimilarity_ctrw}), the scaling function $\Phi$ is 
also given by the L\'{e}vy distribution with $\alpha=1/\kappa$. Because of this intimate relation of the particles� density to the L\'{e}vy distribution, the model of random walks with instantaneous jumps received the name of L\'{e}vy flight \cite{mandelbrot1982}\footnote{\textcite{shlesinger19862} proposed an alternative schematization of 
L\'{e}vy flights: The are no waiting pauses and the duration of each step is constant so that the velocity of a flight is proportional
to the step length drawn from a L\'{e}vy distribution.}. We can now ask about the behavior of the second moment of the density profile given by the L\'{e}vy distribution. 
This and higher moments in the Fourier (or the Fourier-Laplace) space are:
\begin{equation}
\left<x^n\right>=\int\limits_{-\infty}^{\infty}x^nP(x,t)dx=(-i)^n\frac{d^n}{dk^n}\left.P(k,t)\right|_{k=0}.
\label{moments_formula}
\end{equation}
By substituting Eq. (\ref{levy_flight_tk}) into this formula, 
we immediately see that all moments with $n \geqslant 2$ diverge. 
It is thus impossible to compute the MSD as a function of time for a 
particle performing L\'{e}vy flights for a simple reason: it does not exist. 
Already after the first jump all 
particles will be distributed as $g(x)$ and this distribution has an infinite second moment. 
This feature of L\'{e}vy flights is often referred to as a shortcoming of the model 
when applied to physical processes, in which one expects regular behavior of moments \cite{mantegna1994}. 
%
However, as we have already mentioned, physical intuition points to the possibility to modify the L\'{e}vy flight model 
assuming that longer jumps must have a higher cost; there has to be a certain coupling between the length of the flight and its duration. The simplest coupling is the finite velocity of particles, 
when the time to accomplish a flight is linearly proportional to its length. 
As we show below, the introduction of the finite velocity of particles into the L\'{e}vy flight model 
retains the anomalous character of the transport process while regularizing the behavior 
of moments of the particle density. We believe that the compliance of the new model with physical intuition and ability to account for the velocity of particles explains its success in applications.

\section{L\'{e}vy walks}\label{finite_velocity}
\begin{figure*}
\center
\includegraphics[width=1.0\textwidth]{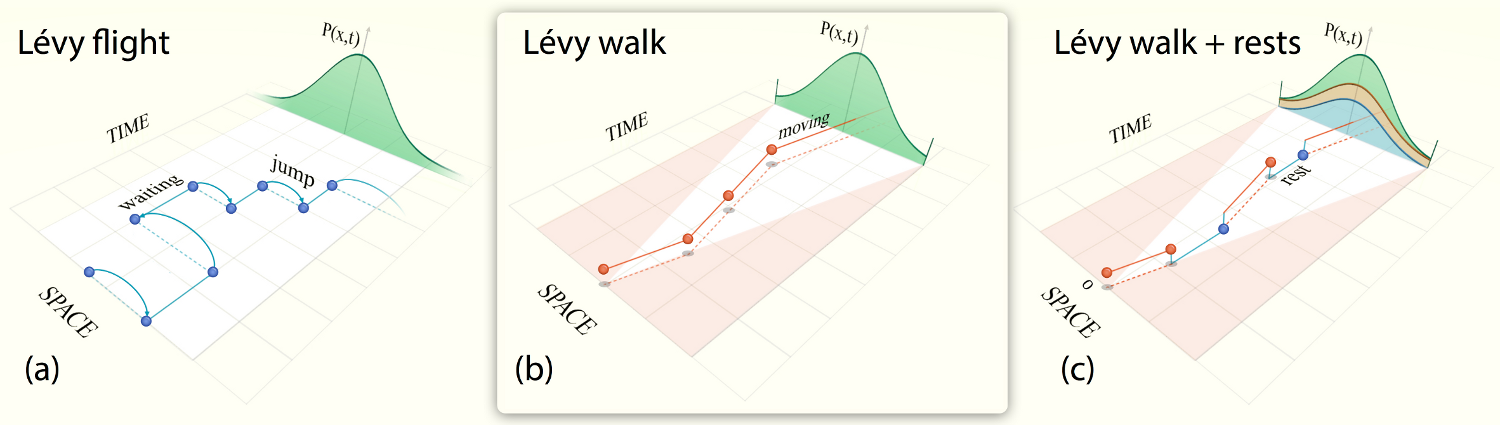}
\caption[Models]
{(Color online) Random walk models of superdiffusion. 
(a) L\'{e}vy flight: A particle performs instantaneous jumps 
alternated with waiting pauses. 
The length of jumps and durations of waiting events are independent random variables. 
The resulting PDF $P(x,t)$ is not local in space at any moment of time. 
(b) L\'{e}vy walk: A particle moves with a constant velocity for a random time and then, at a turning point, 
instantaneously chooses a new direction and moves again. 
In this basic model particle's velocity can assume two values $\pm v$ only. 
As a result the length and duration of each event of ballistic motion  are coupled. 
There is a ballistic cone $x_{\text{f}}=\pm vt$ beyond which no particle can go (shaded area). 
As a result the PDF is bounded in space and its fronts are marked by two delta peaks. 
(c)  L\'{e}vy walk with rests: Ballistic flights of a particle are alternated with pauses during which the particle does not move. 
At any instant of time the statistical ensemble consists of two fractions of particles, flying  and resting. 
The total PDF is the sum of the two.}
\label{figt1}
\end{figure*}

We will discuss several ways to introduce a coupling 
between jump length and time in the upcoming sections. 
Here we start with the conventional dynamical  coupling of 
the particle position and current time via a constant velocity of the  particle. 

There are two closely related models which incorporate finite velocity of random walkers. 
The first one is a direct modification of the CTRW model: After spending its waiting time, a random walker 
does not jump  instantaneously but instead moves with a constant speed to its destination \cite{klafter1994, zaburdaev2002}, see Fig. 1(c). 
Long excursions are still responsible for the anomalously fast diffusion, but now there are well-defined ballistic fronts and 
behavior of all moments is regularized. 
In the second model, waiting periods are eliminated and the particle is always  
on the move \cite{shlesinger1986, klafter1994, Zumofen1993}, see Fig. 1(b). 
The distance of ballistic flights is distributed randomly and each 
flight is performed with a finite speed. 
At the end of the flight, the particle randomly changes direction. 
The latter model is what historically received the name of L\'{e}vy walks. 
Note that the L\'{e}vy walk model has only one distribution function to 
parametrize the motion of the particles as it discards waiting. 
This minimalistic setup remains the most popular model in modern applications. 

\subsection{L\'{e}vy walk model}\label{levy_walk}
The formulation of the microscopic model is very similar to that by Pearson cited in the introduction. 
A particle moves on a straight line with a fixed speed for some random time. 
At the end of the excursion, the particle randomly chooses a new direction of motion 
and moves for another random time with the same speed, see Fig. 1(b). 
There are only two characteristics of this model, that are the PDF of the duration of movements, which we will 
denote by $\psi(\tau)$, and the speed $v$ of the particles. 
Despite its simplicity, this model is able to describe various regimes of stochastic transport,
from classical to ballistic superdiffusion. 

We now derive the transport equations of the L\'{e}vy walk model. 
First we introduce the frequency of velocity changes at a given point (analogue of the flux of particles from a given point in the CTRW model),
\begin{eqnarray}\nu(x,t)&=&\int\limits_{-\infty}^{\infty}dy\int\limits_{0}^{t}\phi(y,\tau)\nu(x-y,t-\tau)d\tau \nonumber\\
&+&\delta(t)P_0(x).\label{nu}
\end{eqnarray}
Here we incorporated a {\em coupled} transition probability density $\phi(y,t)$ which takes care of the fact that only particles 
flying from $x-v\tau$ and $x+v\tau$ can reach $x$ in time $\tau$ and change the direction of their velocities after the flight time of $\tau$:
\begin{equation}
\phi(y,\tau)=\frac{1}{2}\delta(|y|-v\tau)\psi(\tau)
\label{phi_coupled}
\end{equation}
Therefore, a particle changes its velocity if it is at the end of the flight of duration $\tau$ which originated from $x\pm v\tau$. We assume that at $t=0$ all particles at once choose new velocities and hence the second term on the right hand side of Eq.(\ref{nu}) contains a delta-function (note a difference to a gradual leaving of particles from their waiting positions for the CTRW model, Eq.(\ref{Qctrw})).  

To calculate the actual amount of particles at a given point, we write
\begin{equation}P(x,t)=\int\limits_{-\infty}^{\infty}dy\int\limits_{0}^{t}\Phi(y,\tau)\nu(x-y,t-\tau)d\tau,\label{Plevywalk}
\end{equation}
where
\begin{equation}
\Phi(y,\tau)=\frac{1}{2}\delta(|y|-v\tau)\Psi(\tau). \label{Phicoupled}
\end{equation}
is the probability density to travel a distance $y$ and remain in the state of flight (note that with respect to $\tau$ Eq. (\ref{phi_coupled}) has the meaning of the probability density whereas Eq. (\ref{Phicoupled}) is probability). Therefore, a particle is at the point $(x,t)$ if it has started some time $\tau$ ago at $x\pm v\tau$ and is still in the state 
of the flight, taken care of by multiplication with $\Psi(\tau)$, Eq.(\ref{survival}). Note that in Eq.(\ref{Plevywalk}), the influence of the initial condition appears only indirectly, through the frequency of velocity changes $\nu(x,t)$ (cf. Eq.(\ref{Pctrw}) for the CTRW model with an extra term for immobile particles survived from the start). By taking the limit $t\rightarrow 0+$ we can substitute $\nu(x,t)$ by $P_{0}(x)\delta(t)$ and recover $P(x,t)\rightarrow P_{0}(x)$.

The equations can be solved by using the combined Fourier-Laplace transform, but an additional technical complexity due to the coupling of space and time variables occurs  \cite{klafter1987,Zumofen1993}. We resolve it by using the shift property of the Laplace and Fourier transforms; as a result the corresponding Laplace transformed functions hold a linear combination of Fourier/Laplace coordinates $s\pm ikv$ as its argument:
\begin{equation}
P(k,s)=\frac{\left[\Psi(s+ikv)+\Psi(s-ikv)\right]P_0(k)}{2-\left[\psi(s+ikv)+\psi(s-ikv)\right]}\label{Plevywalkks}
\end{equation}
This is a formal solution of the problem and, as in the case of L\'{e}vy flights, the next step is to perform 
the asymptotic analysis. 
Due to the simple ballistic  coupling $x=v\tau$, the possible scaling regimes of diffusion 
are governed by the power law tail of the flight time distribution, which we again take in the form given by  Eq. (\ref{psi}). 

\subsubsection{Telegraph equation}\label{telegraph}
If the mean squared flight distance is finite, $\gamma>2$,  the 
classical diffusion takes place in the asymptotic limit. 
However, the effects of finite velocity can be seen in this regime too. 
Consider for a moment an exponentially distributed flight time $\psi(\tau)=(1/\tau_0)\exp(-\tau/\tau_0)$. 
By taking its Laplace transform and substituting it into Eq. (\ref{Plevywalkks}), 
we can invert the Fourier and Laplace transforms to obtain the telegraph equation \cite{goldstein}:
\begin{equation}
\frac{\partial P}{\partial t}+\tau_0\frac{\partial^2P}{\partial t^2}=D\triangle P(x,t),
\end{equation}
where $D=v^2\tau_0$. On very short times, it describes almost ballistic spreading of particles. 
As time goes, ballistic fronts run away much faster than spreading of the diffusive evolution ($\overline{x}\propto t$ vs. $\overline{x}\propto t^{1/2}$, for $t\gg\tau_0$) which starts to dominate the central part of the density profile. Finite velocity ensures, however, that there are no particles beyond the ballistic fronts.

\subsubsection{Superdiffusion}\label{sub_ballistic}
As the flights get longer, $1<\gamma<2$, the mean squared flight length 
diverges (but the average flight time is still finite) and we turn 
to the regime of superdiffusion. By using the expansion from Eq. (\ref{psis}) for small $s$ 
and substituting its leading terms in Eq. (\ref{Plevywalkks}) we arrive at the similar answer as Eq. (\ref{levy_flight_ks}) for the propagator
\begin{equation}
G(k,s)\simeq\frac{1}{s+K_v|k|^{\gamma}}\end{equation}
with $K_v=\tau_{0}^{\gamma-1}v_0^{\gamma}(\gamma-1)\Gamma[1-\gamma]\sin(\pi\gamma/2)$. 
Several things to be noted here. 
After the inverse Fourier-Laplace transform we get the stable time-parametrized L\'{e}vy distribution 
with a scaling given by Eq.~(\ref{selfsimilarity_ctrw}) and $\Phi(\xi) = L_{\gamma}(\xi)$, $\alpha = 1/\gamma$. 
For $1<\gamma<2$ the cloud of particles spreads faster than classical diffusion but still slower than the running ballistic fronts. 
Therefore ballistic fronts do not appear in this analysis and affect (like in previous subsection) 
only the far tails of the particle density distribution. Nevertheless, the existence of fronts is 
crucial for the calculation of moments, as we show below. 
Now we take a closer look at what is happening at the ballistic fronts. 
Assume that at the moment of time $t=0$ we start with all particles initiating their flights at $x=0$. 
Ballistic fronts are formed mostly by the particles which are still in their 
very first flights. The probability to remain in the flight is $\Psi(t)$ and therefore 
we can write down the density of particles in the ballistic peaks or ``chubchiks'' \cite{klaftersokolov2011}
\begin{equation}
G_{\text{front}}(x,t)=\frac{1}{2}\Psi(t)\left[\delta(x-vt)+\delta(x+vt)\right]
\label{peaks}
\end{equation}
This gives the first approximation of the whole density of particles as a L\'{e}vy distribution sandwiched between 
two running ballistic delta-like peaks (see Fig.2). A more detailed understanding of the density can be achieved by 
using the so-called infinite density measure \cite{Rebenshtok2014,Rebenshtok2014a}, as we discuss in Section \ref{ID}.
\begin{figure}[t]
\center
\includegraphics[width=0.45\textwidth]{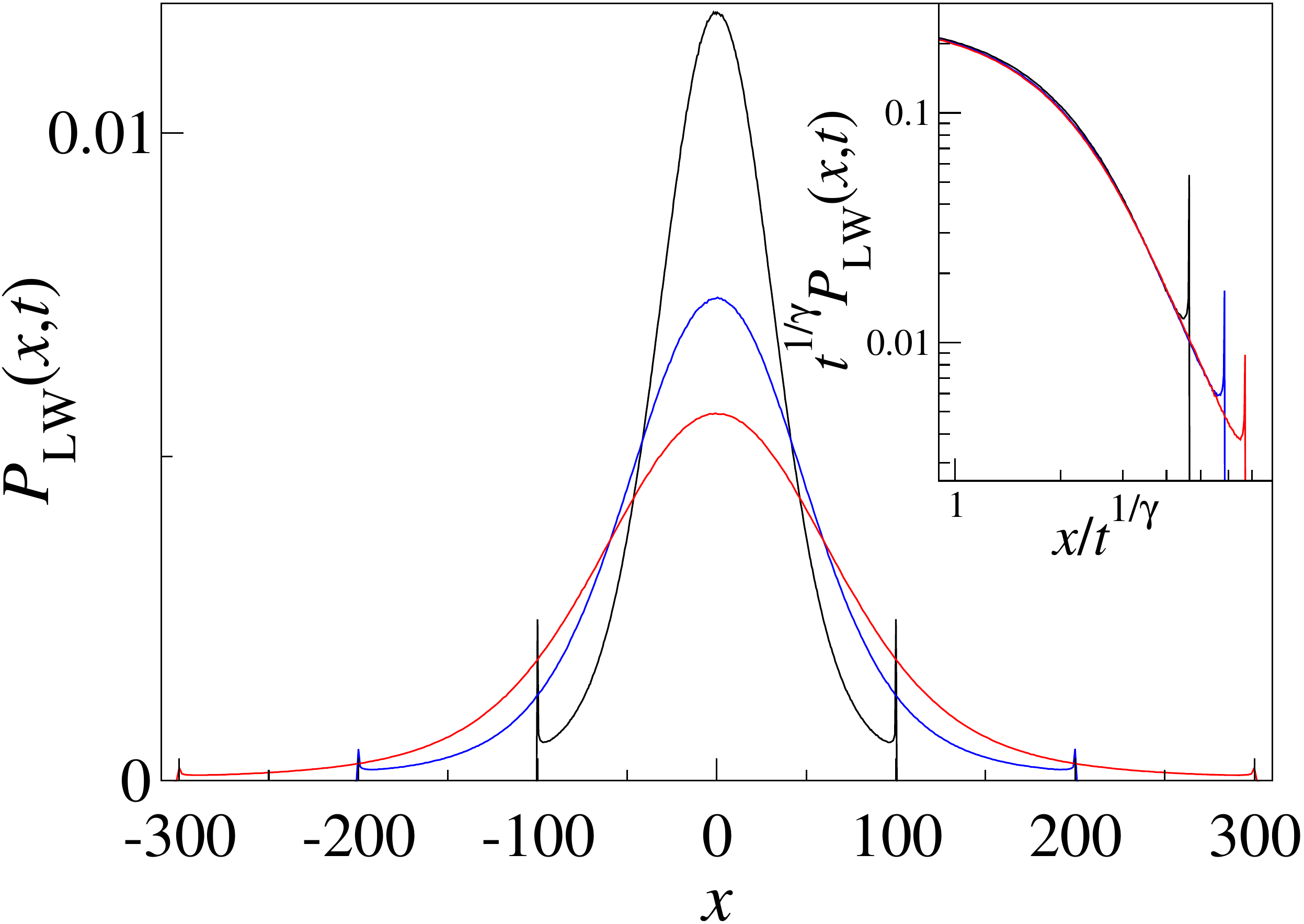}
\caption[Superdiffusive Levy walk]
{(Color online) Propagators of the superdiffusive  L\'{e}vy walk for different times. 
The propagators have a central part of the profile approximated by the L\'{e}vy distribution 
sandwiched between two ballistic  peaks. 
In the inset, the same curves are shown in double logarithmic scale after the rescaling as given on the axis labels. 
A characteristic linear slope on the log-log plot illustrates the power-law tails  of the density. 
The propagators were obtained by numerical simulations with $\gamma=3/2$, $\tau_0=1$, 
and all particles starting their flights at $t=0, x=0$ with velocities $v=\pm1$. 
The PDFs was sampled at $t=100$, $200$, and $300$ (the width increases with time).}
\label{figt2}
\end{figure}

\subsubsection{Ballistic diffusion}\label{ballistic}
In case of even longer flights $0<\gamma<1$, when the mean flight time diverges, 
the diffusion process changes dramatically. 
In the case of L\'{e}vy flight, the scaling $\overline{x}\propto t^{1/\gamma}$ for 
small $\gamma$ results in spreading which is faster than ballistic,
leading to an obvious conflict with the light front limitation in the L\'{e}vy walk setup. 
Clearly, the ballistic front is playing a crucial role here. 
An asymptotic expansion of the propagator [obtained from Eq. (\ref{Plevywalkks}) with $P_{0}(x) = \delta(x)$] has the following form:
\begin{equation}
G(k,s)\simeq\frac{(s+ikv)^{\gamma-1}+(s-ikv)^{\gamma-1}}{(s+ikv)^{\gamma}+(s-ikv)^{\gamma}}.
\label{levy_ballistic_ks}
\end{equation}
Now Fourier and Laplace variables appear 
in the same scaling $s\sim k$ and indicate the ballistic behavior. 
It was suggested to call the inverse Fourier-Laplace transform of $(s+ikv)^{\gamma}+(s-ikv)^{\gamma}$ 
as a fractional generalization of the substantial or material derivative operator,
$(v^{-1}\partial/\partial t \pm
\partial/\partial x)^{1/\gamma}$ \cite{sokolov2003}. 
In the ballistic regime there is again a technical difficulty to
find the inverse Fourier-Laplace transform. The ballistic case is special in that its
analytical solution can be found by the method discussed in Section \ref{exact_ballistic}. 
Here we just illustrate the shape of the propagator 
on a particular example $\gamma=1/2$ [for arbitrary $\gamma$ the answer is given by the Lamperti distribution \cite{lamperti1958, bel2005}, see also below, Eqs. (\ref{lamperti1}) and (\ref{lamperti})]:
\begin{equation}
G(x,t)=\frac{1}{\pi(v_0^2t^2-x^2)^{1/2}}.
\label{Glevywalkballistic}
\end{equation}
Figure 3 shows a $U$-shaped profile (for $\gamma\gtrsim 0.6$ the shape is $W$-like, see Fig. \ref{figt6} b) with a divergent density at the ballistic fronts \cite{Zumofen1993}. 
This divergence is, however, integrable and the total number of particles is conserved. 
Although the density profile is very different from the Gaussian profile of the classical diffusion, 
the ballistic diffusion remains a stochastic transport phenomena where the initial condition is gradually forgotten 
with time and the solution approaches the universal self-similar profile of the corresponding Green's function. 
\begin{figure}[t]
\center
\includegraphics[width=0.45\textwidth]{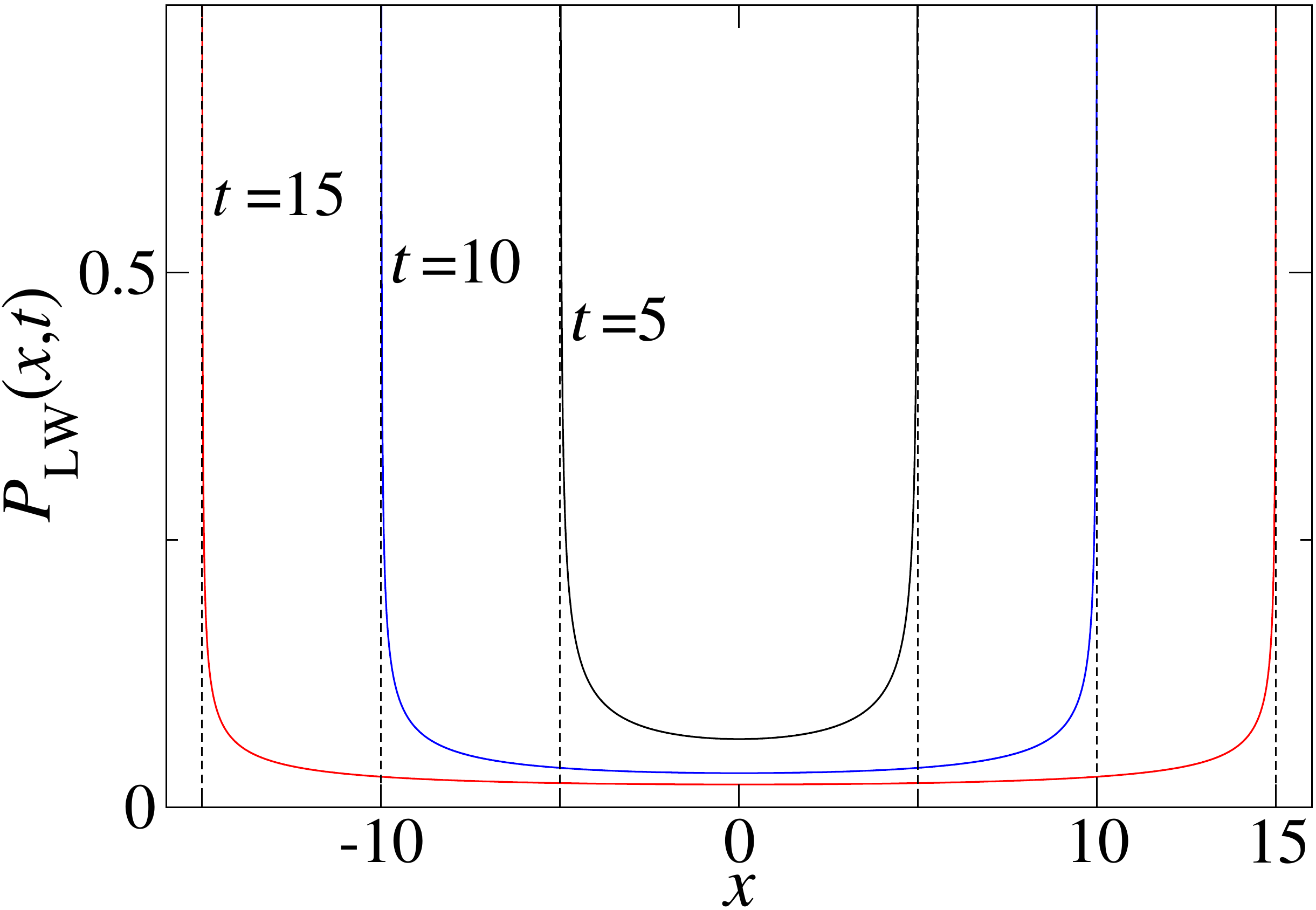}
\caption[Ballistic Levy walk]
{(Color online) PDF of the L\'{e}vy walk model in the ballistic regime. 
This plot shows  the density of particles,  Eq.(\ref{Glevywalkballistic}), at different moments of time. 
The parameters are $\gamma=1/2$, $v=\pm1$, $\tau_0=1$, and $P_{0}(x)=\delta(x)$. 
The densities have integrable divergences at the ballistic fronts 
due to the conservation of  the total amount of particles.}
\label{figt3}
\end{figure}

\subsubsection{Mean squared displacement and other moments}\label{scaling_section}
The PDFs of the L\'{e}vy walk in general do not possess a global scaling where the whole propagator can be 
represented in the form of Eq. (\ref{selfsimilarity_ctrw}), 
unlike the case of the Gaussian profile and normal diffusion. 
A clear example is the superdiffusive regime ($1<\gamma<2$), 
where the middle part of the profile scales as $\overline{x}\propto t^{1/\gamma}$, 
but, at the same time, the fronts exhibit the ballistic scaling. 
As a consequence, in L\'{e}vy walks the scaling exponents of the propagator and MSD are not the same; 
for a summary of these issues we refer to \textcite{schmiedeberg2009,Rebenshtok2014,Rebenshtok2014a}.  

\begin{figure}[t]
\center
\includegraphics[width=0.5\textwidth]{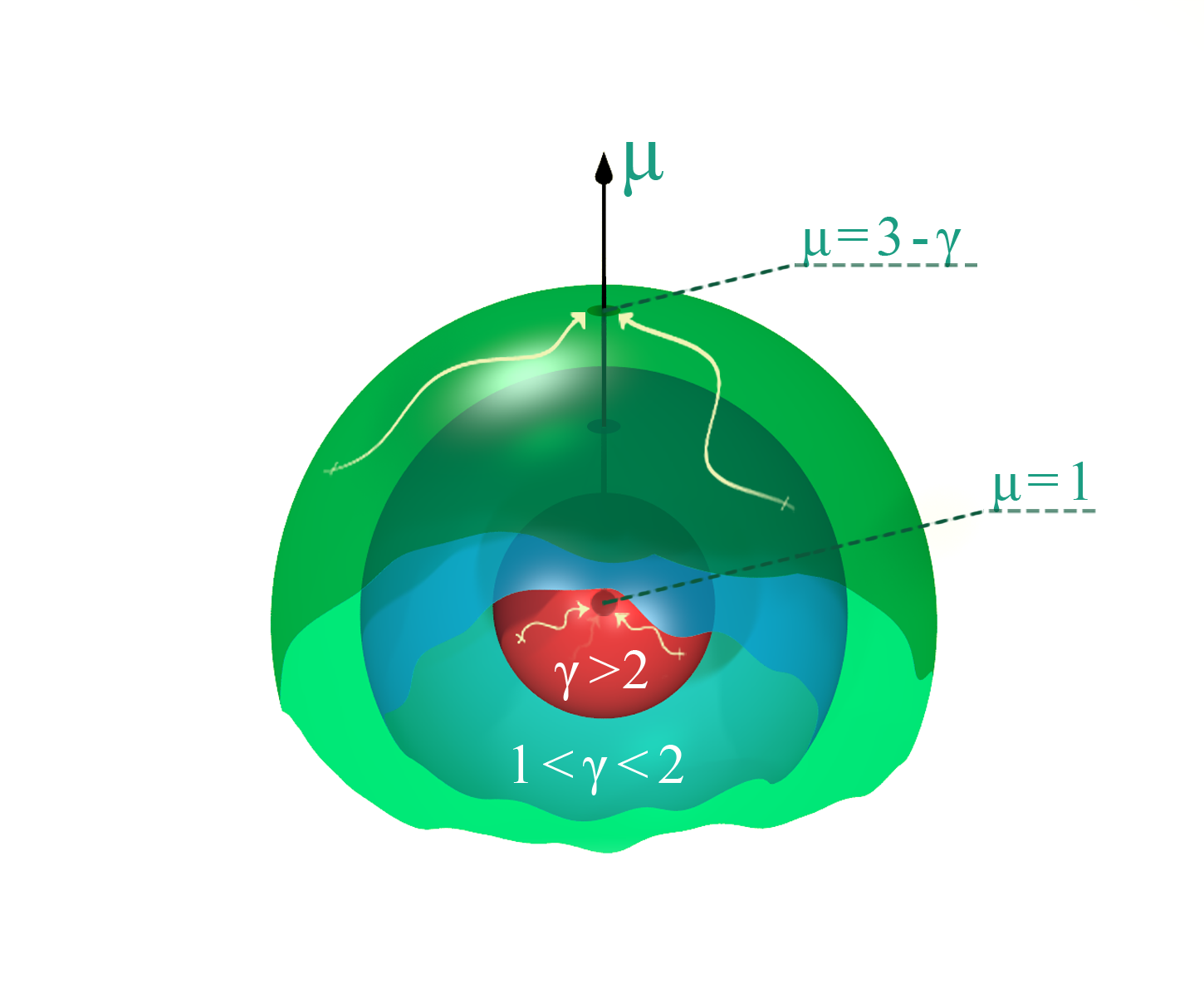}
\caption[scaling]
{(Color online) Scaling exponent of the mean squared displacement for L\'{e}vy walks, Eq.~(\ref{msd_scaling1}). 
When the second moment of the PDF $\psi(\tau) \sim t^{-\gamma-1}$ exists, $\gamma > 2$, the scaling 
$\langle x^2(t)\rangle\propto t^{\mu}$ is universal with the exponent $\mu = 1$. 
When $1 < \gamma < 2$ and the variance $\langle \tau^2 \rangle$
diverges, the mean squared displacement scales with the exponent $\mu = 3 - \gamma$. 
Finally, for very heavy tails $0<\gamma<1$, the scaling is ballistic $\mu=2$.
Inspired by a sketch in \textcite{bouchaud1990}.}
\label{figt4}
\end{figure}

As we have discussed, the asymptotic dynamics of propagators can be analyzed  
by considering the limits $k\rightarrow 0$ and $s\rightarrow 0$. Remarkably, 
in coupled models these two limits do not commute \cite{schmiedeberg2009}. In general, by changing the order 
of these limits we imply which effect is dominating: larger distance or longer time, or maybe the interplay of both. 
The MSD can be calculated by using Eq. (\ref{moments_formula}) via the second derivative with respect to $k$ 
and taking the limit $k\rightarrow 0$. To compute the asymptotic time dependence of the MSD we can take 
the second limit $s\rightarrow 0$ and then calculate the inverse Laplace transform.
By inverting the order of limits and first taking $s\rightarrow 0$, we can follow the behavior of 
the density of particles closer to the origin, from where the the scaling exponent of the propagator 
$\alpha$ [see Eq. (\ref{selfsimilarity_ctrw})] could be obtained. Finally, in order to find the shape of the propagator, both limits have to be taken simultaneously. 
We first provide the results for the scaling of the MSD \cite{Zumofen1993}. To compute the MSD for 
superdiffusive sub-ballistic regime, (the LW processes with the finite mean flight time $\langle \tau \rangle$, corresponding to $1<\gamma<2$), one more term in the expansion of the nominator has to be included, in order 
to capture the effect of the ballistic fronts. This leads to
\begin{equation}
\left<x^2(t)\right>\propto\left\{
                \begin{array}{cc}
                  t^2 & 0<\gamma<1\\
                t^2/\ln t & \gamma=1\\
                  t^{3-\gamma} &1<\gamma<2\\
                  t\ln t & \gamma=2\\
                  t & \gamma>2
                \end{array}
              \right.
              \label{msd_scaling1}
\end{equation}
Figure~\ref{figt4} gives a pictorial view of the MSD scaling regimes.
The scaling exponent $\alpha$ is given by \textcite{Zumofen1993}:
\begin{equation}
\alpha=\left\{
                \begin{array}{cc}
                  1 & 0<\gamma<1\\
                  1/\gamma &1<\gamma<2\\
                   1/2 & \gamma>2
                \end{array}
              \right.
               \label{propagator_scaling}
\end{equation} It is important to note that in the sub-ballistic regimes the scaling exponent $\alpha$ 
refers to the central part of the density profile. 

There is an interesting concept to characterize the  stochastic transport phenomena 
by using a spectrum of fractional moments \cite{castiglione1999,metzlerklafter2000,artuso2003,sanders2006,anna2013,Rebenshtok2014,seuront2014}:
\begin{equation}
\left<|x|^q\right>=\int_{-\infty}^{\infty}|x|^qP(x,t)dx \simeq M_q\cdot t^{q\nu(q)}.
\label{fractionalmoments}
\end{equation}
\begin{figure}[t]
\center
\includegraphics[width=0.45\textwidth]{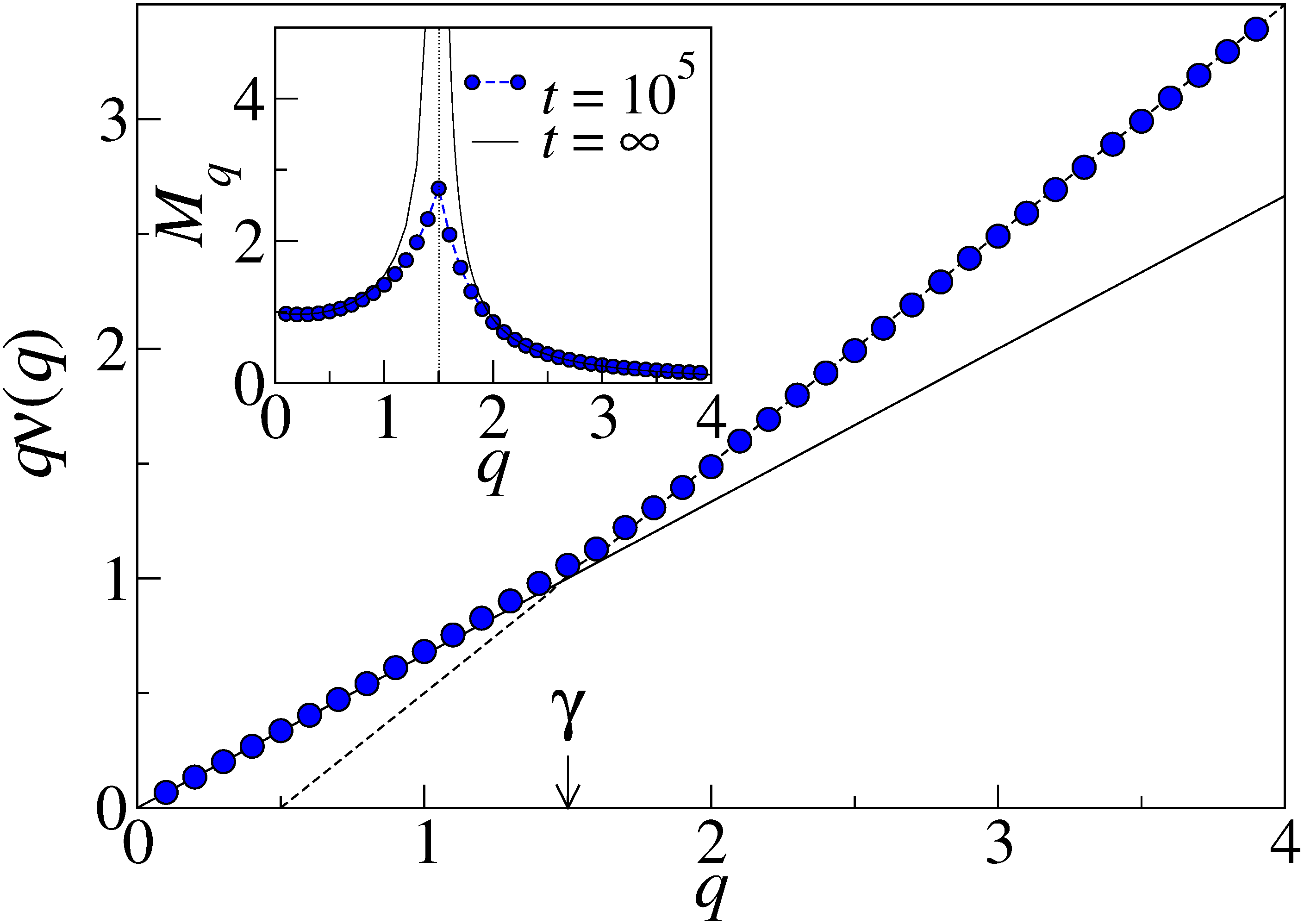}
\caption[Fractional moments]
{(Color online)
Scaling of the fractional moments of the L\'{e}vy walk process. 
Scaling exponent $q\nu(q)$ for the $q-{\text{th}}$ moment of the PDF, 
see Eq. (\ref{fractionalmoments}) as a function
of $q$ for the sub-ballistic L\'{e}vy walk model, $\gamma=3/2$. 
It has a characteristic bi-linear behavior: $q\nu(q)=q/\gamma$ for 
$q<\gamma$ (solid line) and $q\nu(q)=q+1-\gamma$ (dashed line) otherwise. 
The inset shows the pre-factors of fractional moments $M_{q}$. 
Dots correspond to the numerical data sampled for $t = 10^5$, 
whereas lines diverging at $q=\gamma$ are analytical predictions in the limit $t\rightarrow\infty$.
Adapted from \textcite{Rebenshtok2014a}.}
\label{figt5}
\end{figure}
For normal diffusion, because of its self-similar shape and the unique 
scaling $\overline{x}\propto t^{1/2}$, Eq.(\ref{fractionalmoments}) 
leads to a constant value of $\nu(q)=1/2$. If $\nu(q)$ is not constant, this kind of diffusion process is 
referred to as {\em strongly anomalous} \cite{castiglione1999}. Because of its multi-scaling property, 
sub-ballistic  L\'{e}vy walks belong to this class. Figure 5 shows  characteristic be-linear shape 
of $q\nu(q)$ as a function of the moment order $q$. The linear dependencies are $q/\gamma$ for 
small $q$ is replaced by the dependence $q-\gamma+1$ for higher moments. For small values of $q$, 
the dominating contribution to the averaging 
integral comes from the central part of the propagator, where it can be approximated by the self-similar L\'{e}vy 
distribution. For $1<\gamma<2$ the fractional moments of L\'{e}vy 
distribution exist for $q<2$ and  $q\nu(q)=q/\gamma$. For higher moments, the far tails of the propagator 
are important and this is where the ballistic cut-off by the running peaks plays a crucial role. The scaling of the fractional 
moments for large $q$ can be obtained by assuming that the PDF has an asymptotic shape $P(x,t)\sim t/|x|^{1+\gamma}$ and has to 
be integrated with $|x|^{q}$ till the cut-off distance $|x|=v_0t$. That would lead to the $q-\gamma+1$ result or $\nu(q)\sim 1$. 
Exact results on the behavior of fractional moments could be obtained by using the concept of infinite densities
\cite{Rebenshtok2014}, which is discussed in Section \ref{ID}. 

For $q\rightarrow 0$, the $\nu(q)\rightarrow \alpha$ gives a possible way of estimating the scaling from the experimental data. 
Recently \textcite{Gal2010} measured the
spectrum of exponents $q\nu(q)$ for the dispersion of polystyrene bead particles 
internalized by live human metastatic breast cancer epithelial cells and found for large $q$ 
a linear behavior $q\nu(q)\sim cq$ with $c\simeq 0.8-0.6$. That means that the observed spreading
is of the sub-ballistic superdiffusion type. This is probably related to the active 
transport of the beads within a cell. 


\subsection{L\'{e}vy walks with rests}\label{levy_walks_with_rests}
When performing  L\'{e}vy walks, a particle always moves, see Fig. 1(b), 
whereas during CTRW evolution its makes instantaneous jumps alternated with waiting events, see Fig. 1(a). 
By combining both of them, we arrive at the model where waiting periods alternate with periods of 
ballistic motion, see Fig. 1(c). 
One can describe this model as L\'{e}vy walk interrupted by rests \cite{klafter1994, zaburdaev2002,klaftersokolov2011}. 
As in the standard L\'{e}vy walk model, there can not be 
particles beyond the fronts $|x|>vt$. Interestingly, there is a natural
separation of particles into two groups: sitting in a
given point and moving somewhere else. The total density of
particles at a given point $x$ is the sum of two fractions \cite{zaburdaev2002, uchaikin2003}.  
The PDF of resting times we denote $\psi_{\text{r}}(\tau)$ \cite{klaftersokolov2011} 
and the PDF $\psi(\tau)$, as before for L\'{e}vy walks, is used to describe the durations of ballistic phases. 
Both functions are of  the same power-law form but may have different exponents. 
By $\tilde{\nu}(x,t)$ we denote the flux of particles which finished their rest and start moving out 
of a given point $x$ (analogy to the velocity re-orientation points in the standard L\'{e}vy walk model). 
It satisfies the following balance equation:
\begin{eqnarray}
\tilde{\nu}(x,t)&=&\int\limits_{0}^{t}\psi_{\text{r}}(\tau)\int\limits_{0}^{t-\tau}\phi(y,\tau_1)\tilde{\nu}(x-y,t-\tau-\tau_1)d\tau_1 d\tau\nonumber\\&+&\psi_{\text{r}}(t)P_{0}(x)\label{nulwrests},
\end{eqnarray}
where $\phi(y,\tau)$ is the coupled transition probability of the L\'{e}vy walk model. 
The densities of sitting and flying particles are then given by
\begin{eqnarray}
P_{\text{r}}(x,t)&=&\int\limits_{0}^{t}\Psi_{\text{r}}(\tau)\int\limits_{0}^{t-\tau}\phi(y,\tau_1)\tilde{\nu}(x-y,t-\tau-\tau_1)d\tau_1 d\tau\nonumber\\&+&\Psi_{\text{r}}(t)P_{0}(x),\\
P_{\text{fly}}(x,t)&=&\int\limits_{0}^{t}\Phi(y,\tau)\tilde{\nu}(x-y,t-\tau)d\tau,\label{Plwrests}
\end{eqnarray}
where $\Phi(x,t)$ is the coupled survival probability of L\'{e}vy walks (\ref{Phicoupled}). 
The total density of particles is the sum of flying and sitting PDFs, $P_{\Sigma}=P_{\text{fly}}+P_{\text{r}}$. 
In the Fourier-Laplace space it can be expressed as \cite{klaftersokolov2011}
\begin{equation}
P_{\Sigma}(k,s)=\frac{\left[\Phi(k,s)\psi_{\text{r}}(s)+\Psi_{\text{r}}(s)\right]P_{0}(k)}{1-\psi_{\text{r}}(s)\phi(k,s)}
\label{Psigma}
\end{equation}
The first and second terms in the brackets of the nominator correspond to the contributions from the flying and sitting particles, respectively. 
As in the case of the CTRW, 
the long trapping times can compete with long excursions. 
If, however, the mean trapping time is finite, the scaling of the propagator and of the corresponding  MSD is the same as 
in the L\'{e}vy walk model. It is also easy to see from Eq. (\ref{Psigma}) that if both mean resting time and the mean moving time 
are finite, the density of the flying particles is locally proportional to that of the resting particles. The coefficient of proportionality 
is the ratio of times a particle spends on average in each phase \cite{zaburdaev2002},
 $P_{\text{fly}}(x,t)=(\left<\tau\right>_{\text{fly}}/\left<\tau\right>_{\text{r}})P_{\text{r}}(x,t)$. 
In the regime when the mean flight time diverges, there is an irreversible transition of resting particles into flying ones, see Fig. 6, 
and therefore a convergence to the standard L\'{e}vy walk process. 
If at $t=0$ all particles are resting, their total population will decrease in time 
as $\int\limits_{-\infty}^{\infty}P_{\text{r}}(x,t)dx\propto t^{\gamma-1}$. 
\begin{figure}[t]
\center
\includegraphics[width=0.45\textwidth]{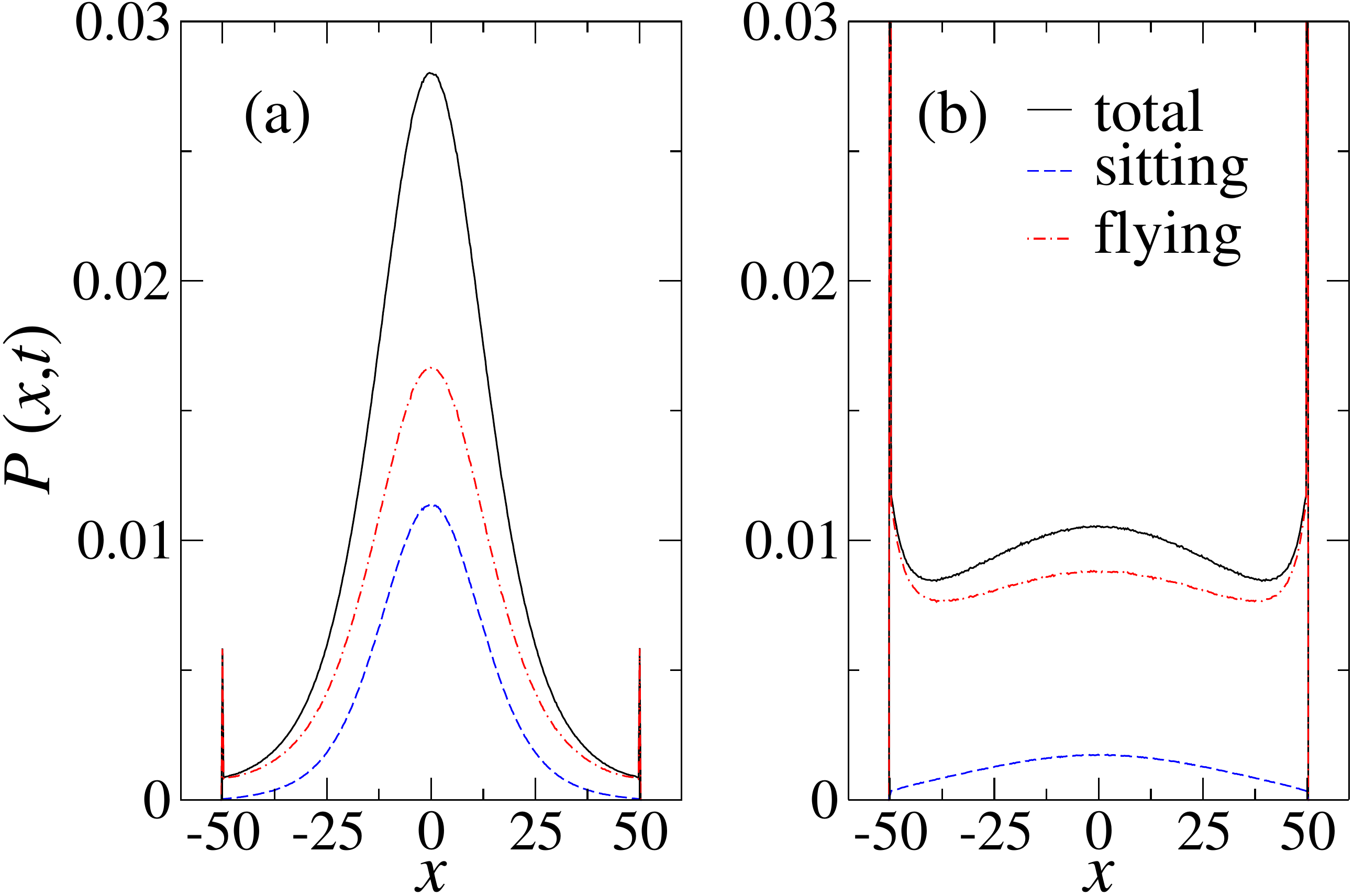}
\caption[Levy walks with rests in ballistic regime]
{(Color online) L\'{e}vy walks with rests. 
Panel (a)  shows  the PDF of the L\'{e}vy walk model with exponentially distributed resting times 
and power-law distribution of flight times with $\gamma=3/2$. 
Both average resting, $\langle \tau_r \rangle$, and flying times are finite. 
The total density of particles (solid line) is the sum of sitting (dashed line) and flying (dash-dotted line) particles. 
Panel (b) shows the PDF for the ballistic regime with $\gamma=0.8$.
Here the number of sitting particles is  greatly reduced. 
In the asymptotic limit $t \rightarrow \infty$  the total PDF and the PDF of flying particles  will coincide. 
The parameters are $v=\tau_0=\langle \tau_r \rangle=1$ and $t=50$.}
\label{figt6}
\end{figure}

With that we close the discussion of the  
relatives of the standard L\'{e}vy walk model. We will 
meet them again in Sections  \ref{physics} and \ref{biology}, when discussing their applications both in physics and biology. 
We now proceed to the generalizations of the L\'{e}vy walk model. 

\section{Generalizations of the L\'{e}vy walk model}\label{generalisations}

\subsection{Random walks with random velocities}\label{rwrv_section}
A natural generalization of the L\'{e}vy walk model is the process in which the velocity of a particle is not fixed but  
is a random variable itself \cite{barkai1998,zaburdaev2008}. 
A number of examples where a random walker has a changing velocity is discussed in \cite{zaburdaev2008}. 
When the velocity of particles is characterized by a heavy tailed distribution, 
the palette of possible diffusion regimes is defined by the interplay of flight time and velocity distributions. 
We denote the velocity PDF by $h(v)$ and write down the corresponding 
transport equations of random walks with random velocities (RWRV) \cite{zaburdaev2008}:
\begin{eqnarray}\nu(x,t)&=&\int\limits_{-\infty}^{\infty}dv\int\limits_{0}^{t}\nu(x-v\tau,t-\tau)h(v)\psi(\tau)d\tau \nonumber\\
&+&\delta(t)P_0(x).\label{nurwrv}
\end{eqnarray}
\begin{equation}P(x,t)=\int\limits_{-\infty}^{\infty}dv\int\limits_{0}^{t}\nu(x-v\tau,t-\tau)\Psi(\tau)h(v)d\tau.\label{Prwrv}
\end{equation}
Despite the fact that equations now are more complicated they can still be resolved by using the integral transforms
\begin{equation}
P(k,s)=\frac{\int_{-\infty}^{+\infty}\Psi(s+ikv)h(v)dv}{1-\int_{-\infty}^{+\infty}\psi(s+ikv)h(v)dv}.
\label{Prwrvks}
\end{equation}
It is easy too see that for $h(v)=\left[\delta(v-u_0)+\delta(v+u_0)\right]/2$ we recover the standard L\'{e}vy walk model result Eq. (\ref{Plevywalkks}).
Because of the additional complexity added through velocity distribution it
 is even harder to find an example where an exact analytical solution can be obtained. 
However, one very remarkable example is the case of the Lorentzian or Cauchy velocity distribution:
\begin{equation}
h(v)=\frac{1}{u_0\pi}\frac{1}{1+v^2/u_0^2}.
\label{cauchy}
\end{equation} 

\begin{figure}[t]
\center
\includegraphics[width=0.45\textwidth]{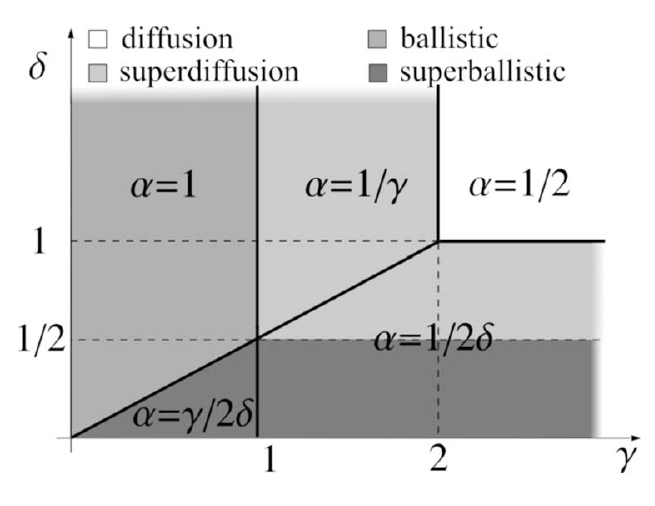}
\caption[Phase diagram or RWRV]
{Scaling regimes of the random walks with random velocities. 
By varying the exponents of the power-law tails of the velocity distribution 
$\delta$ and of the flight time distribution $\gamma$, the model can be tuned into different 
diffusion regimes, from classical diffusion to super-ballistic superdiffusion. 
From \textcite{zaburdaev2008}.}
\label{figt7}
\end{figure}

It appears in physical problems of two-dimensional turbulence \cite{tong1988,min1996,chukbar1999}, 
as a model distribution of kinetic theory \cite{ben-naim2005,trizac2007}, 
and also as a particular case of the generalized kappa-distributions of plasma physics applications \cite{meng1992,hasegawa1985} 
and  statistics \cite{tsallis1988,tsallis1999}. It was also reported for the distribution of velocities of 
starving amoeba cells \cite{takagi2008}. 
In this case the density of particles {\em does not} depend on the flight time distribution at all and also has a shape of the Lorentzian:
\begin{equation}
P(x,t)=\frac{u_0t}{\pi(x^2+u_0^2t^2)}.
\label{Pcauchy}
\end{equation}
To understand the scaling behavior of the RWRV model, we use the scaling regimes of the CTRW model as a guideline. 
For that we calculate the effective jump length distribution, which, due to a simple coupling $x=vt$, can be 
obtained by the following integration:
\begin{equation}
g_{\text{eff}}(x)=\int\limits_{-\infty}^{+\infty}dv\int\limits_{0}^{+\infty}\delta(x-v\tau)h(v)\psi(\tau)d\tau.
\label{geffective}
\end{equation}
For the velocity distribution we assume a generic power-law form, $h(v)\propto |v|^{-1-2\delta}$. 
We can now integrate Eq. (\ref{geffective}) and find the exponent of the tail of the effective jump length 
distribution $\beta(\gamma,\delta)$. 
The waiting time distribution of CTRW model represents the time cost of the flight, therefore we can use the flight 
time distribution exponent. By substituting $\gamma$ and $\beta(\gamma,\delta)$ into the scaling relation for the CTRW,
we find the scaling exponent $\alpha=\gamma/2\beta$ of the RWRV model, see Fig. 7.

Besides the classical diffusive, superdiffusive, and ballistic transport, superballistic scaling is possible. 
In the latter case, the mean absolute velocity has to be infinite ($\delta<1/2$). 
As in the L\'{e}vy walk model, the regime of subdiffusion is inaccessible; 
with non-zero velocities there is no possibility to trap a particle for a long time. 
As we see, the introduction of the velocity distribution significantly increases the flexibility of the model while still keeping it amenable to the analytical approach. 

\subsection{Random walks with velocity fluctuations}\label{lw_with_fluctuations}
In all previous models we neglected interactions of the walker with 
its environment or assumed that it had no effect on the particle as it moved. 
In this section we discuss a model of random walks in active media. 
We assume that a particle can interact with its surrounding which results in the weak fluctuations of particle's velocity. 
The term ``active'' emphasizes the fact that particles not simply lose velocity as a result of 
passive friction but can gain positive and negative velocity increments such that on average their velocity 
remains constant during a single flight event, see Fig. 8. 
This model was applied to reproduce the perturbation spreading 
in Hamiltonian many particle systems by \textcite{zaburdaev2011,denisov2012}, see Section \ref{many_body}. 
\begin{figure}[t]
\center
\includegraphics[width=0.45\textwidth]{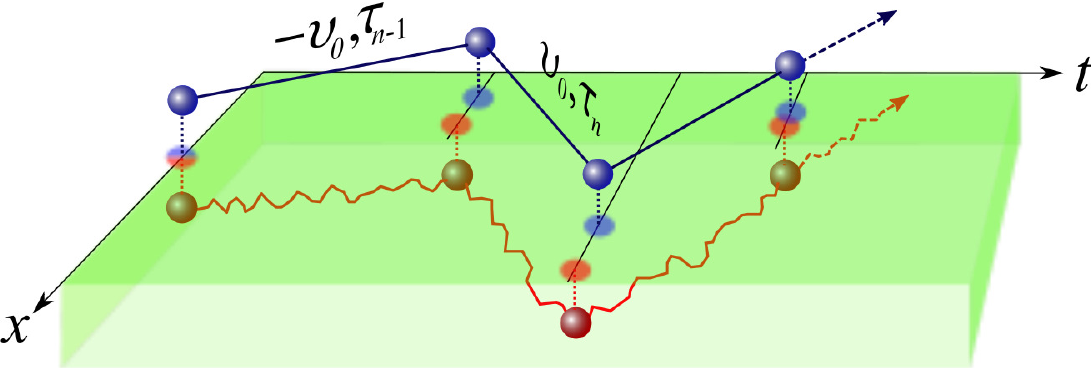}
\caption[Random walks in active media]
{(Color online) Sketch of the random walk in active media. 
Velocity during  flights fluctuates around a fixed averaged value. 
As a result the fluctuations accumulate with time and the final position of the particle, 
passing through the active medium, will differ from that produced by an ideal  L\'{e}vy walk process. 
From \textcite{zaburdaev2011}.}
\label{figt8}
\end{figure}
To setup the model accounting for velocity fluctuations, we modify the L\'{e}vy walk model \cite{zaburdaev2011}. 

During each flight of a particle, its position is described by a simple Langevin equation: $\dot{x}=v_0+\zeta(t)$ \cite{vankampen2011},
where $\zeta(t)$ is a delta-correlated Gaussian noise of zero mean and  
finite intensity $D_v$, i.e.,  $\left<\zeta(t) \zeta(s)\right>=D_v\delta(t-s)$. 
This equation describes the well known biased Wiener process with drift $v_0$ \cite{karatsas}. 
After an integration over a time interval $\tau$, we obtain:
\begin{equation}
x(t+\tau)=x(t)+v_0\tau+w(\tau),
\label{eq:fluct}
\end{equation}
where $w(\tau)=\int_{t}^{t+\tau}\zeta(s)ds$ is characterized by a Gaussian PDF $p(w,\tau)$ 
with the dispersion $\sigma_{\tau}^2=\langle (x(\tau) - v_0\tau)^2\rangle = D_v\tau$. 
Transport equations for this model can also be written and solved in the Fourier-Laplace space. 
When velocity fluctuations are small, $(D_v\langle\tau\rangle)^{1/2}\ll v_0\langle\tau\rangle$, 
the central part of the density profile of particles is given by the same L\'{e}vy distribution as in the case of  the standard model. 
New phenomena appear in the ballistic regions, where fronts, due to fluctuations, now look like humps (see Fig. 9):
\begin{equation}
P_{\text{hump}}(x,t)=\Psi(t)\left[p(x+v_0t,t)+p(x-v_0t,t)\right]/2
\label{P_hump}
\end{equation}
\begin{figure}[t]
\center
\includegraphics[width=0.45\textwidth]{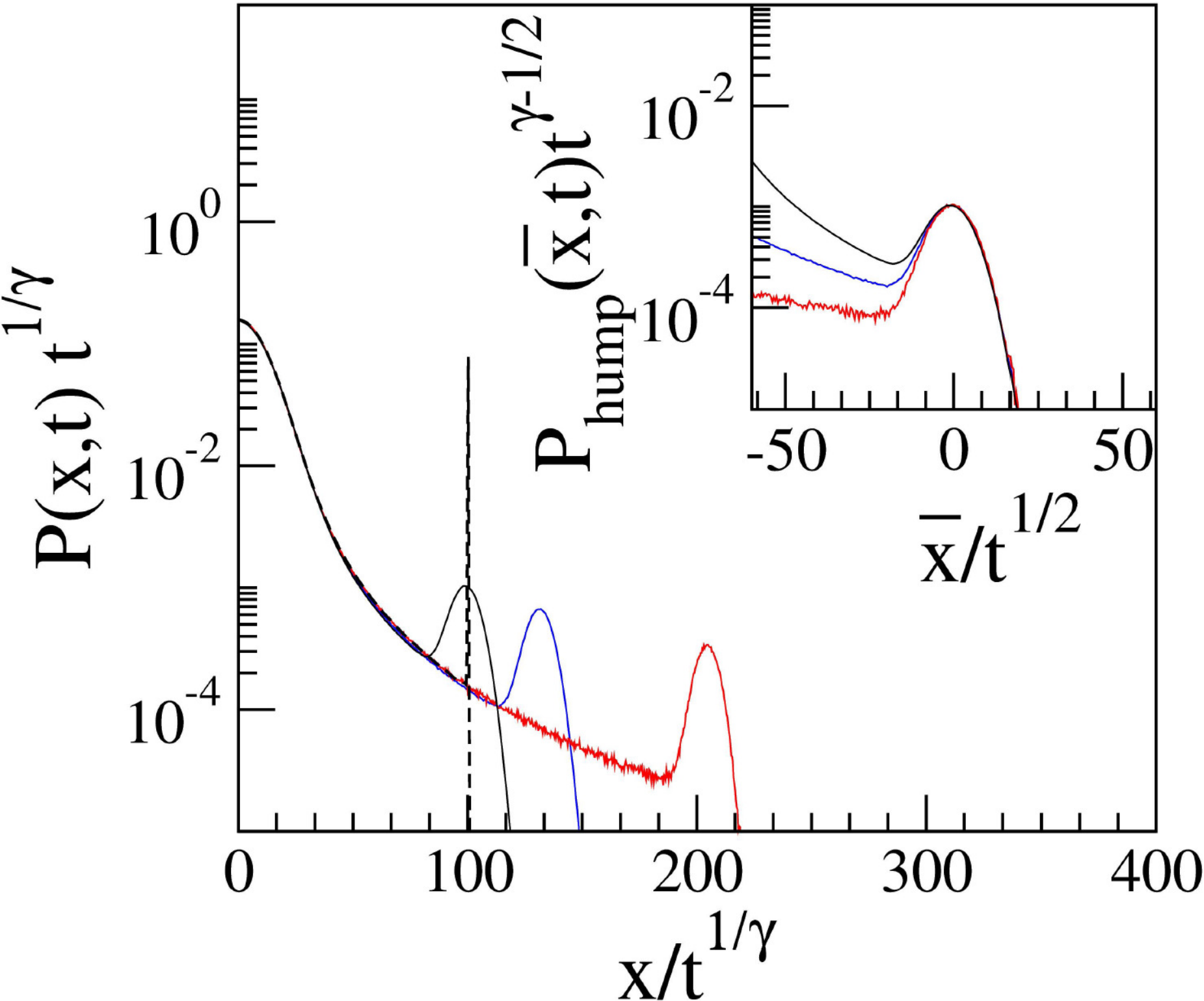}
\caption[Levy walk profile with humps]
{(Color online) Profile of the L\'{e}vy walk model with velocity fluctuations. 
The inset shows the scaling of humps. 
Please note that the figure is adapted from \textcite{zaburdaev2011,Zaburdaev2012err} where a so-called equilibrated 
initial condition was used (see Sec. \ref{memory_effects} for details), 
as a result the height of the hump decays slower than discussed in the text, Eq. (\ref{eq:scaling2}). 
Independent of the type of the initial condition, the shape of humps is Gaussian, with their width growing as $t^{1/2}$.}
\label{figt9}
\end{figure}

As before, $\Psi(t)$ is the probability of not changing the direction of flight during time $t$, Eq. (\ref{survival}), 
and has a power-law asymptotic $\Psi(t)\propto(t/\tau_{0})^{-\gamma}$. 
Consequently, the area under the ballistic humps, Eq. (\ref{P_hump}), also scales as $t^{-\gamma}$. 
During ballistic flights, the particles undergo random fluctuations caused by velocity variations. 
All particles in the hump are in the state of their first flight of duration $t$, thus the dispersion of 
the Gaussian-like humps grows as $t^{1/2}$, and we arrive at the following scaling for the particles density in the
humps:
\begin{equation}
P_{\text{hump}}(\bar{x},t') \simeq u^{-\gamma-1/2}P_{\text{hump}}(\bar{x}/u^{-1/2},t),
\label{eq:scaling2}
\end{equation}
where $u = t'/t$ and $\bar{x} = x - v_0 t$, see inset in Fig. 9. 
From this result we learn that ballistic humps may carry some additional information about the interactions 
between the random walking particles and their environment.

The two models we discussed  are 
the most frequently used modifications of the original L\'{e}vy walk setup. 
In the next subsection, we are going to mention two more models of coupled random walks which introduce higher time cost for longer jumps, but still have instantaneous jumps as the standard CTRW model.

\subsection{Other coupled models}\label{coupled_models}
\begin{figure}[t]
\center
\includegraphics[width=0.45\textwidth]{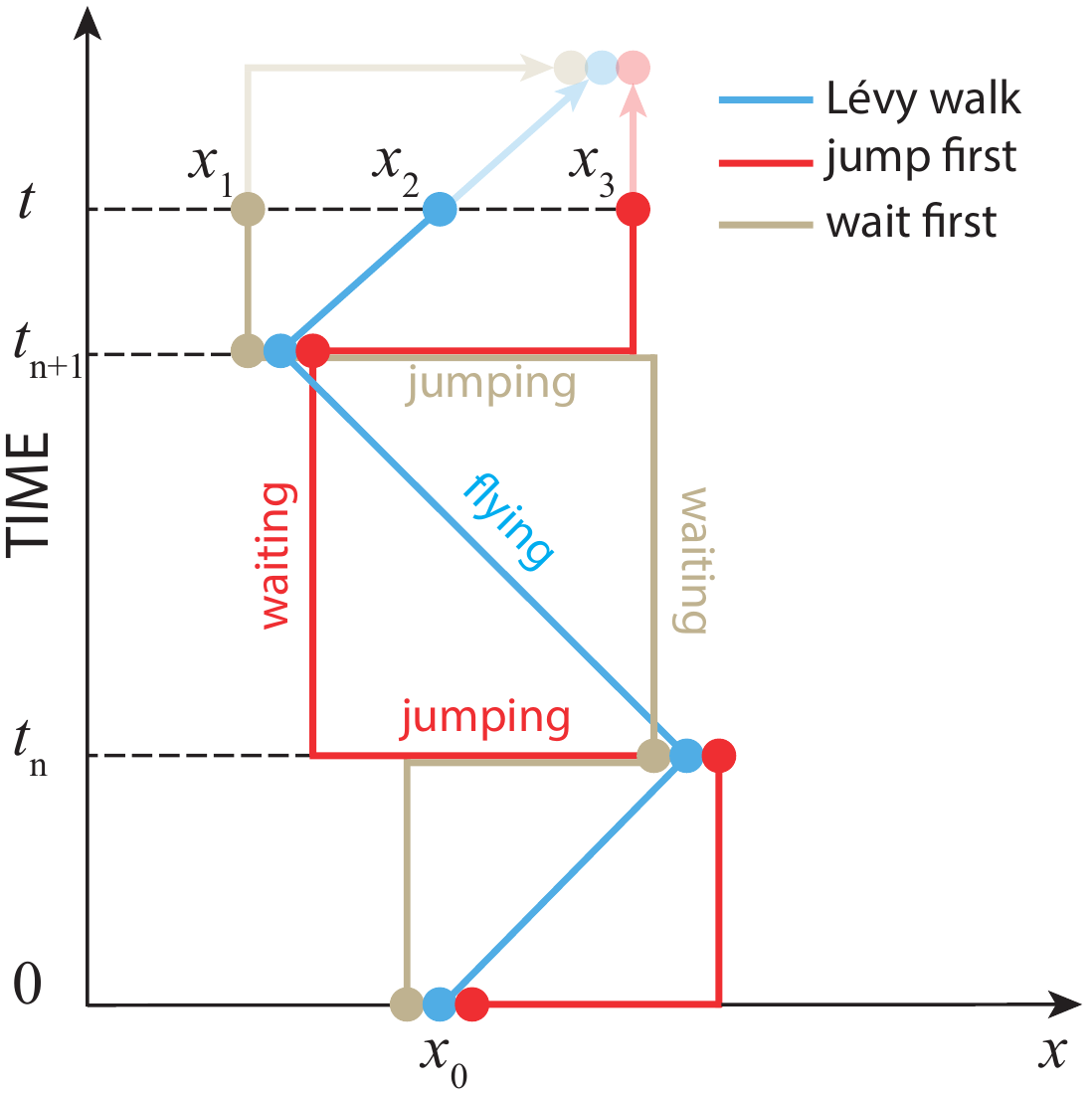}
\caption[Other coupled models]
{(Color online) Comparison of trajectories of 
the L\'{e}vy walk model and two coupled models, ``wait first'' and ``jump first''. 
Trajectories of all three models are passing through the same points on the $(x,t)$ diagram, 
but taking different paths. 
The positions of particles at time $t$ are determined by their last steps.}
\label{fig10}
\end{figure}
There are two modifications of the CTRW model which are very 
similar to the L\'{e}vy walk model, but still do not have a well defined velocity 
of particles \cite{kotulski1995,meerschaert2006,jurlewicz2012,becker-kern2004,straka2011}. 
In both models the waiting time and jump distance are coupled 
but  jumps are instantaneous. 
There are two possibilities: In the ``jump first'' model, a random walker first jumps to a random distance $y$ 
and then waits at the arrival point for a time $\tau=a|y|$, 
where $a$ is a positive coupling constant \cite{zaburdaev2006}. In 
the ``wait first'' model, a particle first waits for a random time $\tau$ 
and then makes a jump of the length $|y|=\tau a^{-1}$ \cite{shlesinger1982,barkai2002}. 
Figure 10 shows trajectories of these two models compared to the one of the standard L\'{e}vy walk model. 
We see that in $(x,t)$ plane the turning points of all three random walks are identical, 
and it is the paths which are different. 
At time $t$ the models differ only by their last step. However, this difference is crucial.
The transport equations for jump first and wait first models can be written down and solved in the Fourier-Laplace space, 
see \textcite{schmiedeberg2009}. The MSD of the jump first model for the anomalous diffusion regime (long jumps) is diverging, 
which is clear if one looks at the distribution after the first jump. 
The wait first model resembles a L\'{e}vy walk model in that it also has a defined light front and therefore finite 
moments \cite{schmiedeberg2009}. All three models have the same scaling properties 
of their propagators, but the shapes of propagators are model specific. 
Figure 11  shows  the density profiles for the three models with similar (leading to the same scaling) 
jump, waiting, and flight time distributions, and with all remaining proportionality constants set to one, $a=v=1$.  
\begin{figure}[t]
\center
\includegraphics[width=0.45\textwidth]{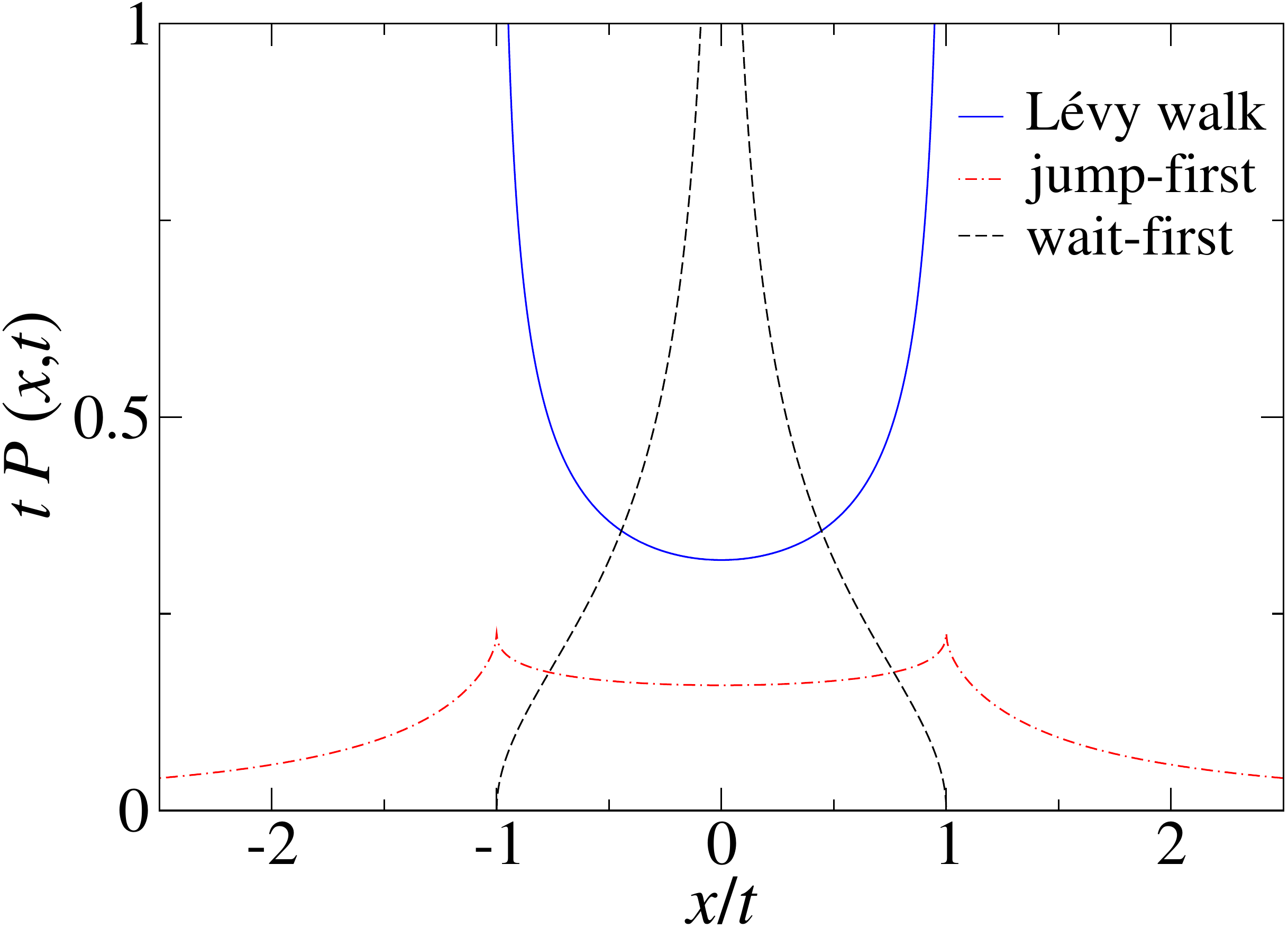}
\caption[Rescaled Green's functions]
{(Color online) Propagators of coupled models. 
Analytical solutions for the propagators of the wait first (dashed line) 
and jump first (dash-dotted line) models are compared with a L\'{e}vy walk propagator (solid line) 
in the ballistic regime with $\gamma=1/2$. The results are obtained by the method discussed in Section \ref{exact_ballistic}, see \textcite{froemberg2014}.}
\label{figt11}
\end{figure}

These two simple coupled models can serve as a starting point for further generalizations. 
Here we looked only at linear couplings but it can be extended to the case when the flight distance 
is some power law function of the flight time \cite{metzlerklafter2004}; 
it will then effect the scaling of the propagator and its moments. 
One interesting example of the wait first model assumes that the jump length is distributed 
as a Gaussian function with a variance which linearly depends on the waiting time. 
In this case, for the power-law distributed waiting times, the resulting equation contains 
the diffusion operator but in a fractional power \cite{shlesinger1982,becker-kern2004}. 
In the spirit of the L\'{e}vy walk model with rests, one can also add velocity to the above 
two coupled models. It is interesting to see how a new effective velocity arises as a 
combination of the coupling, $a^{-1}$ and actual velocity of the particles $v$, $v_{\text{eff}}=v/(1+av)$ \cite{zaburdaev2006}. 
In this direction, one could consider more general models. For example, variation of the L\'{e}vy walk, in which the waiting times 
are power law distributed but jump lengths scale non-linearly with waiting times $\psi(x|\tau)=\psi(\tau)\delta(|x|-u\tau^{\beta})$, 
with $\beta>0$ and $\beta\neq 1$. A nonlinear coupling provides a way to step beyond ballistic scaling regime of L\'{e}vy walks.  It appears in the description of  anomalous Richardson diffusion \cite{shlesinger1986,shlesinger1987,klafter1987} with $\beta=3/2$ or, in a slightly more involved form, in the context of cold atom dynamics with the same $\beta=3/2$, which we discuss in Section  \ref{cold atoms}.

With this subsection we finalize the review of the existing random walk models 
which embrace the concept of finite velocity of particles and the coupled nature of the spatiotemporal transport process. 
The following sections are dedicated to more sophisticated properties of L\'{e}vy walks and more advanced tools for their analysis.

\section{Properties of L\'{e}vy walks}\label{advanced}
While the results of the previous sections can be rated as basic  tools 
needed for  applications of L\'{e}vy walks in practice, 
the following material goes in more details and as a result is more involved. 
However, it touches upon fundamental concepts of physics, 
such as aging, ergodicity, and space-time correlations. 
Most of these results are very recent thus indicating that the properties of L\'{e}vy walks 
are still being explored. 

\subsection{Space-time velocity auto-correlation function}\label{space_time_corr}
Finite velocity of walking particles brings a random walk 
model closer to the basic physical principles and makes its more suitable
for the description of real-life phenomena. 
However, the presence of the well defined velocity in random walks brings additional 
properties to these stochastic process. 
In the realm of the continuous mechanics, the space-time velocity auto-correlation function 
is a fundamental quantity characterizing the dynamics of a fluid or other media. 
It reveals the relation between the velocities at two distant points and two different instants of time. 
Remarkably, the notion of continuous theory can be adapted to the single particle process of random walks. 
We can ask how the velocity of a random walker is correlated to its own velocity at some later moment 
of time but also at a certain distance from the starting point. 
A naive expectation for a random walk, where each next step is independent from the previous one, 
is that the correlations will be zero at a distances larger than a single flight. 
It was shown, however, that even for the regime of classical diffusion, but with finite velocity, the space-time velocity 
correlation function is different from zero and has a non-trivial space-time dependence \cite{zaburdaev2013}.

The space-time velocity auto-correlation function for a single-particle process
can be redefined from the conventional expression  \cite{monin2007}
\begin{equation}
\mathcal{C}_{vv}(x,t)=\left<v(0,0) v(x,t)\right> .
\label{cor_def}
\end{equation}
We assume that the particle starts  its  walk with an initial velocity $v(x=0,t=0) = v_0$.
After a time $t$ the particle is found at the point $x$ with some velocity
$v(x,t)$. To estimate $\mathcal{C}_{vv}(x,t)$, an observer at time $t$ averages the product of the actual and the initial
velocities of all particles that are located within a bin  $[x, x+dx]$.
It can be formalized in the following way:
\begin{equation}
\mathcal{C}_{vv}(x,t) = \int\limits_{-\infty}^{\infty}\int\limits_{-\infty}^{\infty} v v_0 \frac{P(v,x,v_0,t)}{P(x,t)}dv_0dv,
\label{cor_form}
\end{equation}
where $P(v,x,v_0,t)$ is the joint PDF for a particle to start with velocity $v_0$ and to be in the point $x$ at 
time $t$ with velocity $v$. The particle has first to arrive at the point $x$ for the measurement to occur, 
therefore we use Bayes' rule  \cite{Grinstead1997} for the conditional probability to obtain the integral above. 
The spatial density $P(x,t)$ is usually a known quantity for a given random walk model. 
In contrast, a challenging quantity to tackle is the joint probability of particles' positions and velocities. 
To focus on its role, the spatial density of the velocity correlation function can be introduced:
\begin{equation}
C(x,t)=\int\limits_{-\infty}^{\infty}\int\limits_{-\infty}^{\infty} v v_0 P(v,x,t|v_0)h(v_0)dv_0dv.
\label{vacf_integral}
\end{equation}
Here $h(v_0)$ is the distribution of the initial velocities, which also signals that we are using the formulation of 
the random walk with random velocities. 
The integration over $x$, Eq.~(\ref{vacf_integral}) yields the standard temporal velocity auto-correlation function 
$C(t)=\left<v(0)v(t)\right>$. Normalizing $C(x,t)$ by the particle density $P(x,t)$, we return to the
 original velocity-autocorrelation function:
\begin{equation}
\mathcal{C}_{vv}(x,t)=C(x,t)/P(x,t).
\label{connection}
\end{equation}
As for every model we considered so far, 
the integral transport equations can be derived for $P(v,x,t|v_0)$ and solved by using the Fourier-Laplace transforms. 
The definition of the velocity auto-correlation function contains two additional integrals 
with respect to the final and initial velocities, which makes the final answer more involved \cite{zaburdaev2013}.

\begin{figure}[t]
\includegraphics[width=0.48\textwidth]{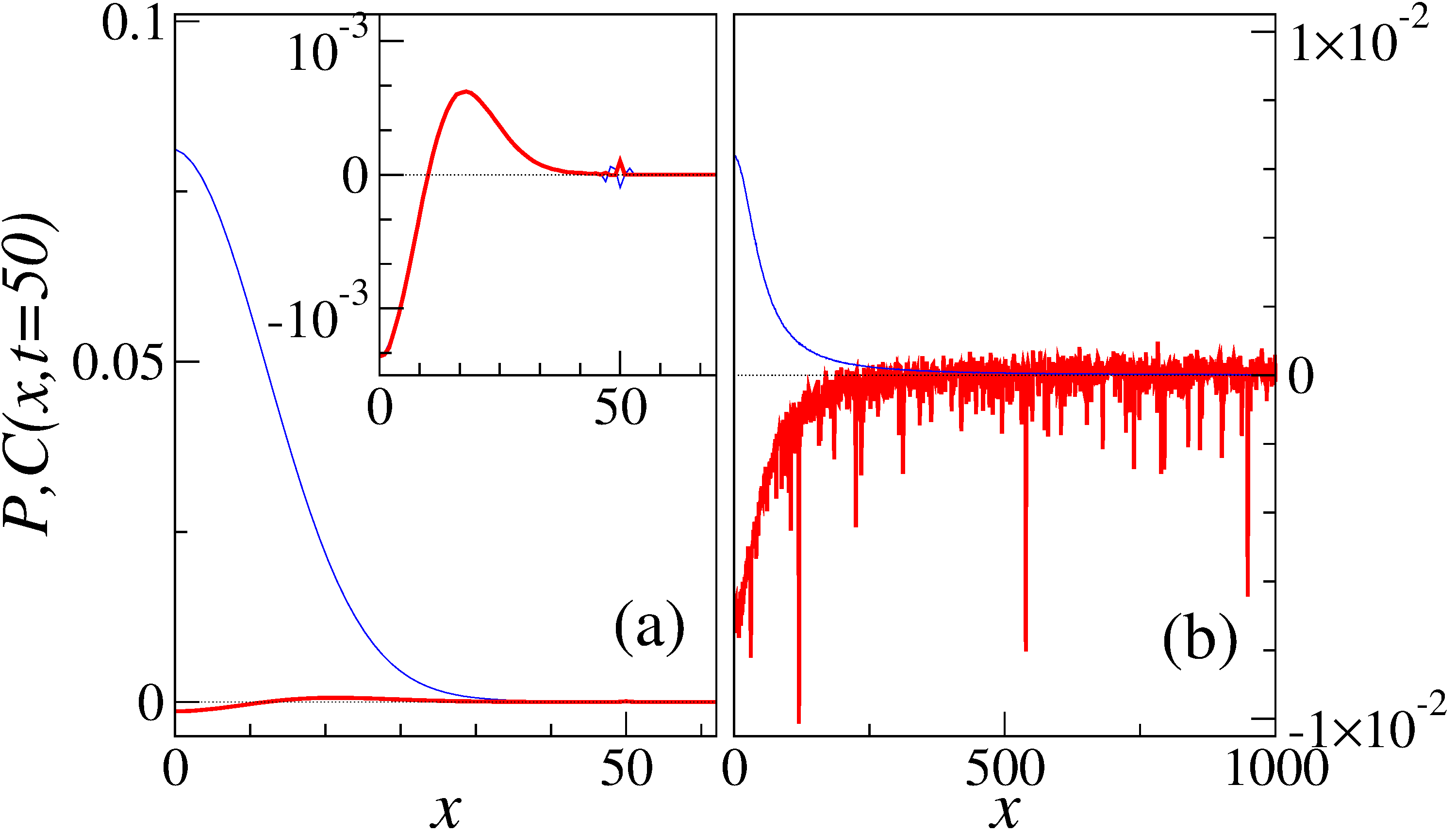}
\caption{(Color online) Propagators (thin blue line) and space-time velocity autocorrelation function (thick red line)
at time $t=50$ for two extreme cases of the random walk with random velocities, see Section \ref{rwrv_section}, 
with the flight-time PDF $\psi(\tau)=\delta(\tau-1)$, for
(a) $h(\upsilon) = [\delta(\upsilon-u_0)+\delta(\upsilon-u_0)]/2$ and (b)  $h(\upsilon)$ in the form of Cauchy 
distribution, Eq. (\ref{cauchy}). In the first case, the autocorrelation function
is proportional to the second spatial derivative of the propagator, 
$C(x,t) \propto \triangle P(x,t)$ [see the inset where both functions are plotted together, with
$\triangle P(x,t)$ weighted as in  Eq. (\ref{cor2})], while in 
the second case $C(x,t)\sim -P(x,t)$. The functions were sampled with $N = 10^{7}$ realizations. 
The parameter $u_0$=1.}
\label{figt12s}
\end{figure}

However, the asymptotic analysis of the general answer in the limit of large time and space 
scales retrieves some surprisingly simple results. 
Consider first the L\'{e}vy walk regime of the velocity model, $h(v)=\left[\delta(v-\upsilon_0)+\delta(v+\upsilon_0)\right]/2$, 
with a power-law distributed flight time, Eq. (\ref{psi}). 
The density of particles is sandwiched between the two ballistic peaks. 
For the peaks we get $C(x=\pm \upsilon_0t,t)=\upsilon_0^2\Psi(t)\delta(x\pm \upsilon_0t)/2$ and $\mathcal{C}_{v,v}=\upsilon_0^2$. 
For the central part of the propagator, in the regime of classical diffusion, $\gamma>2$, 
the density of the correlation function is proportional to the first time derivative of the particle's density:
\begin{equation}
C_{\text{centr}}(x,t)=\upsilon_0^2D\langle\tau\rangle\triangle P(x,t)=\upsilon_0^2\langle\tau\rangle\frac{\partial P(x,t)}{\partial t}.
\label{cor2}
\end{equation}
The above asymptotic result is valid for any flight time distribution with a finite second moment, 
see Fig. \ref{figt12s}(a).
Lets consider this result more closely. As mentioned above, by integrating the density of space-time 
velocity autocorrelation function over the coordinate we should obtain the standard temporal correlation 
function $C(t)$. In the long time limit, the integral of the central part is approaching zero [because of the Laplacian operator in Eq.(\ref{cor2})]. The only non-zero contribution comes from the ballistic peaks which lead to $C(t)=u_0^2\Psi(t)$: velocities of particles remain correlated only during the flight time. If, for example, the flight time is exponentially distributed,
the standard temporal velocity auto-correlation function decays very fast, 
$C(t)=\upsilon_0^2e^{-t/\tau_0}$. If we look at the density of velocity-velocity autocorrelation at zero $x=0$, 
from Eq.(\ref{cor2}) it follows that $C(x=0,t)=-\upsilon_0^2\langle\tau\rangle(\pi D)^{-1/2}t^{-3/2}$. 
First of all, correlations at $x=0$ are negative, and secondly, they decay in time algebraically like $t^{-3/2}$. This decay to zero is faster than that of the density of particles, $P(x=0,t)\propto t^{-1/2}$, but still much slower than the exponential decay of the temporal correlation function. It highlights the fact that the space-time velocity correlation function provides access to long-lived correlations and
therefore increases the chance of their detection in experiments.
\begin{figure}[t]
\center
\includegraphics[width=0.45\textwidth]{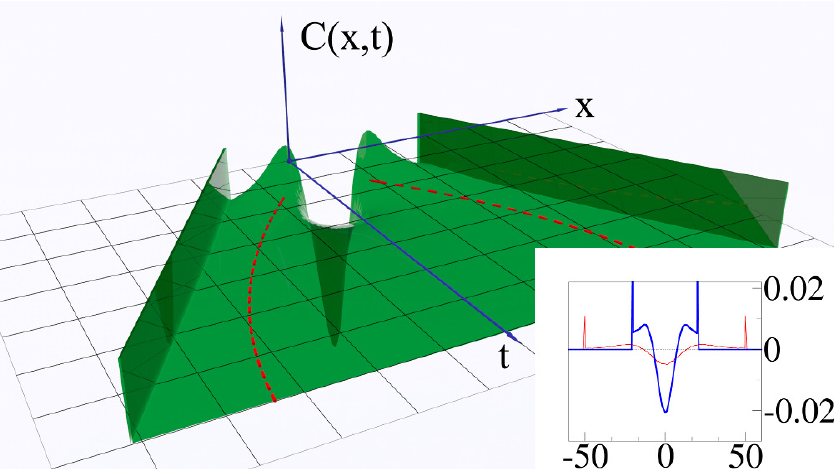}
\caption[Velocity autocorrelation function]
{(Color online) Space-time velocity autocorrelation function of L\'{e}vy walks in 
the superdiffusive regime. The space-time evolution of correlations shows a 
negative dip near $x=0$ and two spreading maxima. The integral of the central part with respect to the coordinate is equal zero. The ballistic peaks carry the correlations of the particles which are still in their first flight. The red dashed lines indicate the positions of local maxima $x^\pm_{\text{m}}$ on the
$x - t$ plane which follow the power-law scaling $x^\pm_{\text{m}}\propto \pm t^{1/\gamma}$, while  the 
height of the maxima decays as
$t^{-1-1/\gamma}$. The inset depicts spatial profiles of $C(x,t)$ for two different instants of time, 
$t = 20$ (heavy blue line)  and $50$ (light red line), $\gamma=3/2$. From \textcite{zaburdaev2013}.}
\label{figt12}
\end{figure}

For the regime of the superdiffusive L\'{e}vy walk, $1<\gamma<2$, the formula with the first time derivative remains valid. 
By further exploiting the properties of the time derivative we see that the velocity auto-correlation function 
is negative near  the point $x=0$, see Fig. 13.
Upon the departure from the origin the correlation density becomes 
positive and produces two local maxima. 
These maxima are traveling with the power-law scaling $x^\pm_{\text{m}}\propto \pm t^{1/\gamma}$, 
while  the height of the maxima decays as $t^{-1-1/\gamma}$. 

As we continue to move toward more anomalous behavior, for example for 
the ballistic regime of L\'{e}vy walks, 
the correlations decay in time even slower \cite{zaburdaev2013}. 
Finally, an example was given, 
when the velocity distribution 
of the particles was Lorentzian, see Section \ref{rwrv_section}. 
In that case the density of the space-time velocity autocorrelation function was proportional 
to the particle density, $C(x,t)\sim -P(x,t)$, see Fig. 12(b). 

In all considered regimes there is a region of negative correlations 
at the vicinity of the starting point. 
This means that majority of particles found there are flying in the direction 
opposite to that of their initial motion. 
The shape of the ``echo'' region and the time-scaling of its width are model-specific
characteristics. Interestingly, simulations of a stochastic process described by a system of Langevin 
equations in the regime of classical Brownian diffusion show analogous results (see \cite{zaburdaev2013} and its Supplementary Material for additional plots), thus suggesting
that these findings are applicable to a broad class of stochastic transport processes characterized by 
finite velocity of moving particles. 

Here we considered only a simple initial condition, when all particles instantly change 
their velocity at $t=0$. As a result, the temporal correlation function $C(t)$ obtained 
by the integration of the density of the space-time velocity correlations describes 
velocity correlations in a specific setting: initial velocity is always taken right 
after the reorientation event and the second velocity is measured after the lag time $t$. 
In general, the temporal velocity autocorrelation function depends on two arbitrary 
times $t_1$ and $t_2$, when the corresponding velocities of particles are measured. 
Such two-point correlation function was considered before in the context of 
CTRW \cite{baule2007,barkai2007,zaburdaev20082,Dechant2014}, and also 
for the case of L\'{e}vy walks (see Section \ref{run_and_tumble} and \cite{taktikos2013,froemberg20132}). 
This results call for the generalization of the space-time velocity correlation function to 
a broader class of  initial conditions.

\subsection{Exact solutions for ballistic random walks}\label{exact_ballistic}
The asymptotic analysis  is very useful 
but in many cases it is still impossible to obtain the expression for the propagators 
in real time and space analytically. 
Interestingly, for random walk models which have the ballistic scaling, 
there is a particular method to calculate the inverse Fourier-Laplace transform without performing it directly. 
It immediately gives the shape of the scaling function $F(\xi)$, see Eq. (\ref{selfsimilarity_ctrw}). 
The method is similar to the one proposed by \textcite{godreche2001},  
and used for the analysis of the renewal process and the inversion of the double Laplace transform. 
The problem of finding the PDF in ballistic regimes is intimately related to the problem 
of time averages \cite{rebenshtok2008}, as the scaled position of the particle after time $T$ is given by
\begin{equation}
x/T=\frac{1}{T}\int\limits_{0}^{T}v(t)dt,\label{timeaverage}
\end{equation}
which is a time average of particle's velocity. 
In application to the random walk concept, 
the method of Godr\`{e}che and Luck has the following formulation. 
Assume that the propagator of the random walk model has the following scaling form:
\begin{equation}
G(x,t)=\frac{1}{t}F\left(\frac{x}{t}\right).
\label{ballisticscaling}
\end{equation}
In the Fourier-Laplace space it can be represented as:
\begin{equation}
G(k,s)=\frac{1}{s} f\left(\frac{ik}{s}\right)=\frac{1}{s} f\left(\zeta\right); \quad \zeta=\frac{ik}{s}.
\label{ballisticscalingfourier}
\end{equation}
Finally,  by using the Sokhotsky-Weierstrass theorem (see \textcite{froemberg2014} for more details):
\begin{equation}
F(\xi)= -\frac{1}{\pi \xi}\lim\limits_{\epsilon\rightarrow 0} \text{Im} f\left(-\frac{1}{\xi+i\epsilon}\right),
\label{GL}
\end{equation}
where $\xi=x/t$ is the scaling variable. This formula allows us to compute the shape of the propagator without calculating of the inverse Fourier-Laplace transforms. 

We first give an example corresponding to the standard L\'{e}vy walk model with a constant speed $v$ and set it to unity for simplicity, $v=1$. The asymptotic profile of the Green's function in the Fourier-Laplace space is given by Eq.(\ref{levy_ballistic_ks}) from which we can easily identify:
\begin{equation}
f(\zeta)=\frac{(1-\zeta)^{\gamma-1}+(1+\zeta)^{\gamma-1}}{(1-\zeta)^{\gamma}+(1+\zeta)^{\gamma}}.
\label{fks}
\end{equation}
Now by using Eq.(\ref{GL}) we can find the shape of the scaling function to be:
\begin{eqnarray}
&&\Phi(\xi) = \frac{\sin\pi\gamma}{\pi} \times\nonumber \\ 
&&
\frac{|\xi-1|^{\gamma}|\xi+1|^{\gamma-1} + |\xi+1|^\gamma|\xi-1|^{\gamma-1}}
{|\xi-1|^{2\gamma} + |\xi+1|^{2\gamma} + 2|\xi-1|^\gamma|\xi+1|^\gamma \cos\pi\gamma}
\label{lamperti1}
\end{eqnarray}
which is the Lamperti distribution \cite{lamperti1958}. 
\begin{figure}[t]
\center
\includegraphics[width=0.45\textwidth]{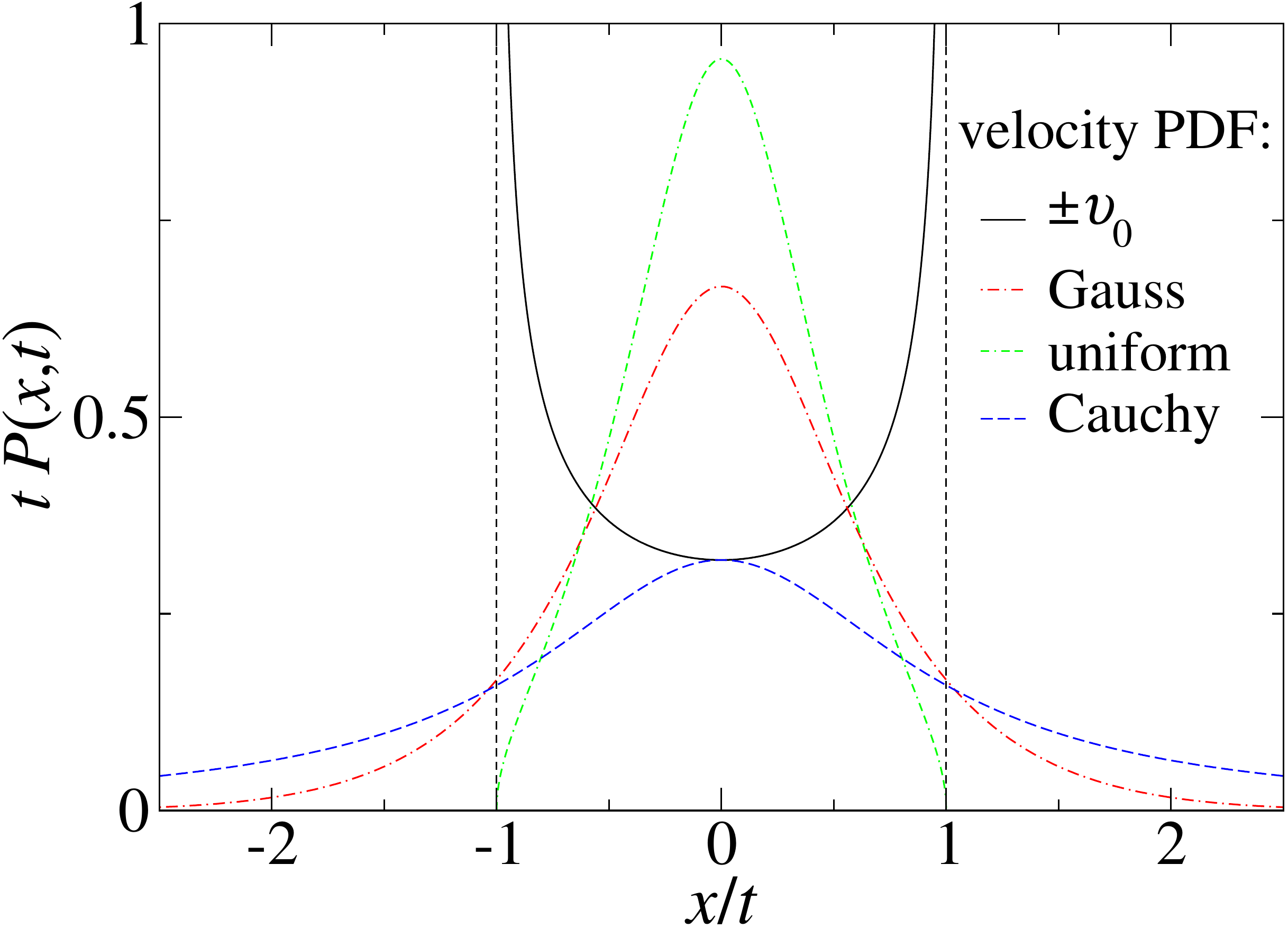}
\caption[Goderche and Luck formula]
{(Color online) Exact solutions of ballistic L\'{e}vy walks. 
Here we plot analytical results obtained from Eq. \ref{GL} for the scaling functions 
of the random walks with random velocity model in the ballistic regime ($\gamma=1/2$), for four different 
velocity distributions, two delta peaks corresponding to the L\'{e}vy walk (solid line), 
Gaussian (dash-dotted), uniform on an interval (double dash-dotted), and Cauchy (dashed). 
Adapted from \textcite{froemberg2014}.}
\label{figt13}
\end{figure}
As another illustration we ask how the shape of the velocity distribution 
of a random walking particle affects the shape of the corresponding propagator. 
For that consider the model with random velocities, Section \ref{rwrv_section}, 
in the ballistic regime ($\gamma=1/2$) with four different velocity distributions, 
$h(v)$: a) two delta peaks (L\'{e}vy walk regime), b) Gaussian, 
c) uniform on a symmetric bounded interval, and d) Lorentzian (Cauchy). 
Equation (\ref{GL}) gives analytical answers for the scaling function $F(x/t)$ for all four cases. 
Figure 13 shows that the velocity distribution has a pronounced effect on the shape of the particles' density. Namely, we see 
the familiar $U$-like profile for the standard L\'{e}vy walk model, 
more of a bell-shaped profile but still bounded by fronts for the uniform velocity distribution, 
and unbounded bell-shaped profile for the Gaussian velocity distribution. 
The Lorentzian case is special as it does not depend on the flight time distribution and has diverging moments. 

We note here that this approach does not require the finite velocity of particles, 
it only relies on the existence of the ballistic scaling. 
Therefore it can be also applied to the coupled setups as wait/jump-first models 
(this is how the plots on Fig. 11 were obtained). 
We present these analytical results 
to emphasize that even a model of random walks with random velocities can be thoroughly analyzed 
with the help of the combination of asymptotic analysis and elegant mathematical machinery. 
It would be challenging to try to extend or find similar approaches to other scaling regimes of random walks.


\subsection{Infinite densities of L\'{e}vy walks}
\label{ID}

Many statistical properties of a L\'{e}vy walk process can be evaluated from the corresponding propagator,
Eqs. (\ref{Gksctrw},\ref{nG}).  For superdiffusive  sub-ballistic regimes the central part of the propagator
is subjected to the generalized central limit theorem and thus it is given by the symmetric
L\'{e}vy distribution $L_{\gamma}[x,\sigma(t)]$, with $\sigma(t) = (K_{\gamma}t)^{1/\gamma}$. However, 
this fundamental fact 
does not allow for calculations of the moments, starting from the second one, 
simply because they do not exist for  L\'{e}vy distributions. The confinement of the process 
to the ballistic cone should be taken into account and in order to calculate higher-order moments one needs to known
the behavior of the propagator at the vicinity of the ballistic fronts. These regions are out of the 
validity  domain of the gCLT and a complementary theory is needed.
The concept of infinite measure \cite{Aaronson1997,Thaler2006} 
provides with such  theoretical framework \cite{Rebenshtok2014,Rebenshtok2014a}
\begin{figure}[t]
\center
\includegraphics[width=0.45\textwidth]{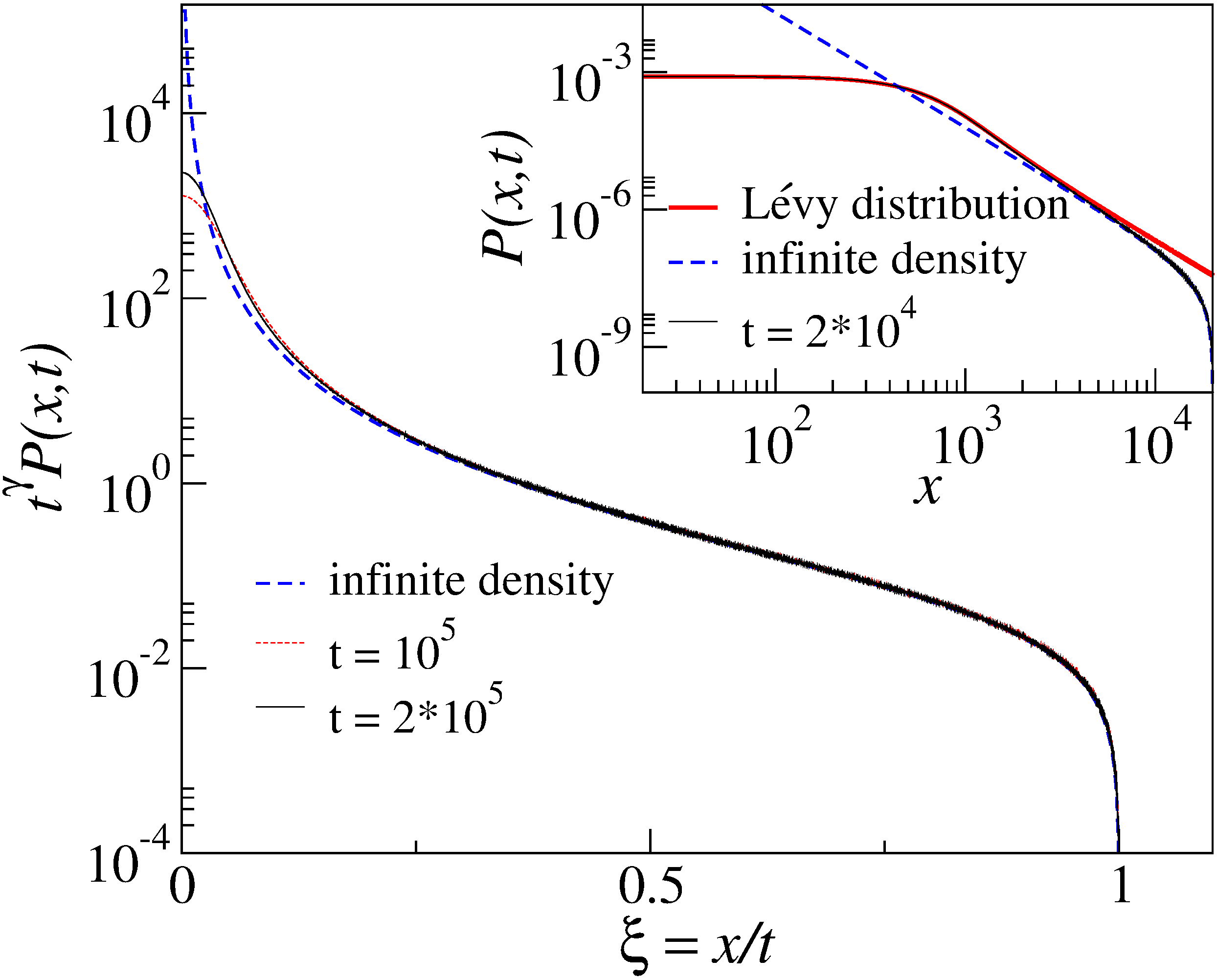}
\caption[Infinite density and L\'{e}vy distribution meet]
{(Color online) Ballistic scaling of the propagators of the  L\'{e}vy walk model 
with random velocities, see Section \ref{lw_with_fluctuations}, for  
the  velocity distribution $h(v)$ uniform on the interval $\in [-1,1]$ and $\alpha = 3/2$.
The corresponding infinite density (dashed blue line), Eq.~(\ref{id1}), matches the tails of the rescaled
propagators. Inset shows the propagator for a shorter time plotted together 
with a rescaled L\'{e}vy distribution (solid red line) and infinite density (dashed blue line).
The propagator is barely visible due to the perfect matching of the theoretical curves at the central region 
near $x = 5000$. The propagators were sampled with $N = 10^{10}$ realizations.
Adapted from \textcite{Rebenshtok2014a}}
\label{figtID1}
\end{figure}

In the intermediate region $t^{1/\gamma} < |x| < t$ the LW propagator scales as $P(x,t) \sim t/|x|^{1+\gamma}$.
Since a power-law is a  scale-free function, there is infinitely many different scaling transformations
which match  power-law tails of the propagators for different times, 
$t^{\xi_{\text{PDF}}}P(x/t^{\xi_x},t)$, with linearly related scaling exponents $\xi_{\text{PDF}}=\xi_x(1+\gamma)-1$.
When $\xi_x = 1/\gamma$ we have the familiar L\'{e}vy scaling, Eq. (\ref{selfsimilarity_ctrw}), 
with $\alpha = 1/\gamma$, which matches the central
parts of the propagators. For $\xi_x = 1$ we have a ``ballistic''
scaling, $t^{\gamma}P(x/t,t)$,  which matches the outmost fronts of the propagators. 
Therefore, this scaling is suitable for the analysis of the asymptotic 
evolutions of the high-order moments which is determined by the propagator tails.
We next introduce a scaled ballistic variable, $\xi = x/t$ and define a density over
this variable  as \cite{Rebenshtok2014}
\begin{equation}
{\cal I}(\xi) = \lim_{t \rightarrow \infty} t^\gamma P(x/t,t).
\label{id1}
\end{equation}
This function scales differently from the L\'{e}vy distribution $L_{\gamma}(x,\sigma(t))$ and
it is non-normalizable\footnote{The non-normalizable 
density function ${\cal I}(\xi)$
should not be thought as a \textit{probability} density function (the latter is always normalizable).
The ID relates to a mathematical concept of a  spatially varying function, 
defined on a smooth manifold that is locally integrable almost everywhere.
Non-normalizable densities are no strangers to physics; check, for example, for
non-normalizable energy densities in black holes and non-normalizable densities of states 
in relativistic quantum dynamics.}, $\int_{-\infty}^{\infty}{\cal I}(\xi) d\xi = \infty$, because of
the power-law singularity in the limit $\xi \rightarrow 0$, 
${\cal I}(\xi) \simeq c_{\gamma}K_{\gamma}|\xi|^{-1-\gamma}$. 
For the general case of the L\'{e}vy walks with
random velocities, the infinite density is given by the following formula \cite{Rebenshtok2014a}:
\begin{equation}
{\cal I} (\xi ) = B \left[ { \gamma {\cal F}_\gamma \left( | \xi| \right)  \over |\xi|^{ 1 + \gamma} } 
- { \left(\gamma -1\right) {\cal F}_{\gamma -1} \left( |\xi | \right) \over |\xi|^\gamma} \right],
\label{eq20}
\end{equation}
where
\begin{equation}
{\cal F}_\gamma (\xi ) = \int_{ |\xi|} ^\infty dv\, v^\gamma  h(v).
\label{eq21}
\end{equation}
It is not surprising that, in contrast to the universal L\'{e}vy-like central part of the Green's function,
the infinite density, which describes function's tails, is specific to the velocity distribution $h(v)$ (it is assumed that this PDF is not heavy tailed so the integral (\ref{eq21}) is finite). 
It accounts for the 
particles with highly-correlated flying histories so that the velocity PDF is  imprinted into the tails of the  PDF 
$P(x,t)$. Two ends meet in the intermediate region where both functions scale similarly,
$L_{\gamma}(x/\sigma(t)) \sim {\cal I}(x/t) \sim x^{-1-\gamma}$. 
For large $t$ two densities match near perfectly at the point $x_c(t) = \left[ { c_\gamma \over L_\gamma(0)} \right]^{{1 \over 1 +\gamma}}
\left( K_\gamma t\right)^{1/\gamma}$ so that the propagator $P(x,t)$ can be approximated with high accuracy
by gluing two functions (properly scaled beforehand) at the point $x_c(t)$, see Fig.~14.

Now the fractional  moments can be calculated as
\begin{eqnarray} 
\nonumber
\langle |x|^q \rangle \simeq
\underbrace{ 2 \int_0 ^{x_c(t)} {L_\gamma \left[ {x \over \left( K_\gamma t \right)^{1/\gamma}} \right]  \left( K_\gamma t\right)^{-1/\gamma} } x^q d x  }_{\text{ 
\mbox{inner region}}} ~~~~~~\\
~+ \underbrace{ 2 \int_{ x_c(t) } ^\infty { {\cal I} \left( {x \over t} \right)  t^{-\gamma}} x^q d x}_{\text{outer tails}}. ~~~~~
\label{secCom03}
\end{eqnarray}
In the long-time limit, the lower limit of the second integral $x_c(t)/t\to 0$
while the upper limit of the first integral is a constant.
For
 $q>\gamma$ the second integral is by far larger than the first, hence we may neglect
the inner region and  get
\begin{equation}
\langle |x|^q \rangle \sim 2 t^{ q + 1 -\gamma} \int_0 ^\infty {\cal I} \left( \xi \right) \xi^q d \xi
\label{secCom05}
\end{equation} 
When  $q<\gamma$ the contribution from the second integral
is negligible in the limit $t \to \infty$ and the upper limit of the first integral is taken
to infinity, hence, after a change of variables
$y = x/\sigma(t)$, and using
the symmetry $L_\gamma(y) = L_\gamma(-y)$, we are left with
\begin{equation}
\langle |x|^q \rangle \sim \left( K_\gamma t\right)^{ q \over \gamma} \int_{-\infty} ^\infty L_\gamma(y) |y|^q d y. 
\label{secCom07}
\end{equation}
These results prove  that sub-ballistic L\'{e}vy walks belong to the class of strongly anomalous diffusion processes 
whose  asymptotic moments satisfy Eq.~(\ref{fractionalmoments}) with $q\nu(q) = q/\gamma$ for $q < \gamma$
and $q\nu(q) = q +1 - \gamma$ for $q > \gamma$, see Fig.~5. There is also a phase-transition-like crossover at the point $q_c = \gamma$ 
in terms of the moment prefactors $M_q$, see inset in Fig.~5. 
Finally, the knowledge of the both, the L\'{e}vy PDF and infinite  density, 
allows one to calculate observables that are not integrable with
respect to either of the two densities, for example $f(x) = 1+x^2$ \cite{Rebenshtok2014a}.

\subsection{Memory effects and ergodicity breaking in L\'{e}vy walks}\label{memory_effects}
In general, a continuous time random walk models is a non-Markovian process, 
meaning that the future of a particle depends not only on particle's 
current state, namely its position and velocity, 
but also on its pre-history, 
like how long it was waiting, or how long it was flying already. 

As a result, a CTRW process can not be fully characterized by its PDF, 
but requires the knowledge of all higher order correlation functions \cite{hanggi1982}. 
However, CTRWs and L\'{e}vy walk models we considered so far, where each next step is independent of 
the previous, represent the so-called semi-Markov processes. In a semi-Markov process, the  points where jumps or velocity changes occur
 form a Markov chain; the renewal events at those points erase all previous memory. In between the renewal points, 
to predict the future of the particle, we need to know how long it was in its current state. In some cases, 
the limiting transport equations are consistent with Markovian dynamics, like in the case of normal diffusion 
Eq. (\ref{diffusion_equation}) or superdiffusion Eqs. (\ref{levy_flight_ks}), and (\ref{levy_flight_tk}) preserve the continuity of evolution for the times exceeding the average waiting times.  
In other cases, when the mean time diverges, the asymptotic transport equations are obviously of a non-Markovian nature, as in the 
case of CTRW in the subdiffusive regime, when the corresponding transport equation has a fractional time derivative.  
A large body of work addressing the semi-Markov property and its consequences for the CTRW models exists and below we will look 
only at those of them which are pertinent to L\'{e}vy walks. 

In the context of L\'{e}vy walks, there are two important interrelated issues which relate 
to the power law distributed flight times. 
The first issue concerns the effects of the initial distribution of particles with respect to their flight 
times on future evolution of the PDF $P(x,t)$ \cite{aquino2004,barkai2003,zaburdaev2003,barkaicheng2003,sokolov2001,zaburdaev2009}. 
The second issue is the weak ergodicity breaking and it points to the fact that time 
and ensemble averaged quantities can be different from each other \cite{bel2005,rebenshtok2007,rebenshtok2008}.

The problem of the initial preparation of the system of particles is important for all random walk models. 
So far we always assumed that all particles were introduced to the system at $t=0$, that 
is they had no history. In this case, the probabilities to make the first jump after a certain waiting time, 
or to make the first turn after a flight time, are governed by the same waiting time or flight time PDFs, 
$\psi(\tau)$. However, if at $t=0$ a particle has already collected some ``history'', 
for example, it was sitting at a given point or it was in the state of flight for some time $\tau_1$, 
then the PDF for it to make the first jump (first turn) at time $\tau$ is in general different from $\psi(\tau)$. 
This case is handled by the so-called renewal theory (in this simple case, it is just the implementation of the 
conditional probability formula), see \textcite{haus1987,tunaley1974}:
\begin{equation}
\psi_{\text{first}}(\tau|\tau_1)=\frac{\psi(\tau+\tau_1)}{\Psi(\tau_1)}.
\label{renewal}
\end{equation}
The only distribution function which is not affected by the pre-history is the exponential distribution and because of 
that is often called memoryless. 
In general, the initial distribution of particles over the flight or waiting times may affect the following evolution. 
An approach to incorporate this distribution was developed for CTRW model and can be extended to the L\'{e}vy walk 
case \cite{zaburdaev20082,zaburdaev2003}. Here we mention one important example of the memory effects. 
Assume that before starting observation we let the system evolve for time $t_1$ and then require that $t_1\rightarrow\infty$. 
This is a so-called {\em equilibrated} [or stationary \cite{klafter1993}] setup where we assume that the system reaches a certain equilibrium before we start 
measuring it. In contrast, the setup where all particles are introduced to the system at $t=0$ 
and do not have pre-histories is called a {\em non-equilibrated} [non-stationary \cite{klafter1993}] setup. 
For the equilibrated case, one has to imagine a system with an infinite number of particles 
in unbounded domain but with a fixed uniform density. 
Particles evolve according to their random walk model for an infinite time. 
At some time point which we denote as $t=0$ we mark all particles located at $x=0$ and then
 follow the evolution of marked particles only. 
It can be shown that in the equilibrated setup, the probability 
of making the {\em first} reorientation event  after the observation started is given by the following PDF 
[see e.g., \textcite{denisov2012} for a simple derivation]:
\begin{equation}
\overline{\psi}(t)=\frac{1}{\langle\tau\rangle}\int\limits_{0}^{\infty}\psi(t+\tau)d\tau.
\label{equilibrated}
\end{equation}
If the flight times are too long, such that the mean flight time diverges, 
there is no sense to speak about the equilibrated setup as it simply does not exist. 
The pre-history affects only the probability of the very first reorientation effect to occur; 
in terms of the transport equations, it will lead to the new terms on the right hand sides of Eqs. (\ref{nu})-(\ref{Plevywalk}) 
\begin{eqnarray}
\nu(x,t)&=&...+\overline{\psi}(t)\delta(|x|-vt)P_0\label{nulevywalkequilibrated}\\
P(x,t)&=&...+\overline{\Psi}(t)\delta(|x|-vt)P_0\label{Plevywalkequilibrated},
\end{eqnarray}
and consequently to different propagators. Here the corresponding probability of not changing the 
direction till time $t$, $\overline{\Psi}(t)$, is given by the similar integration as in Eq. (\ref{equilibrated}): 
$\overline{\Psi}(t)=(1/\langle\tau\rangle)\int_{0}^{\infty}\Psi(t+\tau)d\tau.$ Finally, 
the exponent in the power law tail of $\overline{\psi}(\tau)\propto t^{-\gamma}$  is smaller in the case of equilibrated setup 
than in the non-equilibrated case $\psi(\tau)\propto t^{-1-\gamma}$, 
meaning a  longer lasting influence of the initial distribution on the consequent evolution. 
Understanding of these memory effects is important for the analysis of experimental data or comparison of theory and simulations, as we will exemplify when discussing applications.

The problem of weak ergodicity breaking (WEB) \cite{bouchaud1992weak}
is of a great interest both in theoretical  and experimental communities \cite{he2008,lubelski2008,margolin2005_2,brokmann2003,weigel2011,jeon2011}. WEB, similarly to memory effects discussed above, is found in systems with temporal dynamics governed by power-law distributed time variables with diverging means \cite{barkai2008}. Ergodicity breaking is called weak if the whole phase space of the system can be explored, but the ergodicity is never reached because the characteristic times involved in the corresponding process are always of the order or longer than the total measurement time. In practice it means that the time average of a certain quantity itself is random and can be characterized by a non-trivial distribution. In case of the fully ergodic system the distribution of time averages has a shape of the delta function at the value of the corresponding ensemble average. The most pronounced effects of WEB can be observed for subdiffusive systems \cite{bel2005,he2008,lubelski2008}, however, L\'{e}vy 
walks, as they may 
involve flight time distributions with infinite flight times, also exhibit WEB. In several recent studies \cite{Akimoto2012,froemberg2013,Godec2013}, the effects of WEB in L\'{e}vy walks were investigated on the example of the mean squared displacement calculated as time and ensemble average. The time averaged MSD is defined as:
\begin{equation}
\overline{\delta x^2(\tau)}=\frac{1}{(T-\tau)}\int\limits_{0}^{T-\tau}[x(t+\tau)-x(t)]^2dt,
\end{equation}
where $T$ is the measurement time and $\tau$ is the lag time. 
Several observations were made. In the superdiffusive  sub-ballistic regime, 
the time averaged MSD for finite measurement time $T$ shows different apparent 
scaling for large $\tau$ (but $\tau$ is still much smaller than $T$). 
Some of the individual trajectories can even demonstrate the subdiffusive behavior. \textcite{Godec2013} attribute this effect to the finiteness of the trajectories. Another quantity which can be constructed is the ensemble average of the time averaged MSD $\langle\overline{\delta x^2(\tau)}\rangle$. This quantity now can be compared with the ensemble averaged MSD, $\langle x^2(\tau)\rangle$, namely by calculating the ratio of the former to latter giving the so-called ergodicity breaking parameter, $\mathcal{EB}=\langle\overline{\delta x^2(\tau)}\rangle/\langle x^2(\tau)\rangle$. For the superdiffusive L\'{e}vy walk in the limit $\tau\rightarrow\infty$ it tends to a constant value $\mathcal{EB}=1/(\gamma-1)$. To draw the connection to the previously discussed memory effects we note that the ensemble-time averaged MSD, $\langle\overline{\delta x^2(\tau)}\rangle$, corresponds to the ensemble averaged MSD of the equilibrated setup, whereas simple ensemble average MSD is calculated for the non-equilibrated setup 
\cite{klafter1993}. In \textcite{froemberg2013}, the ballistic regime of L\'{e}vy walks was considered as well. In that case WEB effect is also present, but, surprisingly, is not as pronounced as in the superdiffusive case. The ensemble-time average can be also calculated analytically. Its leading term (assuming $\tau/T\ll 1$) is given by $\overline{\delta x^2(\tau)}\sim (v_0\tau)^2$. Note that the ensemble averaged MSD has a different pre-factor $\langle x^2(\tau)\rangle=(1-\gamma)(v_0\tau)^2$. Furthermore the fluctuations of the shifted quantity $\overline{\delta x^2(\tau)}-(v_0\tau)^2$ can also be quantified, see \textcite{froemberg2013} for details. 

Ergodicity breaking effects discussed here are essential for the analysis of the experimental data, 
particularly in biology, where due to the limited number of measurements one has to resort to the 
time averaging and always deals with trajectories of finite length. 
In addition, understanding of the fluctuations in time averaged observables can help to obtain
more information about the underlying stochastic process \cite{Schulz2014, Dechant2014}. There is also a very recent work on Einstein relation, fluctuation dissipation, and linear response \cite{froemberg2013,Metzler_response}. We refer the interested reader to find out more details about these topics in a recent review by \textcite{Metzler2014}.

\subsection{Langevin approach and fractional Kramers equation}\label{langevin_section}
In the Introduction we mentioned the Langevin equation as 
an approach to stochastic transport phenomena complimentary to the random walk paradigm.  
Random walks have their strength in the flexibility of the model construction 
and amenability of the corresponding transport equations to the analytical treatment. 
Langevin equations utilize the machinery of stochastic differential equations 
and provide a link to the Fokker-Planck equation \cite{risken1996}. 
The Langevin equation was originally proposed in 1908 to describe the Brownian motion \cite{langevin}. 
Since then it has grown into a powerful tool of modern physics \cite{Coffey2012}. 
This success was certainly supported by rigorous mathematical foundations laid by 
mathematicians, such as N. Wiener, It\={o}, and Stratonovich. 
In many cases, the equivalence of random walks and the corresponding Langevin equations can be explicitly demonstrated in a proper limit.
That includes also regimes of normal and anomalous diffusion, with both sub- and superdiffusion. 
This line of research led to the formulation of the fractional 
Fokker-Planck \cite{metzler1999,barkai2000,barkai2001,chechkin2003,heinsalu2007} 
and Klein-Kramers equations \cite{metzler2002,barkaisilbey2000,dieterich2008,friedrich2006,eule2007}, and it
remains an active field of research up to now.

One of the  Langevin pathways to L\'{e}vy walks was proposed recently by \textcite{kessler2012} (we 
discuss it in more detail in Section \ref{cold atoms}).
In brief, the dynamics of the particle is governed by the Langevin equation with a standard white 
noise term but with a non-linear friction, Eq. (\ref{cold_barkai1}). On a mesoscopic scale the trajectory 
of the particle, $\{x(t),p(t)\}$, can be divided into flights, that are events of unidirectional motion. 
Time duration of the $i$-th event, $\tau_i$, is given by the time lag  between two 
consecutive ``turns'' marked by sign alternations of the 
momentum $p(t)$. It was shown by \textcite{Marksteiner1996} that, 
within a certain parameter range,
the PDF of the flight time  scales as $\psi(\tau) \propto \tau^{-1-\gamma}$, 
with a parameter-dependent exponent $\gamma$.
The flight time and flight distance are coupled in a non-trivial way, 
such that the corresponding random walk description of the particle dynamics does not reduce to
the simple L\'{e}vy walk model with linear coupling between $x$ and $t$ \cite{kessler2012,barkai2014}. 

In attempt to model real-life continuous trajectories similar 
to those of Brownian motion  but exhibiting anomalous diffusion, 
a Langevin equation with a special form of multiplicative noise 
term was suggested \cite{lubashevsky2009, lubashevsky20092}. 
A trajectory generated by  this Langevin equation, 
when sampled at fixed time intervals, will reproduce the behaviour 
of the L\'{e}vy flight model. 

Finally, to achieve a one-to-one correspondence 
of the Langevin picture and the L\'{e}vy walk model 
one can use the method of subordination \cite{fogedby1994}. 
In this case an additional variable is introduced, which is 
called an operational time. The dynamics of velocity is 
happening in this operational time and can be tuned to 
produce the desired velocity distributions, $h(v)$. The 
real time is connected to the operational time via its own 
stochastic equation with a noise term which generates long 
traps in real time space. Those traps correspond to the long 
flight intervals as required for the L\'{e}vy walk model, 
see \textcite{eule2012} for more detail. 
This  phenomenological approach allows to connect the world of 
Langevin equations to L\'{e}vy walks where the constant speed 
of a particle during a long time interval is crucial, 
remaining, at the same time, very different from a standard Brownian trajectory 
where the velocity is constantly changing. 


A complementary approach to study anomalous stochastic transport 
is to generalize the Kramers-Fokker-Planck equation.  
Several  versions of generalized Kramers-Fokker-Planck equations were suggested in the literature [they are summarized in \cite{eule2007}]. 
Here we follow a scheme by \textcite{friedrich2006} with a ballistically moving particle subjected to 
random kicks which alter its velocity. 
Provided the times between consecutive collisions are distributed as a power law, 
the fractional Kramers-Fokker-Planck equation for the joint position-velocity distribution, $f(\mathbf{r},\mathbf{v},t)$, can be obtained:
\begin{equation}
\left(\frac{\partial}{\partial t}+\mathbf{v}\nabla_{\mathbf{r}}+\mathbf{F}(\mathbf{r})\nabla_{\mathbf{v}}\right)f(\mathbf{r},\mathbf{v},t)=L_{FP}\mathcal{D}^{1-\gamma}_{t}f(\mathbf{r},\mathbf{v},t).
\label{gFKP}
\end{equation}
Here $\mathcal{L}_{FP}$ is the Fokker-Planck operator  
$L_{FP} f=\tilde{\gamma}\nabla_{\mathbf{v}}\cdot(\mathbf{v} f)+\kappa\triangle_{\mathbf{v}}f$, and $\mathcal{D}^{1-\gamma}_{t}$ 
is the fractional substantial derivative defined through its Laplace transform 
as 
$\mathcal{L}\left[\mathcal{D}^{1-\gamma}_{t}f(t)\right]
=\left(s+\mathbf{v}\cdot\nabla_{\mathbf{r}}+\mathbf{F}(\mathbf{r})\cdot\nabla_{\mathbf{v}}\right)^{1-\gamma}f(s)$. 
Further, $\mathbf{F}(\mathbf{r})$ is the external force, $\tilde{\gamma}$ is the generalized friction coefficient, 
and $\kappa$ is related to the amplitude of the noise in the corresponding Langevin equation for the velocity of the particle. 
For rigorous derivation and many technical details we refer to \textcite{friedrich2006,friedrich20062,Carmi2011}. 
By appropriate modifications, the above equation can be simplified to give 
the equations of the random walk with random velocity model \cite{zaburdaev2008} and of the L\'{e}vy walk model \cite{eule2008}. 
The  genetic link between L\'{e}vy walks, Langevin equations, and fractional Fokker-Planck equations certainly needs to be 
investigated further \cite{lubashevsky2009,lubashevsky20092,magdziarz2012,Turgeman2009}.

At this point it would be timely to mention two relevant approaches (which however fall beyond the scope of this review). 
That is the fractional Brownian motion, 
which is characterized by a Gaussian but time correlated noise \cite{mandelbrot1968}, 
and the generalized Langevin equation, which contains an integral operator with a memory kernel 
on the right hand side of the equation governing the velocity increments \cite{zwanzig2001}. 
Both approaches  are  useful in describing various transport processes across disciplines. 
They possess, however, very distinct features that are different from those of the random walk concept; 
in relation to the questions already discussed in this review, we would like to direct the reader to 
\textcite{magdziarz2009, eliazar2013, meroz2013}.

\section{L\'{e}vy walks in physics}
\label{physics}
\begin{figure}[t]
\center
\includegraphics[width=0.5\textwidth]{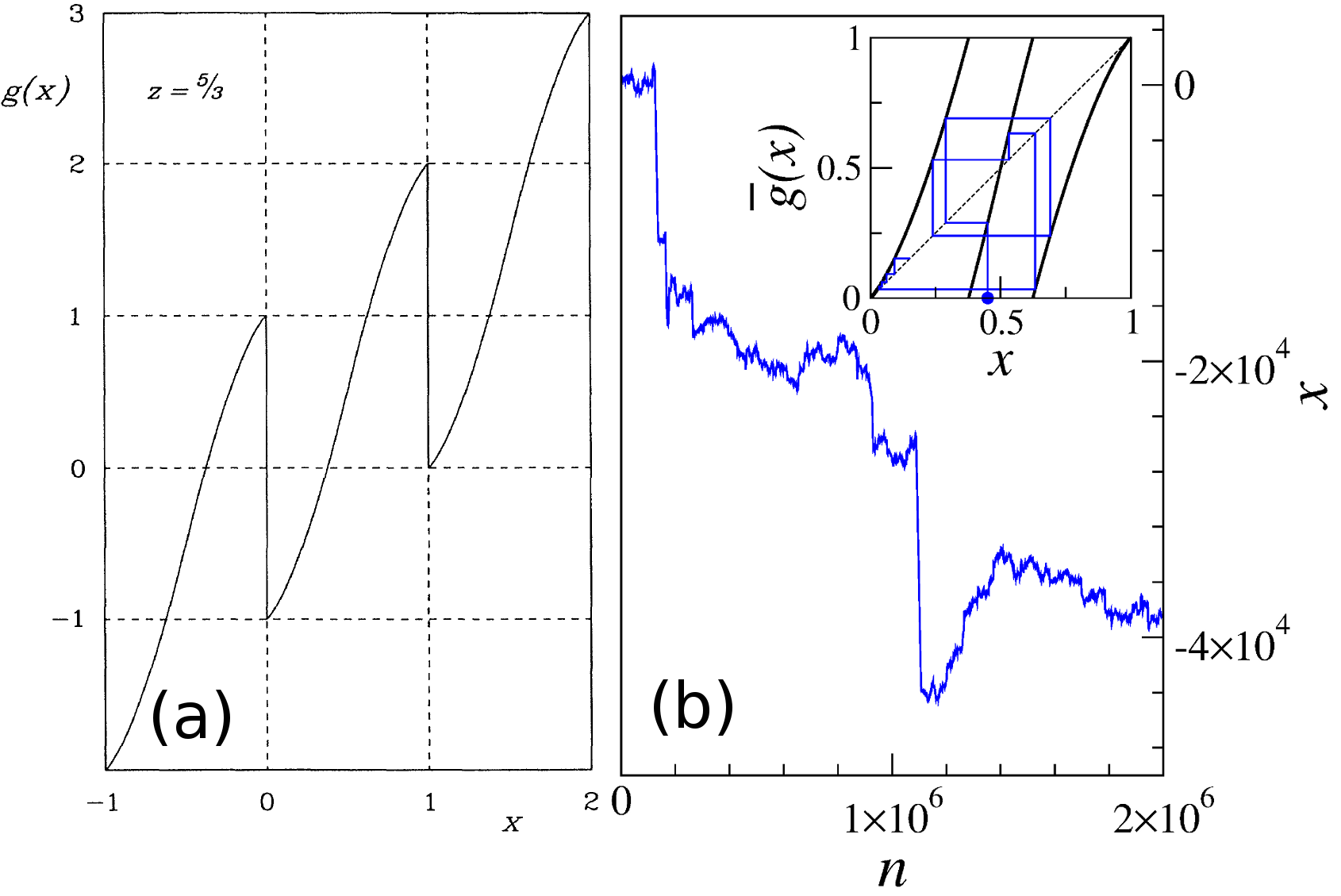}
\caption[The map]
{(Color online) (a) Map $g(x)$, Eqs.~(\ref{geisel0}-\ref{geisel}),
for $z = 5/3$. Adapted from \cite{Zumofen1993}; (b) A trajectory $x(n)$ obtained by iterating the map from the initial
point $x(0) = 0.45$. The inset shows the reduced map $\bar{g}(x)$, Eq.~(\ref{geisel}), with first ten iterations
of the initial point $x(0)$ ($\bullet$). The branch on the left (right) part is responsible for a decrement 
(an increment) of the cell number $N = \mathsf{int}(x)$ produced by the extended map $g(x)$. 
The parameter  $\epsilon = 10^{-4}$.}
\label{figt8a}
\end{figure}

Physics is a natural habitat of random walk models \cite{Gennes1979,Fernandez1992,Weiss1994}. 
During last twenty five years, the
L\'{e}vy walk model has found a number of  applications, mostly in classical chaos and nonlinear 
hydrodynamics \cite{Klafter1996,Shlesinger1999,klaftersokolov2011}.
\textcite{geisel1984} were the first to consider an intermittent 
ballistic motion with power-law flight-time PDFs 
in the context of deterministic chaos. 
Later on, \textcite{geisel1985} studied a model of the  rotational phase dynamics 
in a Josephson junction, by using a  one-dimensional map
\begin{eqnarray}
x_{n+1} = g(x_n),
\label{geisel0}
\end{eqnarray}
assuming discrete translational and reflection symmetries,
\begin{eqnarray}
g(x + N) = g(x) + N, ~~ g(-x) = -g(x).
\label{geisel1}
\end{eqnarray}
where $N$ denotes the number of the unit box, $[N-1,N]$.
With this setup, the definition of the map is required
only for the reduced range $0 \leq x \leq 1$. It can be extended then over 
$x \in [-\infty, \infty]$ by using symmetries in Eq.~(\ref{geisel1}). \textcite{geisel1985}
used a nonlinear map,
\begin{equation}
\bar{g}(x) = \left\{
                \begin{array}{cc}
                 ~~~~~(1+\epsilon)x + ax^z - 1, ~~~~~~~~~~~ 0 \leq x \leq 1/2\\
                2 - (1+\epsilon)(1-x) - a(1-x)^z, ~~ 1/2 \leq x \leq 1
                \end{array}
              \right.
\label{geisel}
\end{equation}
where $\epsilon$ is a small constant and $a = 2^z(1 - \epsilon/2)$. The profile
of the corresponding extended map $g(x)$ for $z = 5/3$ is shown in Fig.~\ref{figt8a}(a).
Because of the power-law form of the second term on the right hand side of Eq.~(\ref{geisel}), the reduced  variable 
$x~\mathsf{mod}~1$ tends to cluster near the semi-stable points $x_N=N$.  
Once entered into the vicinity of one of these points, 
a trajectory performs a uni-rotational motion with a near constant rate $|x_{n+1} - x_{n}| = 1$.
The rotation direction depends on the point to  which the trajectory stuck, $\upsilon = 1$ when it 
stuck to $x^+_{N} = N - 0$ [$x=1$ in the reduced map $\bar{g}(x)$]
and $\upsilon = -1$ when to $x^-_{N} = N + 0$ [$x=0$ in the reduced map $\bar{g}(x)$]. 
It was found that a histogram of the numbers of iterations, 
or the time, if we set $t = n$, spent by the system in a rotation state, yields a 
long-tailed  distribution  $\psi(t) \propto t^{-z/(z -1)}$. 
The results of a numerical sampling reveal that the MSD $\langle x^2(t)\rangle$ 
scales  as in Eq.~(\ref{msd_scaling1}), 
with $\gamma = 1/(z -1)$. It was then shown to be a clear-cut case of a L\'{e}vy walk with the constant 
speed $\upsilon =1$ and the 
exponent  $\gamma$ \cite{Shlesinger1985}, see Fig.~\ref{figt8a}(b). 
A complete evaluation of the diffusion in the intermittent maps
within the L\'{e}vy walk framework was presented by \textcite{Zumofen1993}. 
The next ``dynamical'' realization of the L\'{e}vy walk was found in Hamiltonian 
chaotic systems \cite{Shlesinger1993,Zumofen1994,klafter1994}. 
This was a case when the LW concept perfectly matched a peculiar dynamical effect, 
in  appearance similar to the intermittency in  the dissipative 
maps with power-law singularities  \cite{meiss1984,geisel1987}.
The machinery behind the Hamiltonian stickiness is   related to specific fractal structures living in the phase space of chaotic Hamiltonian systems \cite{MacKay1984,Meiss1992}.
We  will discuss this issue in more detail in the next section. 

With this section we are not up to a comprehensive 
historical review.  We want to present L\'{e}vy walks  in physics as something (re)emergent and promising
rather than something residual and  completed. In the following subsections we will  concentrate on the most 
recent advances and results, both theoretical and experimental,  which underline  the potential and universality of the concept.

\subsection{L\'{e}vy walks in single-particle Hamiltonian systems}

The subject of L\'{e}vy walks in low-dimensional Hamiltonian chaos is already twenty years old \cite{Zumofen1994,klafter1994}.
We  start with a brief outline of it not because of the historical reason but 
because it will help to understand better the recent developments that will be discussed next.

The phase space of a non-integrable single-particle Hamiltonian 
system is \textit{mixed} and consists of different  invariant manifolds, that are chaotic layers, regular
islands, tori, etc. \cite{suz92}. The $i$-th manifold  can be characterized by an  averaged value of any
observable, for example velocity, $\upsilon_{i} = \langle \upsilon_{i}(t) \rangle_{t\rightarrow\infty}$. 
The average velocity of a manifold  might be nonzero and for a regular island it is determined by the winding number of the elliptic orbit 
at the island center. 
A chaotic layer is well separated from regular manifolds by KAM-tori \cite{suz92} so that a trajectory initiated inside the layer
cannot enter a regular island even when the latter is embedded into a chaotic sea.   
A ``coastal area'' near
the island is  structured by  \textit{cantori} \cite{MacKay1984}, which form partial barriers 
for the trajectories. Once entered into the region enclosed by a cantorus, a trajectory
will  be trapped in the vicinity of the corresponding island for a long time. 
During this sticking event \cite{Meiss1986}, the trajectory reproduces the dynamics of the orbits located inside 
the island. If the corresponding island is transporting,  $\upsilon_i \neq 0$,  
the sticking event produces a long ballistic flight with velocity $\upsilon_i$. It has been found that power-law tails of  sticking
time PDFs, $\psi(\tau) \propto \tau^{-1-\gamma}$,  is a general feature of Hamiltonian chaos 
which is related to the self-similar hierarchical structure of cantori  \cite{Meiss1986,Meiss1992,geisel1987}. 
With this finding all needed ingredients were collected and  a link between ``strange kinetics'' of  Hamiltonian chaos and L\'{e}vy 
walks was established \cite{Shlesinger1993}.
\begin{figure}[t]
\includegraphics[width=0.45\textwidth]{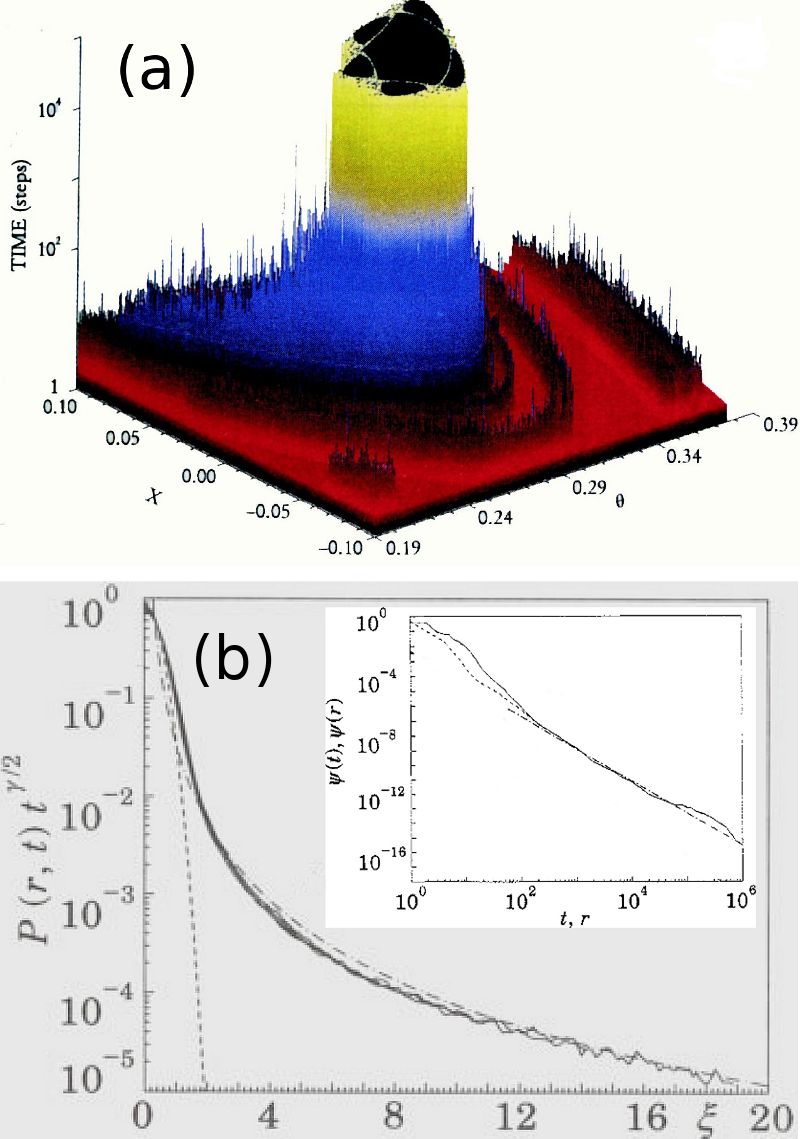}
\caption{(Color online) L\'{e}vy walks in the standard map. 
(a) Time $t_{\text{exit}}(x,\theta)$ it takes for a trajectory initiated at the point $(x,\theta)$
to exit from a vicinity of a regular island around the period-five orbit (black area on the top of the 
distribution). (b)
The rescaled propagators for different time, $t=100, 200, 400, 800, 1600$,
$\xi = x/t^{1/\gamma}$. 
The dashed line is a Gaussian fit and the dashed-dotted line is a power 
law fit $f(\xi) \propto \xi^{-1-\gamma}$. The inset shows the sticking-time (solid line), $\psi(t)$,
and the flight-length (dashed line), $\psi(x)$, PDFs at the vicinity of
the period-five island.
The dashed dotted-line is a power-law with exponent $\gamma = 1.2$. The spatial variable $x$ is denoted $r$
in the original work. The $y$-axis label on (b) should read as $P(r,t)t^{1/\gamma}$.
Adapted from  \textcite{Zumofen1994}.}\label{SM}
\end{figure}

In a Hamiltonian system possessing the time-reversal symmetry,  ballistic islands always exist in pairs 
and have identical sticking time PDFs.
If, in addition,  the long-time dynamics of the system is governed by only two symmetry-related sticky ballistic islands, 
with a sticking time exponent $\gamma$, then the system dynamics will realize the standard L\'{e}vy walk, Fig.~\ref{figt1}(b). 
\textcite{Zumofen1994} considered the kicked-rotor map, 
an archetypical Hamiltonian model \cite{suz92}, 
\begin{eqnarray}
x_{n+1} = x_{n} + K\sin(2\pi\theta),~~~\theta_{n+1}=\theta_{n}+x_{n+1},
\label{sm}
\end{eqnarray}
where $K$ is the stochasticity parameter, as an example.  For $K=1.03$ the system 
phase space represents a chaotic sea which extends over the whole $x$-region. The long-time 
dynamics of the system is governed by two symmetry-related islands enclosing period-five elliptic orbits 
with velocities $\upsilon = \pm 1$. The islands are sticky, see Fig.~\ref{SM}(a), and the locations of the cantori 
are marked by the sudden increase of the sticking times.
The corresponding sticking time PDF, see inset on Fig.~\ref{SM}(b), for $t \gtrsim 10^2$
follows approximately a power law with an exponent $\gamma = 1.2$.  Fig.~\ref{SM}(b) 
shows propagators obtained for different times after they were scaled as in  Eq.~(\ref{selfsimilarity_ctrw}), 
with exponent $\alpha = 1/\gamma$. 
The curves fall on top of each other thus indicating the scaling expected for the propagators of L\'{e}vy walk,
see inset in Fig.~\ref{figt2}.

A L\'{e}vy-walk kinetics has also been found in continuous  ac-driven one-dimensional 
\cite{Gluck1998,Denisov2002}  and 
stationary  two-dimensional  \cite{klafter1994} Hamiltonian systems.
There the key mechanism responsible for the appearance of  anomalous transport
was the cantori-induced stickiness. There are still ongoing debates on the universality of sticking-time exponent(s) 
at the asymptotic limit $t\rightarrow \infty$, with a number of pros and cons for different ``universal'' 
values \cite{Chirikov1999,Cristadoro2008,Shepelyansky2010,Venegeroles2009,Ketzmerick2002}. 
The L\'{e}vy walk can stand  this uncertainty: 
If a PDF of sticking times is well approximated by a power law with 
a particular exponent $\gamma$ over some substantial time interval 
(for example, over several decades in $t$) then the corresponding propagator 
for these times  will scale as in Eq.~(\ref{selfsimilarity_ctrw}), 
with the scaling exponent $\alpha = 1/\gamma$.

It is noteworthy that in Hamiltonian systems only sub-ballistic super-diffusion (see Section \ref{levy_walk} 2)
is possible in the asymptotic limit. This follows from the Kac theorem on the 
finiteness of recurrence
time in Hamiltonian systems \cite{Kac1959,Zaslavsky2002}. Therefore all sticky ballistic manifolds  should have 
finite mean sticking times, so that the corresponding sticking-time PDFs (which are the flight-time
PDFs of the corresponding L\'{e}vy walks), $\psi(\tau) \propto \tau^{-1-\gamma}$,
are characterized by exponents in the range $1 < \gamma \leq 2$  \cite{Denisov2002}.

\begin{figure}[t]
\center
\includegraphics[width=0.45\textwidth]{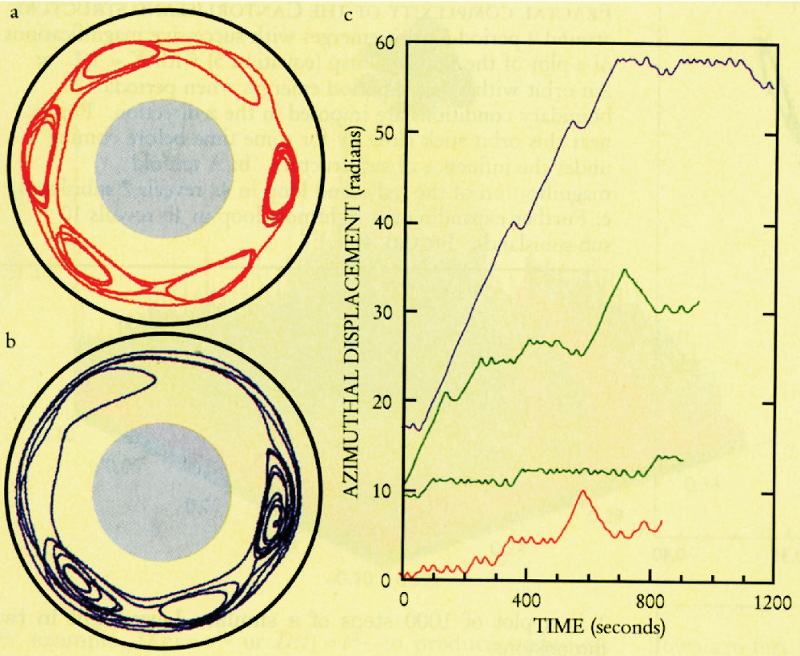}
\caption[Solomon experiment]
{(Color online) Tracer dynamics in a chaotic flow. 
The flow produces a regular chain of six stable vortices which can trap the tracer
as it circles ballistically around the annulus. Two short-time trajectories
are shown on the left part of the figure. The corresponding azimuthal
dynamics is presented on the right panel: while  trajectory (a) (red, lowest on the graph) was trapped
by the chain during all the observation time,   trajectory (b) (blue, upper on the graph) shows that the tracer
produced a long ballistic flight before being trapped. Adapted from \cite{Klafter1996}.
}
\label{figtSol}
\end{figure}

Two-dimensional chaotic advection is another field where the chaotic Hamiltonian phase space, 
with all its trademarks, including cantori and the stickiness phenomenon, appears \cite{aref2014}. 
On the theory level, the dynamics 
of a chaotic flow can be modeled with symplectic equations and the flow stream function
as a Hamiltonian. The path of a passive tracer in the flow can be seen as a
trajectory of the Hamiltonian system. Periodic flow modulations lead to the appearance of mixed 
phase space and regular islands. That idea was behind the first 
experimental observation of the L\'{e}vy walk in a real physical system. In their experiment,
\textcite{Solomon1993} used a rotating annulus tank filed up with fluid. 
The flow was generated by pumping the fluid into and out of the annulus through
the holes in its bottom. This resulted in the appearance of the stable
two-dimensional flow pattern on the surface of the fluid.  Rotational motion of the tracer, monitored by 
by using tracer's azimuthal angle, consisted of ballistic episodes interrupted by trappings of the tracer
by periodic chain of vortices, see Fig.~\ref{figtSol}.

\begin{figure}[t]
\center
\includegraphics[width=0.45\textwidth]{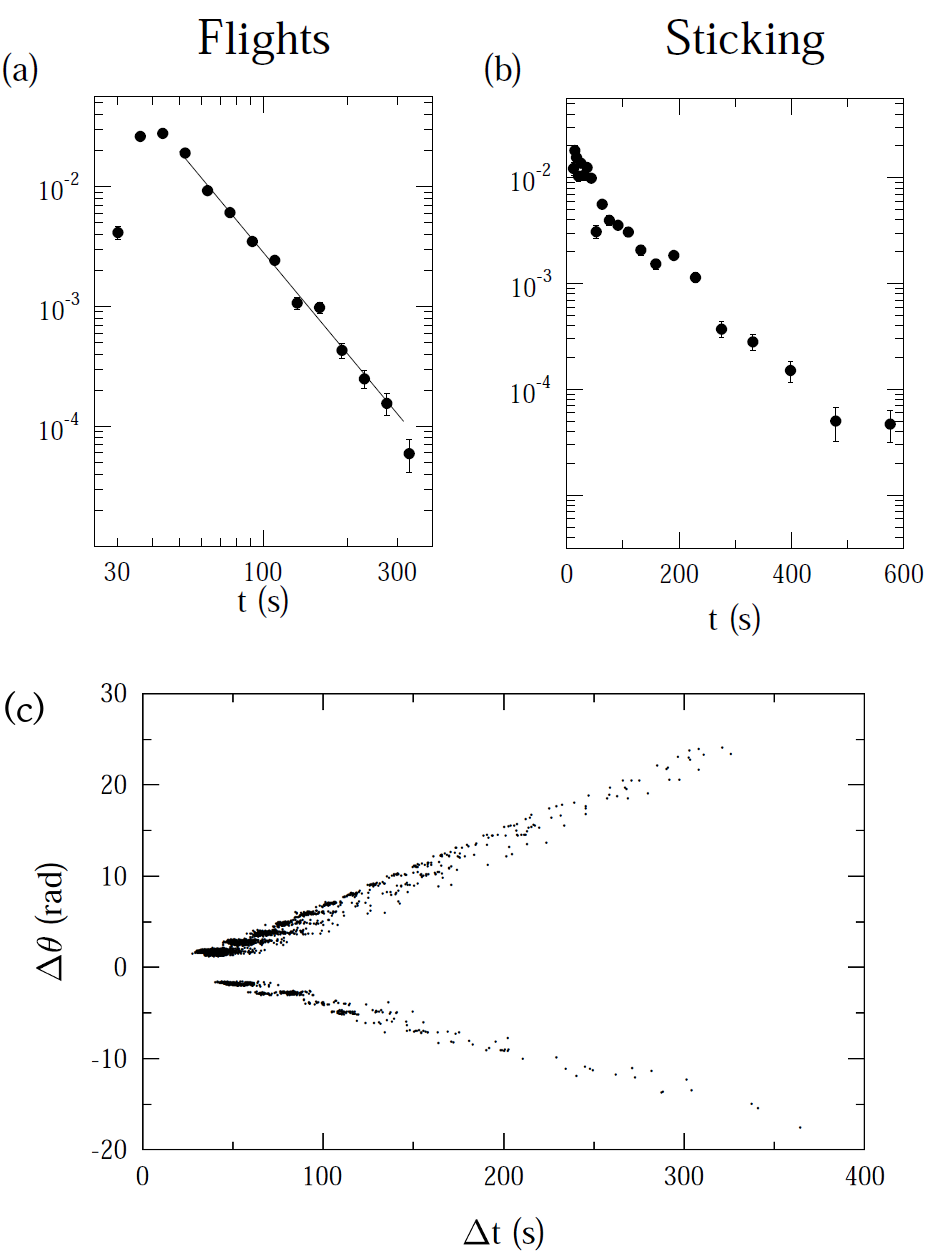}
\caption[Solomon experiment]
{Statistical characteristics of the tracer diffusion in an ac-driven chaotic flow. 
(a) Flight- and (b) sticking-time PDFs obtained for 
a flow with the six-vortex lattice, see Fig.~\ref{figtSol}. Line corresponds to a power law $t^{-1-\gamma}$, 
with $\gamma = 1.5$.  (c) 
Flight length $\triangle\theta$ vs flight duration $\triangle t$. 
The fork-like structure reveals that all flights have near 
constant velocity. Adapted from \cite{Weeks1997}.
}
\label{figtSolb}
\end{figure} 

It followed from the measurements that ballistic flights had near constant velocities and the flight-time PDFs followed 
power-law asymptotics \cite{Solomon1994}, see Fig.~\ref{figtSolb}. At the same time, the sticking-time PDFs 
revealed either an exponential decay or power-law tails with exponent $\gamma_{st} > 1$ so that
the mean sticking time is finite. 
Therefore, following our classification, see Fig.~\ref{figt1}, the process can be 
taken as a L\'{e}vy walk with rests. However, there was a feature: Because of the annulus rotation,
ballistic motion happened predominantly in one direction, clockwise or counterclockwise,  
depending on the rotation direction. This modification of the  walk process can be absorbed into the theory by 
introducing bias in the standard L\'{e}vy walk model, e. g. by making ballistic flights in the positive 
direction less probable than in the negative one. 
The corresponding update was made  by \textcite{Weeks1998} and the model outcomes 
were found to be in a good agreement with 
the experimental measurements. We also refer the reader to the works by \textcite{CastilloNegrete1998} and \textcite{isichenko1992} for a theoretical 
overview of ``anomalous advection'' and dynamical mechanisms behind it.

A   potential of the L\'{e}vy walk model for generalizations can be illustrated with \textit{Hamiltonian ratchets}, 
ac-driven Hamiltonian systems which are able to generate a constant current in the absence of a bias 
\cite{sokd01prl,Schanz2005,df01pre,dfoyz02pre}. 
A directed chaotic transport appears due to  violation of  the time reversal-symmetry with a zero-mean drive \cite{Denisov2014}.
The set of regular islands, submerged into the chaotic layer, 
becomes  asymmetric, so that there are islands with nonzero velocities which do not have symmetry-related twins.  
This leads to the violation of the  balance between ballistic flights 
in opposite  directions and the appearance of  a strong current \cite{df01pre}. The asymmetric 
generalization of the  L\'{e}vy walk process by \textcite{Weeks1998} 
is able to capture many features of the Hamiltonian ratchet dynamics  \cite{dkuf01pd,Denisov2004}.

\subsection{L\'{e}vy walks in many-particle Hamiltonian systems}
\label{many_body}
In many-particle  systems with unbounded interaction potentials, such as nonlinear chains and lattices,
it is no longer reasonable to talk about diffusion of  particles. 
The individual particle dynamics has an oscillatory character due to the confinement induced 
by the interaction with its neighbors. The collective system dynamics 
creates a ``tissue'' which can react to small perturbations locally affecting its dynamics.
The perturbation transport defines overall energy, correlation and information transport 
through a lattice \cite{Helfand1960,Torcini1995,torchini1997,primo2007,Giacomelli2000}.

Consider a many-particle system at microcanonical equilibrium, with a Hamiltonian
\begin{equation}
H_{\text{total}}(\{x_i,p_i\} ) = \sum^{N}_{i=1} H_i,
\label{eq:total}
\end{equation}
where $H_i = H(x_i, x_{i-1}, x_{i+1}, p_i)$ is the energy of the $i$-th particle with position $x_i$ and momentum $p_i$.
It is also assumed that the Hamiltonian guarantees the  preservation of the zero total momentum of the system, $P = \sum^{N}_{i=1} p_i = 0$.
At the initial time $t=0$ one of the bulk particles receives an external perturbation. 
The system gains a small amount of extra energy $E_P$ which is conserved due 
to the Hamiltonian character of the system evolution. However, the perturbation does spread as the 
perturbation energy is shared by a constantly growing number of particles. Remarkably,
the spreading is universally limited by a finite velocity, $v_{0} < \infty$, that  at a given time $t$ the  
perturbation is almost completely confined to the interval $[ -v_{0} t, v_{0} t]$ 
(``almost'' means that outside the cone the perturbation is exponentially  diminished). 
The fundamental fact of the cone's existence, so-called ``Lieb-Robinson bound'' for classical systems, has a status of a
mathematical existence theorem \cite{Marchioro1978, Nachtergaele2009}. Altogether, that was a strong 
hint to consider the perturbation spreading as a diffusion process, treat the normalized local excess of energy 
$\triangle E(i,t)$, $\Sigma_{i=1}^{N} \triangle E(i,t) =  E_p$, as a PDF,
$\varrho (i,t)= \overline{\triangle E(i,t)}/ E_p$ ($\overline{{\cdot \cdot \cdot}}$ 
denotes a microcanonical average), and estimate its second moment.
\begin{figure}[t]
\includegraphics[width=0.5\textwidth]{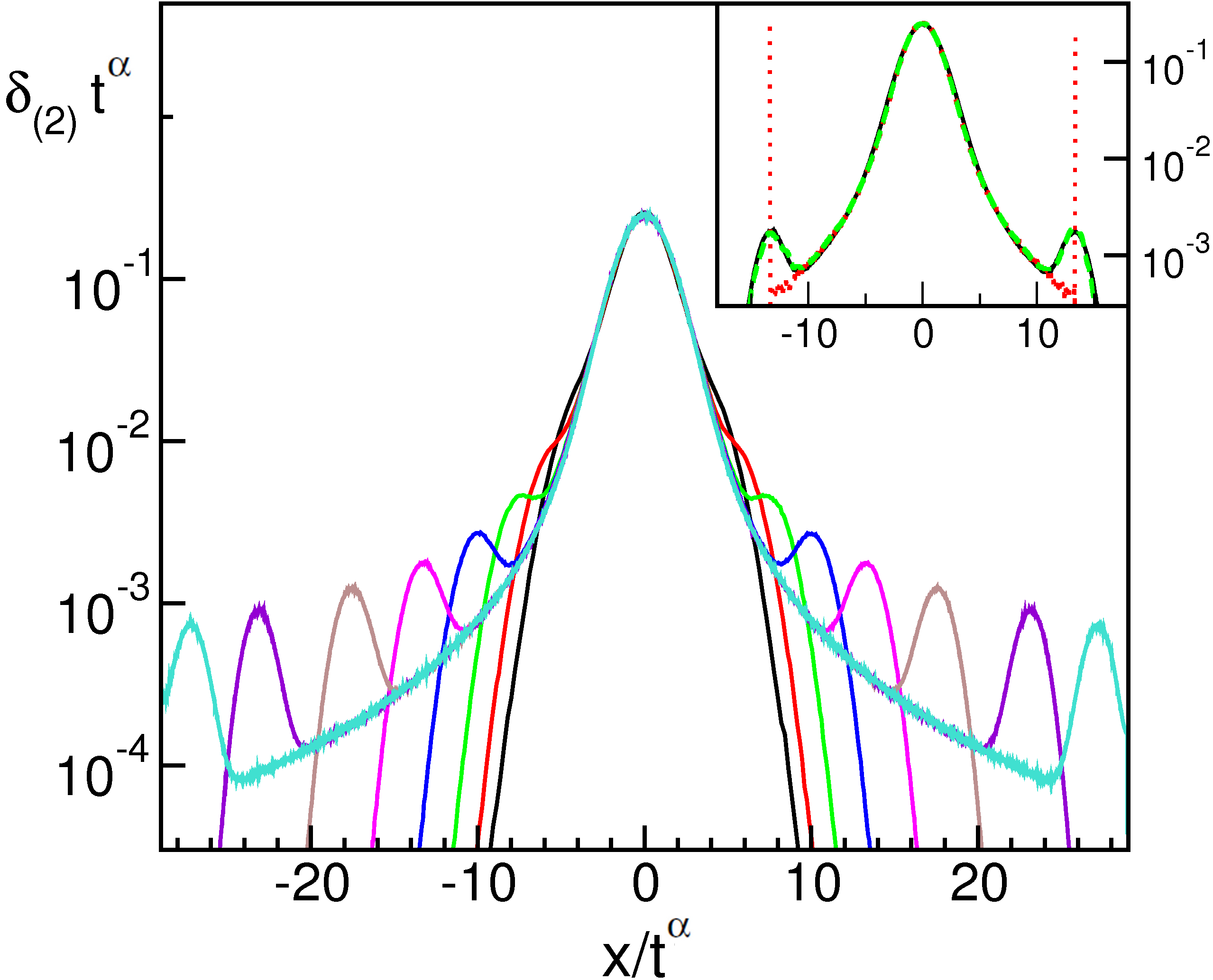}
\caption{(Color online) L\'{e}vy walks in a hard-point gas. 
 Perturbation profiles $\varrho (i,t)$ (denoted by $\delta_{(2)}(i,t)$ in the original publication) 
at $t=40$, $80$, $160$, $320$,
$640$, $1280$, $2560$, and $3840$ (the width increases with time) for the energy per  particle $\varepsilon = 1$, 
rescaled as in  Eq.~(\ref{selfsimilarity_ctrw}) with the exponent $\alpha = 1/\gamma = 3/5$.
In the inset, the profile at $t=640$ (solid line) is
compared with the propagators of the standard L\'{e}vy 
with the velocity $v=1$ (dotted line) and a fluctuating velocities with $D_v = 0.036$ (dashed line). 
Adapted from \textcite{cipriani2005}. 
}
\label{fig:LevyHPG}
\end{figure}

For a one-dimensional hard-point gas  with alternating masses, a protozoan Hamiltonian many-particle model \cite{Casati1976},
it was found that the mean squared displacement $\sigma^2(t) = \Sigma_{i=1}^{N}i^2\varrho (i,t)$ scaled as $\sigma^2(t) \propto t^{\mu}$ 
with the exponent $\mu$ very close to $4/3$ \cite{cipriani2005}. Moreover, a quasi-PDF $\varrho (i,t)$ appeared to 
be the exact propagator of a L\'{e}vy walk with velocity fluctuations, subsection~\ref{lw_with_fluctuations}, 
and exponent $\gamma = 3 - \mu = 5/3$, if we set $i \equiv x$, see Fig.~\ref{fig:LevyHPG}. \textcite{zaburdaev2011,Zaburdaev2012err} further strengthened this finding by showing that
the scaling of the ballistic peaks is identical to that predicted by the L\'{e}vy walk model, Eq.~(\ref{eq:scaling2}).
Perturbation profiles for different values of microcanonical ``temperature'', energy per particle $\varepsilon$, perfectly matched each 
other by assuming that the averaged velocity of the walk and the fluctuations variance both scale 
as $\upsilon_0, D_\upsilon \propto \sqrt{\varepsilon}$.
Similar results were obtained for a FPU $\beta$ chain by using local energy-energy correlation function $e(i,t)$ \cite{Zhao2006} instead 
of a finite perturbation. This switch was induced by the fact that it is not feasible to sample the evolution of 
perturbations in FPU-type systems due to emerging  statistical fluctuations. Although less sharp than in the case 
of hard-point gas, the results obtained for the times $t < 10^4$ revealed the correspondence between 
the correlation function profiles and the 
propagators of a L\'{e}vy walk with fluctuating velocity and exponent $\gamma = 5/3$ \cite{zaburdaev2011,Zaburdaev2012err}.

There is a genetic link between the problem of energy diffusion  and the issue of deterministic heat conduction 
\cite{Helfand1960,liu2012,liu2014}. The latter is typically anomalous in most of nonlinear chains, meaning that the thermal conductivity, $\kappa_T$, 
scales with the length of a chain $L$ as $\kappa_T \propto L^{\eta}$, with $\eta$ between $0$ (normal heat conduction) 
and $1$ (ballistic heat conduction) \cite{Lepri2003}. \textcite{denisov2003} built up a model of a dynamical heat channel
in which energy is carried by an ensemble of non-interacting L\'{e}vy walkers. 
Relatively simple evaluation led to the linear relation between the exponents, 
\begin{eqnarray}
\eta = 3 - \gamma. 
\label{LWheat}
\end{eqnarray}
There are still ongoing debates both on the (non)universality of  Fourier exponent $\eta$ and the 
validity of the single-particle L\'{e}vy walk approach to the heat conduction by many-particle 
chains \cite{eltit2008,liu2012}. 
Meanwhile the L\'{e}vy walk model has been used to reproduce the temperature profiles of 
finite chains  \cite{Lepri2011,dhar2013} and analyze heat 
fluctuations in conducting rings \cite{dhar2013}.  Very recently, \textcite{Vermeersch2014}
proposed an interesting interpretation of 
the interfacial thermal transport through metal-semiconductor interfaces in term of exponentially 
truncated L\'{e}vy flights. Similar to the problem of the light transmission (see Section \ref{LWlight}), 
the set-up of the performed experiments does not allow to differentiate in a clear-cut manner 
between L\'{e}vy flights and L\'{e}vy walks, yet experimentalists could think about new experiments that can do.

\begin{figure}[t]
\includegraphics[width=0.5\textwidth]{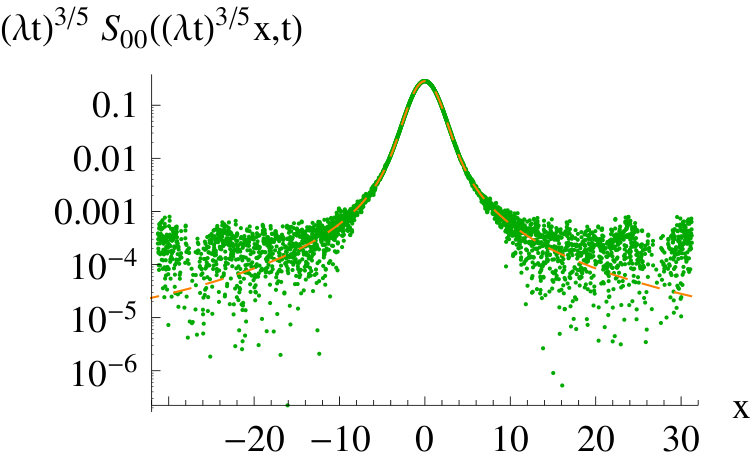}
\caption{(Color online) Heat peak for a hard-point gas with alternating masses at  time $t = 1024$. 
The dashed orange curve is the L\'{e}vy distribution $L_{\gamma}(x)$ with $\gamma = 5/3$.
Adapted from \textcite{Mendl2014}.}
\label{fig:SpohnHPG}
\end{figure}

The findings presented by \textcite{cipriani2005, zaburdaev2011} are phenomenological. To understand the mechanisms which 
sculpt  L\'{e}vy kinetics out of many-particle dynamics, the problem should be considered in a broader context.
\textcite{henk2012} used a hydrodynamic approach to build a mode-coupling theory for the Fourier components of the densities of 
conserved quantities, that are number of particles, total momentum, and energy. A linear transformation  splits the transport 
over three channels facilitated by the three different modes, a heat mode and two sound modes propagating in opposite directions.
Thus, instead of a single energy-energy correlator curve for a given time, as in Fig.~\ref{fig:LevyHPG}, the hydrodynamic approach 
produces three profiles (two of them, for the sound modes, are related by the inversion $x \rightarrow -x$).
The key result by \textcite{henk2012} is  that the scaling of the ballistic sound peaks is of 
the Kardar-Parisi-Zhang (KPZ) universality class \cite{kardar1986}. The anomalous scaling of the central heat peak
with the exponent $\gamma = 5/3$ was predicted, which corresponds to the anomalous scaling of the conductivity
$\kappa_T \propto L^{1/3}$ \cite{Lepri2003}.
Very recently \textcite{spohn2014} presented a complete version of the  hydrodynamic
formalism which addresses also the dynamics of the heat peak in more detail. 
\textcite{das2014,Mendl2014} have found for FPU chains and the hard-point gas the L\'{e}vy scaling for 
the correlator of  their heat modes 
(see Fig.~\ref{fig:SpohnHPG}) and confirmed the  KPZ-type scaling for the correlator of their sound modes.  
The L\'{e}vy-like profiles for the heat mode exhibit cut-offs at the points $x = \pm ct$, where $c$ is the speed 
of sound.
It is tempting to think that further progress in this direction can 
provide with a ``hydrodynamic'' foundation of L\'{e}vy walks.

\subsection{L\'{e}vy flights of light and L\'{e}vy walks of photons}
\label{LWlight}
When passing through a medium, light is subjected to multiple 
scattering by medium inhomogeneities. Physics of this process depends on the characteristic size
of  inhomogeneities and different scattering mechanisms can coexist. For example, scattering 
by molecules (Rayleigh scattering) and by water droplets (Mie scattering) 
work together in a cloudy sky \cite{Kerker1969}.
In some cases one particular mechanism dominates and thus specifies characteristic scales of the path length
between consecutive scattering events (for example Mie scattering dominates inside a cloud). 
If a medium is a fractal \cite{mandelbrot1982} than the structure of 
its inhomogeneities is scale-free
[a stratocumulus cloud is a good example \cite{Cahalan1994}]. 
The path of a photon inside a fractal media can be represented as a random walk  consisting
of free-path segments connecting subsequent scattering points with the PDF of the segment length 
in a power-law form \cite{Davis1997}. The question now is shall we use the L\'{e}vy flight or L\'{e}vy walk  
to  correctly model the process? If we are interested in the stationary
transmission  through the medium only then the answer is ``either'' \cite{Buldyrev2001}. 

Consider a propagation of a light beam through a slab of thickness $L$, with a photon free-path PDF
$p(\ell) \sim \ell^{-1-\gamma}$. A local stationary transmission  on the output surface is defined by all the trajectories 
leading to the corresponding  point from the illuminated spot on the entry side. 
The total transmission is given by the integral over the local
transmission and equals the probability of the absorption 
of a photon that starts at the illuminated spot by the absorption boundary on the opposite side of the slab\footnote{
The absorption effects are neglected within the framework of the approach. A 
photon that entered the slab will appear on the opposite side almost surely
in the asymptotic limit $t \rightarrow \infty$.}.
Within this setup two approaches are equivalent. The total path length of a L\'{e}vy flight corresponds to 
the total traveling time of a L\'{e}vy walk, and results obtained with both models 
are interchangeable \cite{Buldyrev2001}. 
By using a one-dimensional L\'{e}vy flight model\footnote{The problem setup considered by \textcite{Davis1997}
assumed the scattering probability peak in the forward $x$-direction. It was shown that the 
directional correlations in scattering angles can be absorbed into a rescaling of the free path 
within the one-dimensional framework, see Appendix in the cited work.}
with $1 < \gamma \leq 2$, \textcite{Davis1997} 
derived the transmission 
as a function of  $L$,
\begin{eqnarray}
T(L) = \frac{1}{1+(L/\bar{\ell})^{\gamma/2}}~,
\label{trans_scaling}
\end{eqnarray}
where $\bar{\ell}$ is the mean free path and the unity in the denominator regularizes the expression at $L = 0$.
In the continuous limit $L \gg \bar{\ell}$, the problem can be 
recast in terms of the fractional diffusion equation, Eq.~(\ref{FDE}), and, 
by treating the particle PDF as the light intensity, the scaling in Eq. (\ref{trans_scaling}) can be obtained. 
Identically, the scaling could be derived within the L\'{e}vy walk 
framework by using the integral Eqs. (\ref{nu})-(\ref{Phicoupled}) for the PDF of walking photons \cite{Drysdale1998}. 
Photons move with finite velocity in any medium and therefore the L\'{e}vy walk 
is physically more adequate to model the photon dynamics than  the L\'{e}vy flight. 
However, in the context of the transmission problem and from the mathematical  point of view the difference between the two approaches is absent. The difference could become tangible when the problem setup is changed and, for example, 
auto-correlation (Section \ref{space_time_corr}) and/or interference effects are taken into account.  
It remains for future work to set up the corresponding experiments.
Below we overview the up-to-date experimental results.

{\em Solar light transmission by cloudy skies.} The first attempt to get insight into the morphology of 
a scattering media
by utilizing the L\'{e}vy flight concept was made by \textcite{Pfeilsticker1999}. He used 
statistical data obtained by measuring the mean geometrical paths of photons coming from a cloudy sky. 
By assuming the fractal cloud morphology and resorting to the scaling Eq. (\ref{trans_scaling}),
the flight-length exponent was estimated as $\gamma \simeq 1.74 \div 1.78$. Pheilsticker found that the exponent value depends on 
the cloud type: it tends to  $1.5$ for convective clouds and to $2$ for stratiform clouds.

\begin{figure}[t]
\includegraphics[width=0.5\textwidth]{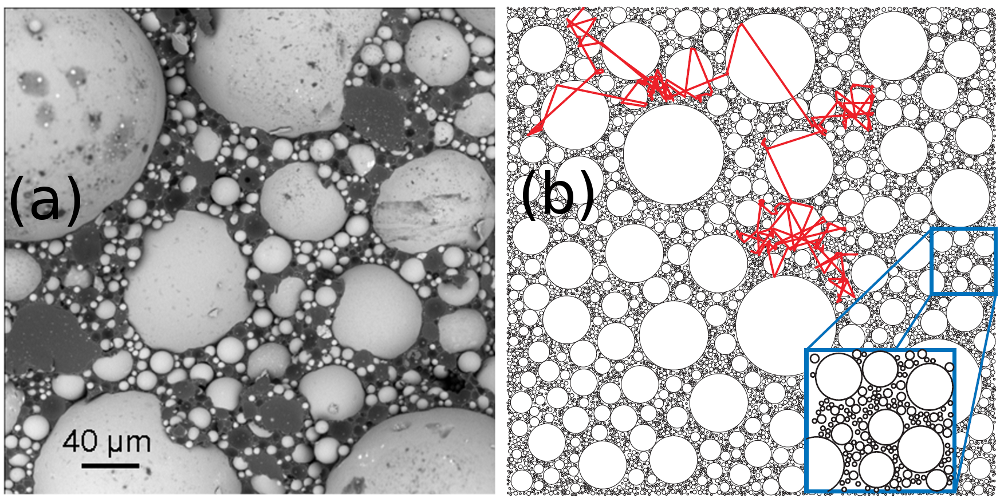}
\includegraphics[width=0.5\textwidth]{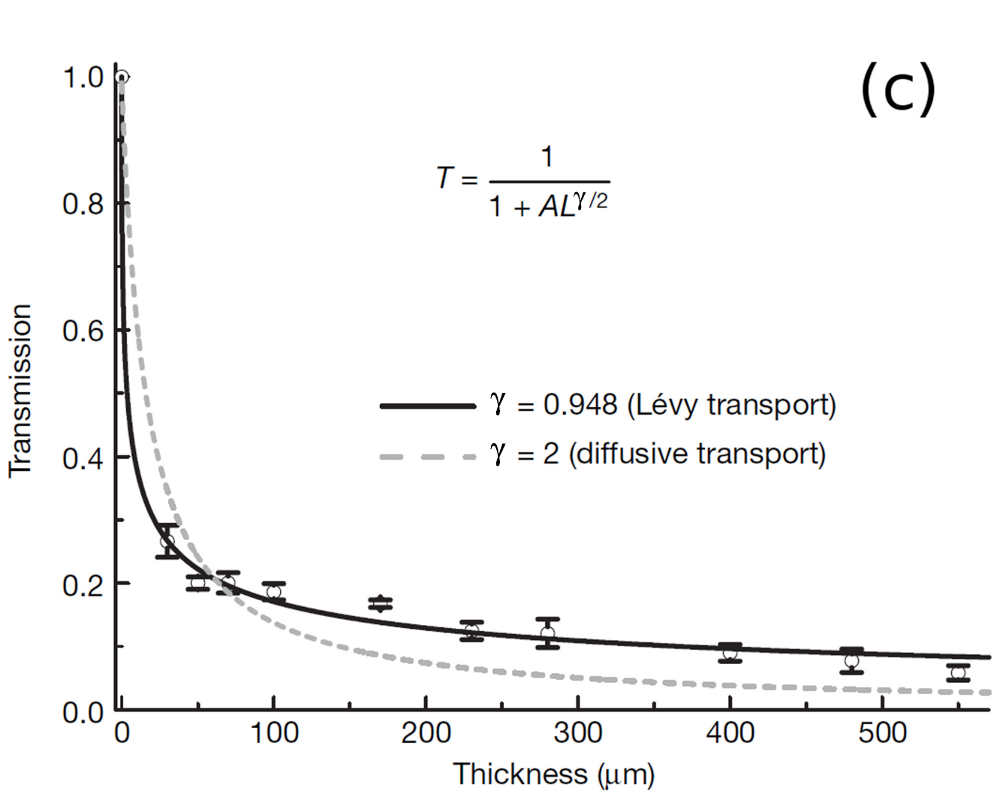}
\caption{(Color online) L\'{e}vy flights of photons in a L\'{e}vy glass. (a) Electron micrograph of a L\'{e}vy
glass. The gray zones are interiors of the glass spheres, whereas the
darker area corresponds to the polymer matrix. Scattering
nanoparticles are too small to be resolved. 
(b) A sketch of a photon walk inside a two-dimensional 
version of a L\'{e}vy glass. Inset shows the scale invariance of the
glass. (c) Measured transmission through a L\'{e}vy glass slab as a function of the slab thickness.  
Gray dashed curve is obtained with Eq.~(\ref{trans_scaling}) for $\gamma = 1$ (normal diffusion)
while black line obtained for $\gamma = 0.948$. Note that the exponent $\gamma$ is denoted $\alpha$ in the original publications.
Adapted from \textcite{Barthelemy2008,burresi2012}.}\label{figLWL}
\end{figure}

{\em Photon transmission through a L\'{e}vy glass.} Modern technologies provide the possibility to synthesize scattering materials with
a tunable fractal structure \cite{Tsujii2008}. One of the recent advances is the creation of \textit{L\'{e}vy glass}, 
a polymer matrix with embedded high-refractive-index scattering nanoparticles  \cite{Barthelemy2008}. 
The matrix also contains a set of glass microspheres with a power-law diameter distribution, 
$p(\varnothing) \sim \varnothing^{-\eta-1}$, Fig.~\ref{figLWL}(a).  
The microspheres do not scatter because their refractive index is the same as of the host polymer and therefore 
scattering happens on   nanoparticles only. The photon transport inside a L\'{e}vy glass is dominated by 
long ``jumps'' performed by the photon when it propagates through the glass spheres, Fig.~\ref{figLWL}(b). 
When the diameter distribution of the spheres is sampled exponentially in $\varnothing$ space,
the jump-length PDF scales as $p(\ell) \sim \ell^{-1-\gamma}$, with $\gamma = \eta -1$  \cite{Bertolotti2010}.
This is a clear-cut case of the  L\'{e}vy walk(flight) of photons. Measurements performed with
a Cauchy glass, $\gamma = 1$, by illuminating the slab with a narrow collimated laser beam, 
corroborated the scaling given by Eq. (\ref{trans_scaling}), see Fig.~\ref{figLWL}(c).

Very recently, \textcite{Savo2014} reported on an experimental retrieval of the scaling exponent 
$\alpha$, Eqs. ~(\ref{selfsimilarity_ctrw}, \ref{propagator_scaling}), by analyzing the scaling
of the \textit{time-resolved} transmission with  $L$. The performed measurements 
verified the universal relation between the three exponents, $\alpha = 1/\gamma = 1/(\eta -1)$,
thus strengthening the position of L\'{e}vy walk (flight) formalism  as an adequate 
theoretical approach to the process of photon (light) diffusion through fractal media.

An interesting aspect of the photon diffusion inside a L\'{e}vy glass is the role of a quenched disorder.
The distribution of glass spheres in a matrix does not evolve in time and so there is a room for correlations 
between flight directions and angles. By using a one-dimensional chain of barriers with a power-law spacing
distribution, \textcite{beenakker2009} found that a walk along the chain is not the standard  uncorrelated L\'{e}vy walk 
because of the strong correlations of subsequent step sizes. Similar results have been obtained by
\textcite{burioni2010} and \textcite{vezzani2011}. However, by using a two-dimensional model of a L\'{e}vy glass, 
\textcite{Barthelemy2010} demonstrated that the influence of the quenched disorder can be neglected 
(in a sense that it can be accounted by a simple parameter tuning) when stepping into higher dimensions. 
Therefore the transport of photons in two- and three-dimensional L\'{e}vy glasses is close to an uncorrelated 
Levy walk [although this could also change when a scattering media is perfectly self-similar and represents a 
regular fractal, as shown by \textcite{Buonsante2011}].

\begin{figure}[t]
   \centering
\includegraphics[width=0.5\textwidth]{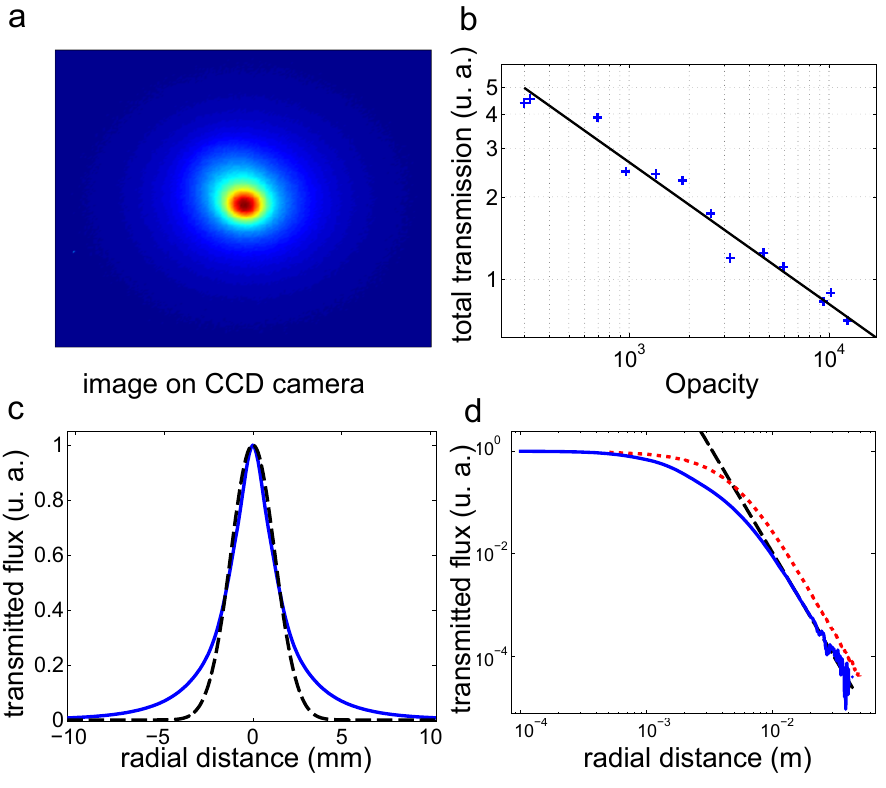}
\caption{(Color online) Photon transmission through a hot rubidium vapor.
(a) The radial profile of the outcoming light on the charge-coupled device camera. 
(b) Blue crosses show experimentally measured transmission 
as a   function of the opacity. The power-law fit $T \propto O^{-0.516}$ is shown by solid curve.
(c)  Experimental radial profile of the light transmitted through the vapor chamber (blue solid line) is compared with a Gaussian distribution with the same width at half maximum (black dashed line). 
(d) Radial profile of the transmitted light.  Black dashed line
is a power-law fit $I(r)\propto r^{-4.03}$. Adapted from \textcite{Baudouin2014}.}
 \label{figKaiser}
\end{figure}

{\em L\'{e}vy flights of photons in atomic vapors.} The two considered realizations of L\'{e}vy walks of light assumed the 
elastic scattering of photons. The needed power law distributions are produced by 
the  fractal spatial inhomogeneity of scattering media and the corresponding flight-time exponents are linearly related 
to characteristic fractal exponents. In the case of inelastic scattering,   the distance 
traveled by a photon depends on its frequency, which changes after every scattering  
event (which is in fact an absorption/emission). The spatial inhomogeneity is no longer needed and 
a power law distribution of flight-length can be obtained 
from the \textit{spectral} inhomogeneity of the medium \cite{radiation}. In an atomic vapor \cite{Baudouin2014review} the absorption probability of a photon 
with a frequency $\omega$ at a distance $r$ from its emission point is $p(\omega|r) = \Phi(\omega)\exp[-\Phi(\omega)r]$, 
where $\Phi(\omega)$ is the absorption spectrum of the atoms. The average absorption probability can be obtained as a frequency-average of $p(\omega|r)$
weighted with an emission spectrum $\Theta(\omega)$ \cite{holstein1947,radiation},
\begin{eqnarray}
p(r) = \int_{0}^{\infty} \Theta(\omega)\Phi(\omega)e^{-\Phi(\omega)r}d\omega,
\label{mfp}
\end{eqnarray}
When emission and absorption spectra are identical, the Doppler spectrum $\Phi_D(\omega) = \exp(-\omega^2)/\sqrt{\pi}$ 
leads to $p(r) \sim r^{-2}[\ln(r)]^{-1/2}$, while the
Cauchy spectrum $\Phi_C(\omega) = 1/[\pi(1 + \omega^2)]$ yields $p(r) \sim r^{-3/2}$. \textcite{pereira2004} proposed this as a 
means to realize a three-dimensional L\'{e}vy flight of photons in a hot atomic vapor where the spectra-equality condition may hold. 
They have also raised two important points. Firstly, in a high opacity atomic vapor many elastic scattering events 
happen before an inelastic scattering event occurs. This is a natural call for an extended intermittent model
in which  L\'{e}vy walks are alternated with periods of Brownian diffusion 
(see also Section~\ref{intermittent} where such processes appear in the 
context of animal search). Secondly, they pointed out that in lab vapors the time of flight
is negligible compared to the waiting time between absorption and emission events. Therefore 
the use of L\'{e}vy flight model, Fig.~\ref{figt1}(a), 
is well-justified. However, as noted by \textcite{pereira2004}, in interstellar gases the flight time can be larger than the characteristic 
absorption/emission time and the L\'{e}vy walk will be more appropriate in the astrophysical context.

By using a specially designed experimental set-up, \textcite{Mercadier2009} measured the 
first step length distribution of Doppler-broadened photons which enter a hot rubidium vapor. The obtained PDF follows
a power law with $\gamma = 1.41$. The step-length PDF changes after each scattering event 
while remaining a near perfect power-law. The dependence of the exponent $\alpha(n) = \gamma(n) + 1$ on the number $n$ 
of scattering events saturates to a value close to $2$, as expected from the theory \cite{pereira2004}. 
Recently, \textcite{Baudouin2014} measured transmission through a hot rubidium vapor by changing the opacity
of the media, $O = L/\bar{\ell}$, over two decades. This was realized by controlling the density 
of atoms (and thus $\bar{\ell}$) by adjusting the temperature inside the vapor chamber.
The results fit the dependence predicted by Eq.~(\ref{trans_scaling}) with the exponent $\gamma \simeq 1.01$, Fig.~\ref{figKaiser}(d).
The radial profile of the transmitted light has a power law tail, $I(r) \propto r^{-3-\mu}$, as expected from the 
L\'{e}vy-based theory developed in the paper, Fig.~\ref{figKaiser}(c-d). 
By comparing these results with the single step length PDFs obtained before, 
\textcite{Baudouin2014} stated an excellent agreement with a L\'{e}vy-walk approach.

\subsection{Blinking quantum dots}
Blinking quantum dots serve another realizations of the ballistic L\'{e}vy walk. 
Similar to L\'{e}vy flights of light, photons are again involved but in a different way.

\begin{figure}[t]
\center
\includegraphics[width=0.45\textwidth]{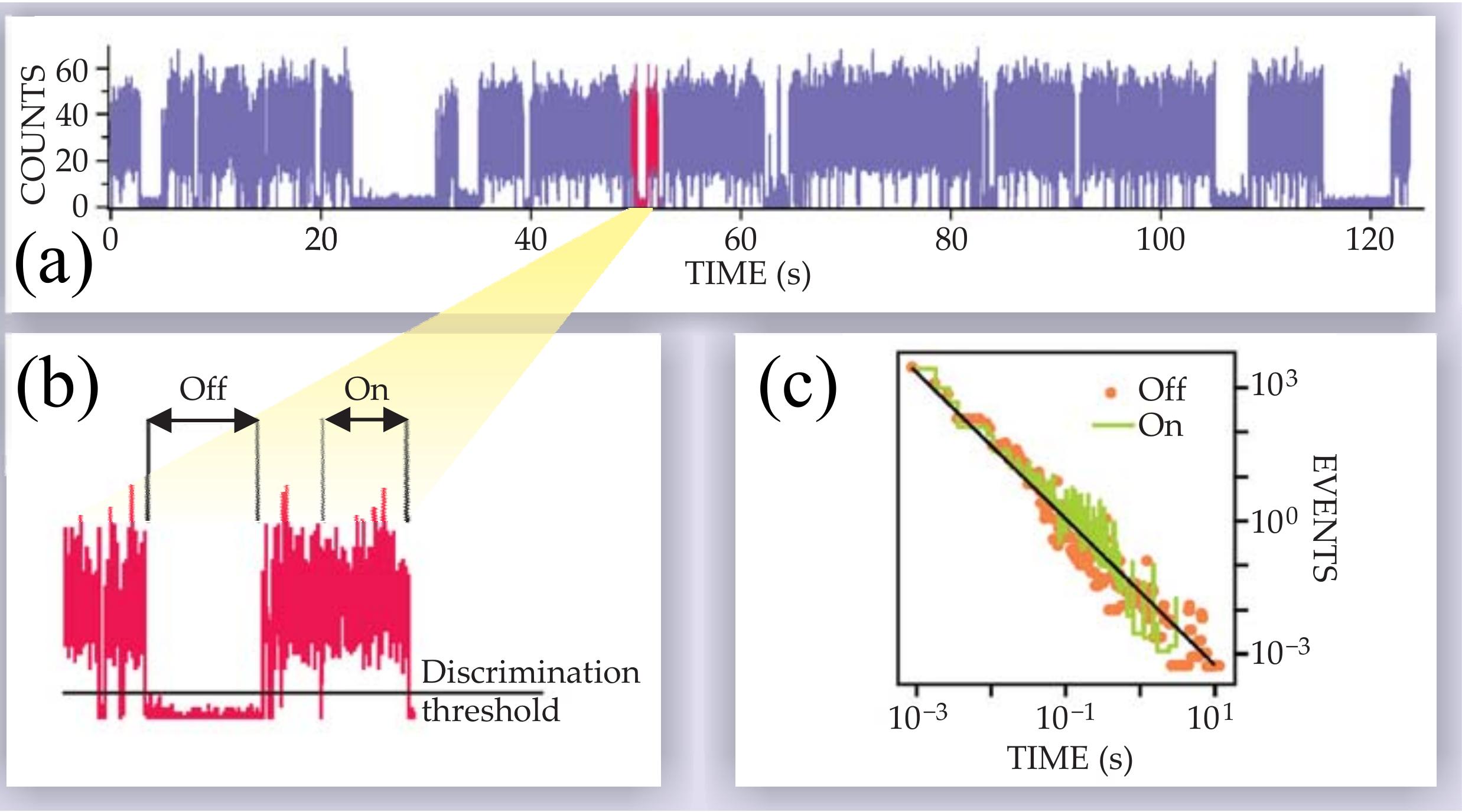}
\caption[Blinking Quantum Dots]
{(Color online) Blinking quantum dots. Panel (a) shows a sample trace of a quantum dot fluorescence. 
A zoom in part (b) shows how a threshold is defined, which allows 
to identify the periods of on and off times. The PDFs of those times 
are shown on (c) and have a clear power-law dependence with the same 
tail exponent for on and off times. The solid black line indicates a slope of 1.65, i. e. $\gamma=0.65$. Adapted from \textcite{stefani2009}.}
\label{figp_qd}
\end{figure}

A quantum dot (QD) is a nano crystal made out of semi-conducting material and is several 
nanometers in size \cite{Alivisatos1996}. The size is crucial for determining the specific properties of QDs which are 
governed by quantum effects and are on the border between bulk and molecular behavior. One of the important features 
of QDs is the so-called quantum confinement, when the exciton Bohr radius is of the order of the object size, leading to 
the discrete energy levels and a band gap which depends on the size of the object. When under the laser light with 
energy above the band gap, QDs can adsorb light by creating and exciton pair and then re-emit a photon when the exciton 
decays. The frequency of the emitted light is increasing with decreasing QD size and can be accurately tuned in applications. 
One of the important QDs applications is bio-imaging; in addition to their small size, QDs have higher brightness 
as compared to organic fluorescent dyes and show minimal photobleaching. 
However, there is one interesting effect: QDs blink \cite{Nirmal1996}. 
Experimentally it was found that quantum dots alternate periods of fluorescence with no emission of photons, 
and the durations of these periods are not exponentially distributed but instead have a fat tailed power-law distribution 
with diverging  average time. A quantum dot can fluoresce or be completely dark during the whole measurement time, 
which can be on the order of hours. 
Current experimental results provide the {\em on} and {\em off} times statistics spanning four orders of magnitude. 
In Fig. \ref{figp_qd} one realization of the QD fluorescence intensity track is shown. 
By defining a certain threshold in the intensity, a sequence of on and off times can be 
identified and characterized. Many experiments with different QD materials and at different temperatures show power 
law distributions of those times, and in many cases the exponent is nearly the same for both of them: $\gamma\simeq0.5$ \cite{margolin2005}. 
Therefore the blinking dynamics of a QD can be described as a two state model, 
where the durations of phases in each state (on, $I(t)=1$, and off, $I(t)=0$) are 
distributed as power laws with diverging means. As we discussed in Section \ref{memory_effects}, 
systems with such distributions exhibit memory effects, aging and weak ergodicity breaking. 
Interestingly, the problem of blinking nanodot can be mapped onto the L\'{e}vy walk model.
Consider the fluorescence intensity $I(t)$ and define its time average as:
\begin{equation}
\overline{I}=\frac{\int_{0}^{T}I(t)dt}{T}
\end{equation}
As we learned from Section \ref{memory_effects}, for the weak ergodicity breaking problems the
 time average is itself a random variable with a certain distribution. 
It can be shown, that the PDF of the time averaged intensity $P(\overline{I})$ is 
given by the Lamperti distribution \cite{margolin2005, lamperti1958}:
\begin{equation}
P(\overline{I}) = 
\frac{\pi^{-1}\sin\left(\pi\gamma\right)\overline{I}^{\gamma-1}\left(1-\overline{I}\right)^{\gamma-1}}
{\overline{I}^{2\gamma} + \left(1-\overline{I}\right)^{2\gamma} + 2\overline{I}^{\gamma}\left(1-\overline{I}\right)^{\gamma} \cos\left(\pi\gamma\right)}
\label{lamperti}
\end{equation}
We know this distribution from the analysis of the  ballistic L\'{e}vy walk model, and it is easy to draw an analogy. 
In the ballistic regime of the L\'{e}vy walk, we can define a time averaged position of the particle $x/T$ as 
an integral from $0$ to $T$ of the particles velocity $v(t)$, see Eq. (\ref{timeaverage}). 
As in quantum dots, the time spent in each velocity state has a power-law distribution with infinite mean. 
The only difference to the QD blinking problem is that the velocity of particles can have values of $v(t)=\pm v_0$, 
while the intensity switches between $0$ and $1$. As a result, in case of L\'{e}vy walks, the PDF $P(x/T)$ is symmetric around zero, 
whereas for the time average intensity it is shifted and has a support from $0$ to $1$. 
One particular example of $\gamma=1/2$ gives a simple particular case of the Lamperti distribution, 
see Eq. (\ref{Glevywalkballistic}) and Fig. \ref{figt3}. An intuitive expectation 
that the QD will be half time on and half time off appears to be least probable: the corresponding PDF 
has a minimum at $\overline{I}=1/2$. Instead the $P(\overline{I})$ has a divergent behavior 
at $\overline{I}=0$ and $\overline{I}=1$ (similarly to the divergence of the PDFs at ballistic fronts in case of  L\'{e}vy walks). 
Therefore a quantum dot is either on or off for most of the observation time.

The particular mechanism responsible for the appearance of the power-law 
distributed blinking times in quantum dots remains unknown. There are several working models which relate 
the statistics of on and off times to the dynamics of exciton pair including its transport, diffusion, and trapping 
[see \textcite{stefani2009} for an overview], but none of the models are able to describe all available experimental observations. 
For the context of this review it is important that the experimental data on blinking QD can be directly mapped to the model of L\'{e}vy walks  in the ballistic regime \cite{margolin2006photon}. 
We can speculate that the analytical results available for the L\'{e}vy walk model with random velocities could be useful for 
the interpretation of experiments with QDs with a whole distribution of intensities and not just two levels. Reciprocally, 
a possible correlation between consequent long on (off) times, when a long on (off)  time is followed with higher probability 
by another long on (off) time \cite{stefani2009}, calls for further generalizations of the L\'{e}vy walk model. 

\subsection{L\'{e}vy walks of cold atoms}
\label{cold atoms}
\begin{figure}[t]
\includegraphics[width=0.5\textwidth]{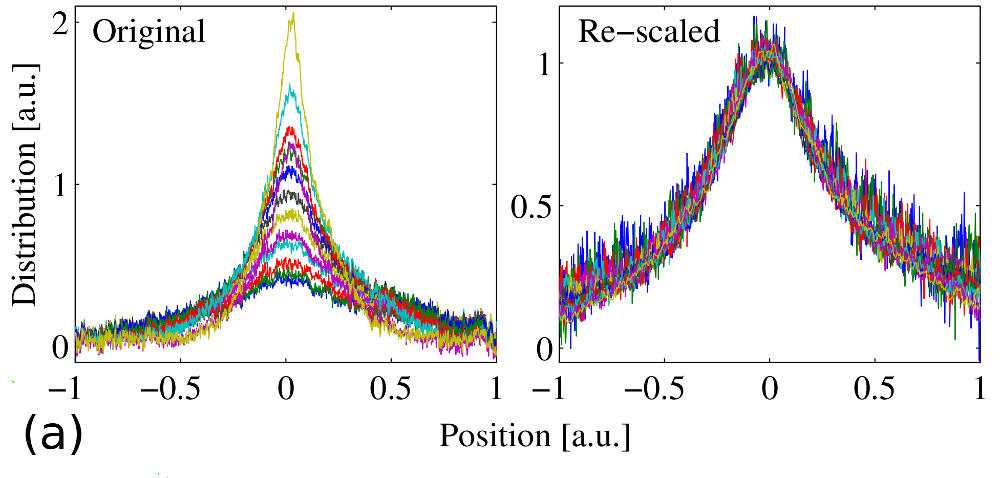}
\includegraphics[width=0.5\textwidth]{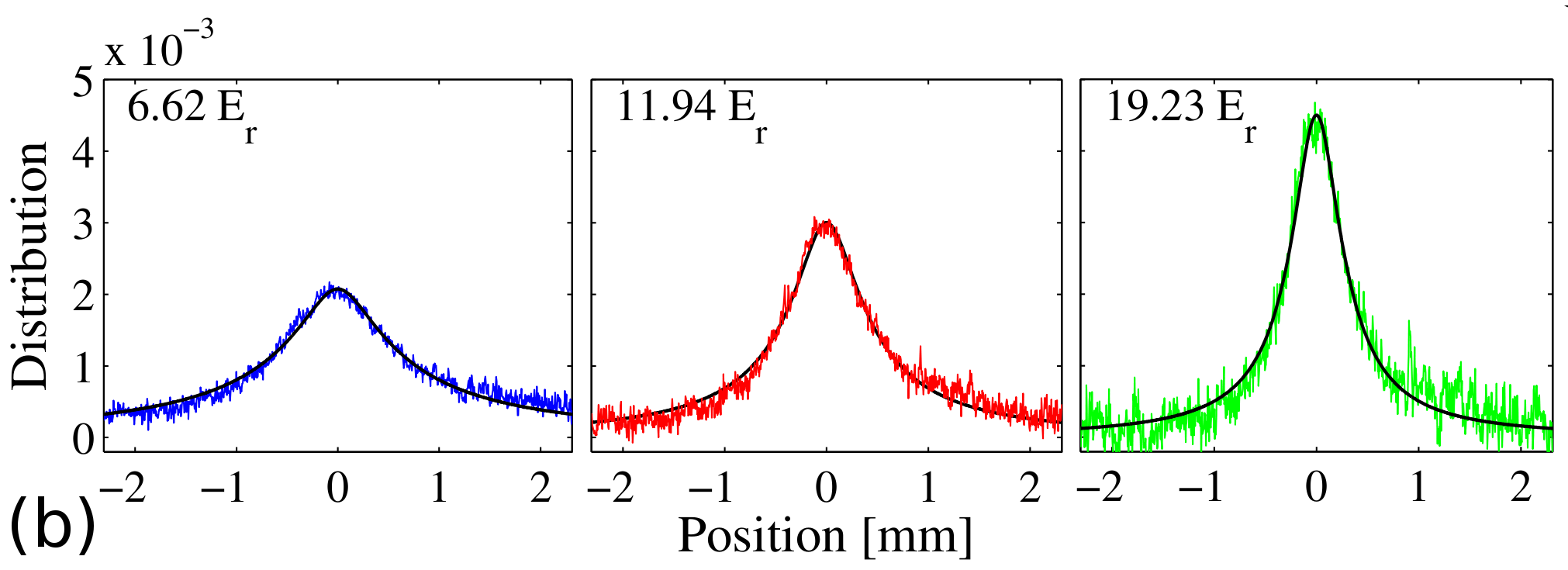}
\caption{(Color online) L\'{e}vy walks of atoms in optical lattices. (a) Atomic distributions obtained for different times,
$t \in [10 - 40]$ ms, before (left) and after the scaling transformation  (\ref{selfsimilarity_ctrw}) with exponent
$\alpha = 0.8$ (right). (b) Atomic distributions after $30$ ms of spreading for three different optical potential depths.
Lines correspond to L\'{e}vy distributions with depth-specific exponents. Adapted from \textcite{sagi2012}. 
}
\label{fig:davidson}
\end{figure}

L\'{e}vy distributions are known in the field of cold atom optics since $1990$s, when
\textcite{Bardou1994} and \textcite{reichel1995} discussed the relation between the process of the so-called subrecoil
laser cooling \cite{aspect1988} and  anomalous diffusion  in terms of L\'{e}vy flights.
L\'{e}vy walks appeared in cold atom optics in the context of Sisyphus cooling  of atoms
loaded into an optical bi-potential 
created by two counterpropagating linearly polarized laser beams \cite{Dalibard1989}. There are two 
internal atomic states and atoms with different internal states 
feel potentials of different polarizations. The laser-induced  transitions of an atom between its internal states 
influence translational motion of the atom along the bi-potential.  Elaborated within the 
Monte Carlo wave function framework, this connection was shown to be responsible for L\'{e}vy walk-like dynamics 
of atoms \cite{Marksteiner1996}.
\textcite{katori1997} measured the mean squared displacement of Mg ions in 
a bi-potential optical lattice and found the scaling $\sigma^2(t) \propto t^{\mu}$ with the exponent $\mu > 1$ 
for potential depths below the critical value.  \textcite{sagi2012} performed more sophisticated experiments and 
measured the spreading 
of a packet of cold Rb atoms in optical lattices of different depths. The obtained atomic distributions scaled with the characteristic 
scaling, Eq.~(\ref{selfsimilarity_ctrw}), Fig.~\ref{fig:davidson}(a),
and their shapes could be nicely fit by L\'{e}vy distributions, Fig.~\ref{fig:davidson}(b).
However, a  L\'{e}vy walk description did not work well in this case, because in both, experiments and  Monte Carlo wave function simulations,
strong correlations between  velocities and durations of atom flights were found.

\textcite{Marksteiner1996} derived that on the semi-classical level the distribution $W(x,p,t)$ of atoms 
can be described by the Kramers equation 
\begin{equation}
{\partial W \over \partial t} + p {\partial W \over \partial x} = \left[
D {\partial^2 \over \partial p^2 }  - {\partial \over \partial p } F(p) \right]W,
\label{eqKramers}
\end{equation}
with the cooling force \cite{Castin1991}
\begin{equation}
F(p) = - { p \over 1 + p^2},
\label{cooling}
\end{equation} 
where momentum is expressed in dimensionless units $p/p_c$,  with the  capture momentum $p_c$ set to unity.
For small momenta the force is of the conventional linear form, $F(p) \sim -p$, while 
$F(p) \sim -1/p$  for large $p$ so that the atom becomes frictionless at the high-momentum limit.
The diffusion constant $D$ combines all  relevant parameters such as the depth of the optical potential, recoil energy, see below.
The equilibrium momentum-momentum correlation function corresponding to Eq.~(\ref{eqKramers}) scales as $t^{-\lambda}$, 
with the exponent 
\begin{equation}
\lambda = (1/2D) - 3/2.
\label{cool_exponent}
\end{equation} 
The control parameter,
\begin{equation}
D = cE_R/U_0.
\label{cool_D}
\end{equation} 
depends on the recoil energy $E_R$ and the depth of the optical lattice potential $U_0$. The constant $c$
is specific to the type of atom/ion cooled, with the typical value around $10$. Therefore, one could,
by tuning the potential depth $U_0$ while keeping all other parameters fixed, control the exponent $\lambda$ and 
switch between the regimes of normal and anomalous atom diffusion.

When $\lambda < 1$, the  integral of the correlation function over time diverges and an anomalously fast diffusion 
appears. Equilibrium velocity distribution for Eq.~(\ref{eqKramers}) has a form of the Tsallis distribution \cite{douglas2006,Lutz2003}.
Although interesting as an indication of a strong deviation from the Boltzmann-Gibbs thermodynamics  \cite{Lutz2013}, 
these distributions themselves do not provide sufficient insight into the diffusion of atoms in the real space.

\begin{figure}[t]
\includegraphics[width=0.5\textwidth]{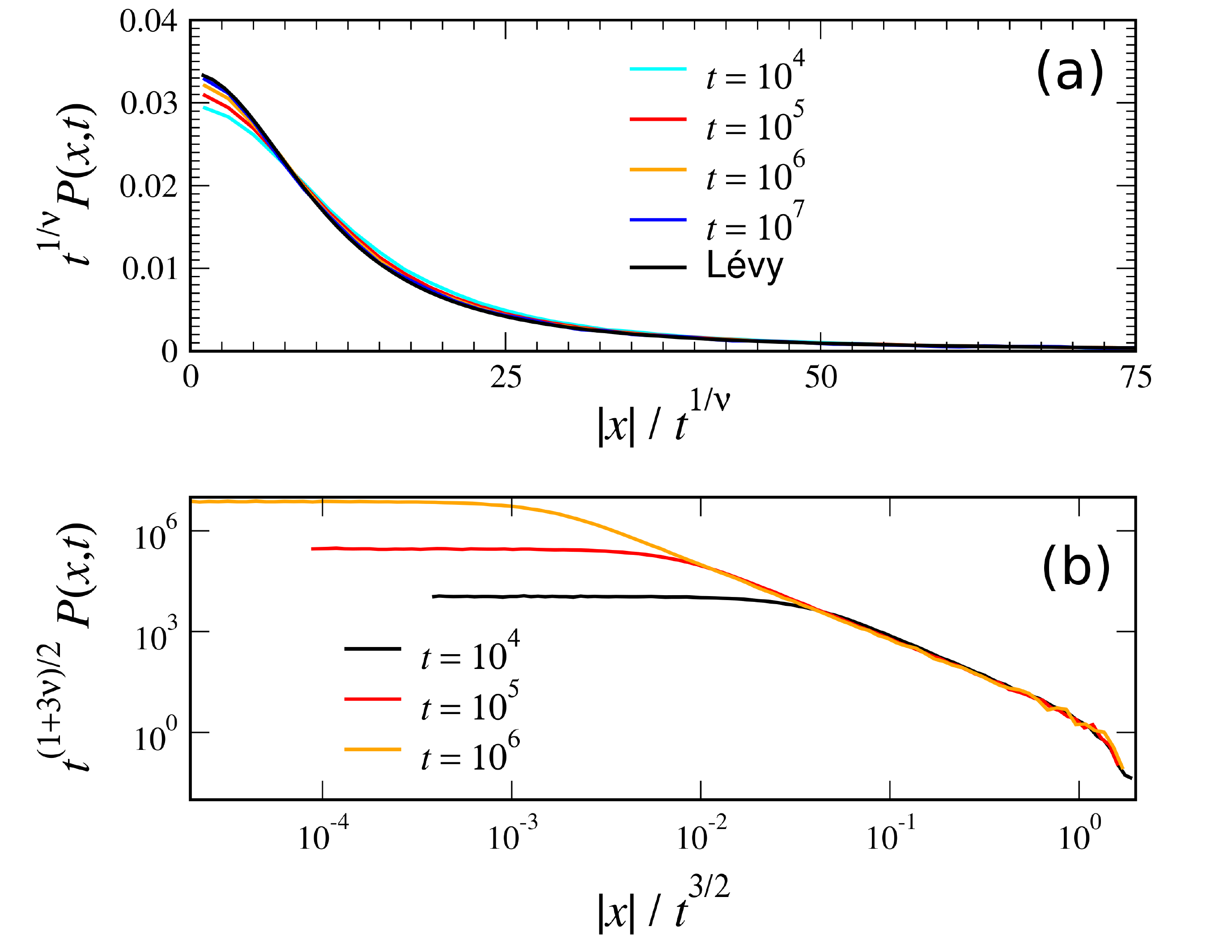}
\caption{(Color online) Scaling of the propagators $P(x,t)$ of the stochastic process, Eqs. (\ref{cold_barkai1},\ref{cold_barkai2}), modelling the 
diffusion of cold atoms. The scaling exponent $\gamma$ (denoted by $\nu$ in the original publications) is
$(1+D)/3D$. (a) With the increase of time the rescaled profiles start to fall onto the L\'{e}vy distribution $L_{\gamma}(\xi)$.
(b) Another scaling reveals the  the cutoff induced by a nonlinear space-time coupling, 
Eq. (\ref{phi_coupled_Barkai}).
The parameter $D = 2/5$. Adapted from \textcite{kessler2012}.}
\label{fig3}
\end{figure}

By  unraveling the Kramers equation (\ref{eqKramers}) into a Langevin equation with a 
white Gaussian noise as a drive, 
\begin{eqnarray} 
\dot{p} = F(p)+ \sqrt{2D}\zeta(t), \label{cold_barkai1}\\
\dot{x} = p, ~~~~~~~~~~~~~~~~~~~~\label{cold_barkai2}
\end{eqnarray}
\textcite{kessler2012} analyzed 
the atom diffusion  from the microscopic point of view. 
The theory developed by \textcite{barkai2014} predicts the existence of three phases in the dynamics, 
generated by Eqs. (\ref{cold_barkai1}-\ref{cold_barkai2}), depending on the value of $D$. Namely, it can exhibit 
normal diffusion,
L\'{e}vy-walk superdiffusion, and Richardson's diffusion \cite{Richardson1926}, when 
the MSD scales as $\sigma(t) \propto t^{3}$.
The existence of the L\'{e}vy walk regime 
was  proved analytically for the  range  $1/5 < D <1$. In the asymptotic 
limit of large $t$, the central part of the propagator $P(x,t)$ scales with the distinctive scaling, Eq.~(\ref{selfsimilarity_ctrw}), where
$\Phi(\xi) = L_{\gamma}(\xi)$, and $\gamma = (1+D)/3D$. The coupled transition probably is different form that for the standard L\'{e}vy walk,
Eq. (\ref{phi_coupled}), and has the form \cite{kessler2012,barkai2014}
\begin{equation}
\phi(y,\tau)=\psi(\tau)p(y|\tau),
\label{phi_coupled_Barkai}
\end{equation}
with the conditional PDF $p(y|\tau) \sim \tau^{-3/2}B(y/\tau^{3/2})$. The normalized nonlinear 
function $B(x)$ is responsible for the cutoff of the propagator tails, so that all moments of the process 
are finite. Together with the mode-coupling theory of the energy transport in classical nonlinear chains (Section \ref{many_body}), 
these results pave the way toward physical foundations of L\'{e}vy walks.
 
\section{L\'{e}vy walks in biology}\label{biology}
From rather complex but objective systems of physics, we are shifting to the field of biology and biophysics 
where effects and phenomena are much harder to quantify because of their intrinsic diversity and variability. In recent years, the 
topic of L\'{e}vy walks resonated in research communities working on motility of living organisms, their foraging, and 
search strategies. By the level of debates, the  topic may even be called controversial. 
Fortunately our review is preceded by two very recent monographs devoted to these subjects \cite{viswanathan2011,mendez2014}. 
Here we are presenting our point of view through the prism of the L\'{e}vy walk framework and 
point to the examples that are directly relevant to this model. 
Before passing to particular examples we would like to outline the general complexity of the problem in question. 

\subsection{Motility is a complex issue across many scales}\label{complex_issue}
Motility spans many scales, ranging from swimming micron-sized bacteria to albatrosses 
which can travel hundreds of kilometers at a time. 
Motility involves interactions of moving animals with their environment and habitats, 
which in most cases is hard to quantify or predict. 
In ecology, the interest in motility usually does not arise {\em per se} but in relation to some greater issues, 
for example, questions of how animals search for food, how they navigate home, how they find each other to mate or 
to agglomerate into colonies, and  others. 

In a very interesting twist, L\'{e}vy walks are involved in a particular topic of 
effectiveness of search  and foraging strategies. 
L\'{e}vy walks are argued to be the most efficient search strategy \textit{under 
certain conditions} imposed on the distribution and properties of targets. 
There is a constantly growing number of accounts where L\'{e}vy statistics is reported 
for the trajectories of animals. 
Quite often these results get criticized or disputed, 
based on insufficient data, 
an inconsistent analysis, or just out of different beliefs. 
As a side effect of these still ongoing discussions, new papers constantly appear where  researchers report the analysis 
of the motion patterns of yet another living species and claim that the patterns {\em do} or {\em do not} 
look as  L\'{e}vy flight or L\'{e}vy walk trajectories. 
There is even a philosophical flavor in this discussion \cite{Baron2014}. 
The possible reasons of this controversy are manifold. Below we summarize them from rather evident to more complex levels.

i) {\em Difference in sizes, forms of locomotion, habitats etc.} All these difference 
dictate different experimental techniques 
and also call for different statistical techniques. As pointed out by \textcite{mendez2014}, 
on the micron scales of single cells, positional data can be acquired with  high 
space and time resolution leading to almost continuous recorded trajectories. 
Such observations are common in a lab since 1970s. 
Tracking of big animals in their habitats is a much harder task due to complex interactions of 
the animals with the environment 
and large spatial scales they travel over. This field advanced only recently, 
to a greater extent due to the miniaturization and growing accuracy of the 
portable GPS devices.  Therefore there is
much less and sparser statistical data for big animals. Still, while 
it is possible to follow $1500$ individual sperm cells at a time \cite{Su02102012}, 
this number remains unrealistic for sharks or deers. 
There is a data-driven gap in the applied methodology. 
Some researchers are trying to use Langevin-type equations for continuous tracks while others prefer 
more coarse-grained random walk models for the trajectories recorded with limited resolution. 
Currently the gap is narrowing, as there are examples of random walks 
used to model the motility of bacteria and attempts to apply the 
Langevin machinery to analyze  the trajectories of bumblebees and beetles.

ii) {\em Complex trajectories.} Some trajectories  resemble 
neither L\'{e}vy flights nor L\'{e}vy walks but are still modeled as such. 
These are usually almost smooth continuos tracks of cells or other organisms \cite{levandowsky1997,dieterich2008,deJager2011}. 
There is often a problem of how to define a flight or a step of a 
random walk for such tracks, to resolve which several methods were 
suggested  \cite{Turchin1998,humphries2013,raichlen2014,rhee2011}. 
The proposed random walk models often are of academic interest only, 
since most of the information encoded in continuous trajectories is lost or disregarded. 
These approaches provide, however, some statistical characteristics of the foraging patterns that 
can be compared with those for the known search strategies. 
An alternative approach is to look into the microscopic details of motility patterns 
by using, for example, Langevin dynamics \cite{selmeczi2008,lenz2013,zaburdaev20112}, 
and then pose a question of how it can lead to the appearance of the L\'{e}vy like behaviour on larger spatial scales \cite{lubashevsky2009}. 
In a few cases, the information provided by trajectories was sufficient to suggest biological  mechanisms of 
the motility, as was demonstrated for some cells and bacteria \cite{Gibiansky2010, Jin2011, Marathe2014, Zaburdaev2014, li2008}.

iii) {\em L\'{e}vy flight vs. L\'{e}vy walk.} Although it is evident that living organisms can 
only move with a finite speed, 
there is a big subset of studies where the L\'{e}vy flight is used to 
model the observed trajectories. 
Some papers mention both approaches, walks and flights, interchangeably, but then they
mostly consider the statistics of displacements at fixed time intervals 
or the MSD. The distribution of displacements at fixed time intervals in fact yields a velocity distribution \cite{lopez-lopez2013}, 
and therefore suggests a very different model of random walks with random velocities;
as we have seen already, its properties are different from both the L\'{e}vy flight and L\'{e}vy walk models. 
We also know that the MSD of a L\'{e}vy flight diverges and therefore the corresponding model is not suitable for the analysis of the MSDs obtained from the experimental data. 
The ignorance to the difference between the L\'{e}vy flight and L\'{e}vy walk concepts does not add positively to the clarity of the issue. 

iv) {\em Other biological reasons.} 
As we have  mentioned, 
whether a bacterium or a deer, both interact with the environment. 
The more complex the organism is, the more rich and unpredictable are the effects of this interaction. 
While lab conditions for bacteria or cell experiments can be controlled to a high degree, 
the question of how much of deer's motion is influenced  by the type of a forest the deer moves in, 
is much harder to disentangle. Individuals may have different responses to the same stimuli, 
because, for example, they can be at  different developmental stages. 
Therefore it should not be forgotten that some effects which look like anomalous behaviour for the ensemble of organisms, may come about only as a result of variability between the individuals, where each individual behaves quite normally but on its own scale, see a book by \textcite{mendez2014} and original works by \textcite{Petrovskii2011,hapca2009}.

It is certainly beyond our goals and abilities to resolve all these challenging issues in this review. 
We can only welcome attempts to summarize and critically address 
these points by \textcite{mendez2014, selmeczi2008}. 
We  hope that the theoretical background provided in 
this review will help to introduce the L\'{e}vy walk model (and its appropriate modifications) 
to the community of biologists and biophysicists with more rigor so that it can be applied 
to the collected data in a proper way. 

\subsection{Soil amoeba}\label{amoeba}
\begin{figure}[t]
\center
\includegraphics[width=0.25\textwidth]{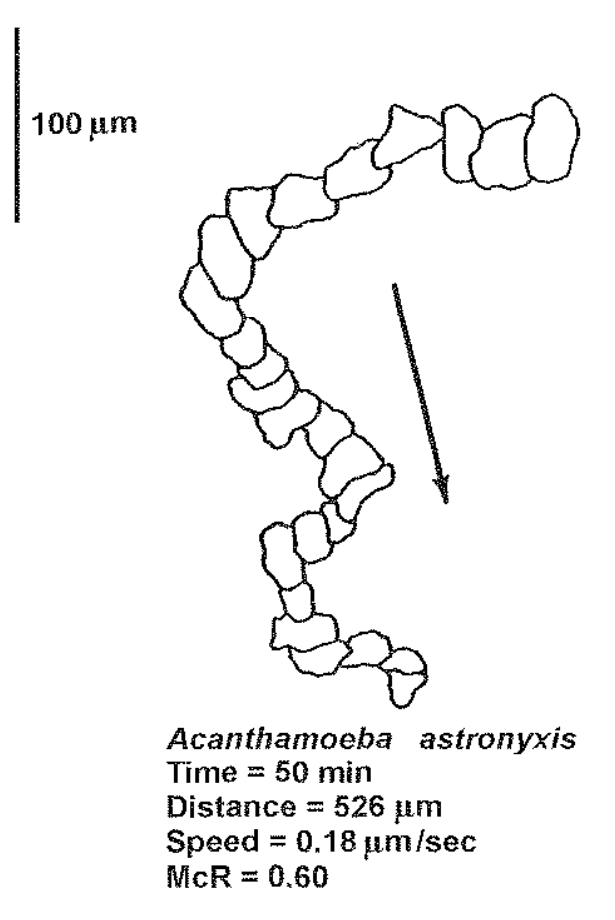}
\caption[Amoeba trajectory]
{A sample trajectory of soil amoeba showing outline of the cell at 1 minute intervals. Adapted from \textcite{levandowsky1997}.}
\label{figb1}
\end{figure}

One of the first mentions of the L\'{e}vy walk model in biological context 
was made in the work on crawling amoeba by \textcite{levandowsky1997}. 
Amoeba are uni-cellular organisms which can move on surfaces and three-dimensional media by growing cell protrusions called pseudopodia. 
In \textcite{levandowsky1997}, 17 amoeba isolates were tracked with a help of a microscope and a video recorder. 
Different traces of the overall duration  $15-60$ min were recorded with a time step of 1 or 2 minutes. 
Considered species represented a range of sizes $10$ to $100$ microns and average speeds $0.16$ to $1.3$ $\upmu\text{m}/\text{s}$. 
This means that cells roughly moved about one cell size per one step (one minute), see Fig. \ref{figb1}. 
After each step, the authors measured turning angles, velocity distribution and the MSD. 
For all observed cells the MSD scaled as  $\langle x^2\rangle\propto t^{\mu}$ with $\mu\sim1.5-1.9$, 
which led the authors to the conclusion that the L\'{e}vy walk could be a good candidate for a model. 
The obtained histograms of turning angles indicated little directional change. 
The authors also stated that their tracking was not long enough to check whether cells switch to the normal diffusion at longer times. 
Although not a clear-cut example of the L\'{e}vy walk, 
this was the first and balanced assessment of the experimental observations. 
After two decades of similar research one could suggest that the Ornstein-Uhlenbeck process, i. e. 
a Langevin equation for velocity increments containing friction and random force \cite{risken1996}, 
could be a reasonable alternative approach. 
The corresponding stochastic process is characterized by an exponentially decaying velocity auto-correlation function, 
and if the observation time is less or of the order of the correlation time the MSD behaves almost ballistically 
and only at later times switches to diffusive behavior, see similar results for beetles \cite{Reynolds2013}. 
A recent comprehensive study of {\em Dictyostelium discoideum} amoeba motility considered several possible mechanisms, 
including the generalized Langevin equation with a memory kernel, non-trivial fluctuations, 
and a more microscopic, zig-zag motion strategy \cite{li2008}. 

\subsection{Run and tumble of bacteria}\label{run_and_tumble}
Until recently, motion of a swimming {\em E.coli} bacteria was considered as a clear 
example of the standard diffusion. The diffusive dynamics
naturally follows from the mesoscopic picture of random walks describing 
the run and tumble motion \cite{Berg2004, Berg1993}. 
{\em E.coli} have multiple flagella, helical filaments which rotate and thus propel the cell in the fluid. 
Because of the microscopic size of the cell, the swimming occurs at low Reynolds numbers, 
which has its implications on the physics of the process \cite{purcell1977,Lauga2012}. 
Molecular motors can rotate flagella in two opposite directions, clockwise (CW) and counterclockwise (CCW). 
In CCW mode multiple flagella form a bundle and the cell swims following almost a straight path, which is called a ``run''. 
When one or several motors switch the direction to CW, the bundle dissolves and the cell rotates almost 
on the same spot, so called ``tumble'' phase. 
When the bundle forms again in the CCW mode, the cell begins its next run. 
The angle between the directions of the two consequent runs is not completely random but has a non-uniform distribution 
with a mean around 70$^{\circ}$. 
{\em E.coli} are rod shaped bacteria of about two microns long, 
the average run time is one second and the corresponding almost constant speed is $\sim 20\upmu \text{m}/\text{s}$. 
\begin{figure}[t]
\center
\includegraphics[width=0.45\textwidth]{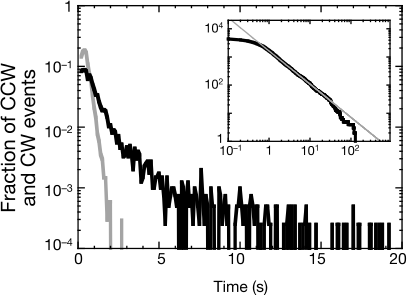}
\caption[Distributions of run and tumble times of E.coli bacteria.]
{Distributions of run and tumble times of {\em E.coli} bacteria. Counterclockwise (CCW) rotation (black line) corresponds to the run of bacteria, whereas clockwise (CW) rotation (gray line) corresponds to tumbling. Measurements are presented for a single bacterium. While tumbling times are shorter and well described by the exponential distribution, the durations of CCW rotations exhibit a long non-exponential tail which can be fitted by a power law. The inset shows the cumulative distribution function for CCW rotation times and the gray line is the power law with an exponent $\sim2.2$. From \textcite{Korobkova2004}.}
\label{fig_ecoli}
\end{figure}
Therefore the length of a run is roughly ten times the cell body length. 
Tumbles are approximately ten times shorter 
than the runs and usually neglected in theoretical models. 
Since the motion occurs in a fluid, runs are not entirely straight but are subjected to the effects of the rotational diffusion. 
If, for a moment, we neglect the rotational diffusion, the swimming cell can be seen as a biological realization of the L\'{e}vy walk model: 
it moves with an almost constant velocity, then tumbles and chooses a new swimming direction. 
The experimentally measured run time distribution of {\em E.coli} was usually described by an exponential distribution \cite{Berg2004}. 
However, in a recent experiment with individual tethered cells by \textcite{Korobkova2004}, 
it was shown that the PDFs of durations of CCW rotation of flagella (corresponds to run of the cell) fit
the power law distribution with an exponent $\gamma=1.2$, see Fig. \ref{fig_ecoli}. 

It was also shown theoretically that the genetic circuit responsible for the duration of motor rotation in CCW direction can generate 
power-law distributed times in the presence of chemical signal fluctuations \cite{dobnikar2011}. Experiments with tethered cells suggest that power-law distributed run times could be also observed in individual swimming cells, but there is no experimental confirmation of this yet. 

To encompass the possibility of the power-law distributed run times, L\'{e}vy walk model is the natural choice to describe the dispersal of idealized {\em E.coli} bacteria (by neglecting the effects of rotational diffusion during the runs) in two or three dimensions.

\begin{figure}[t]
\center
\includegraphics[width=0.45\textwidth]{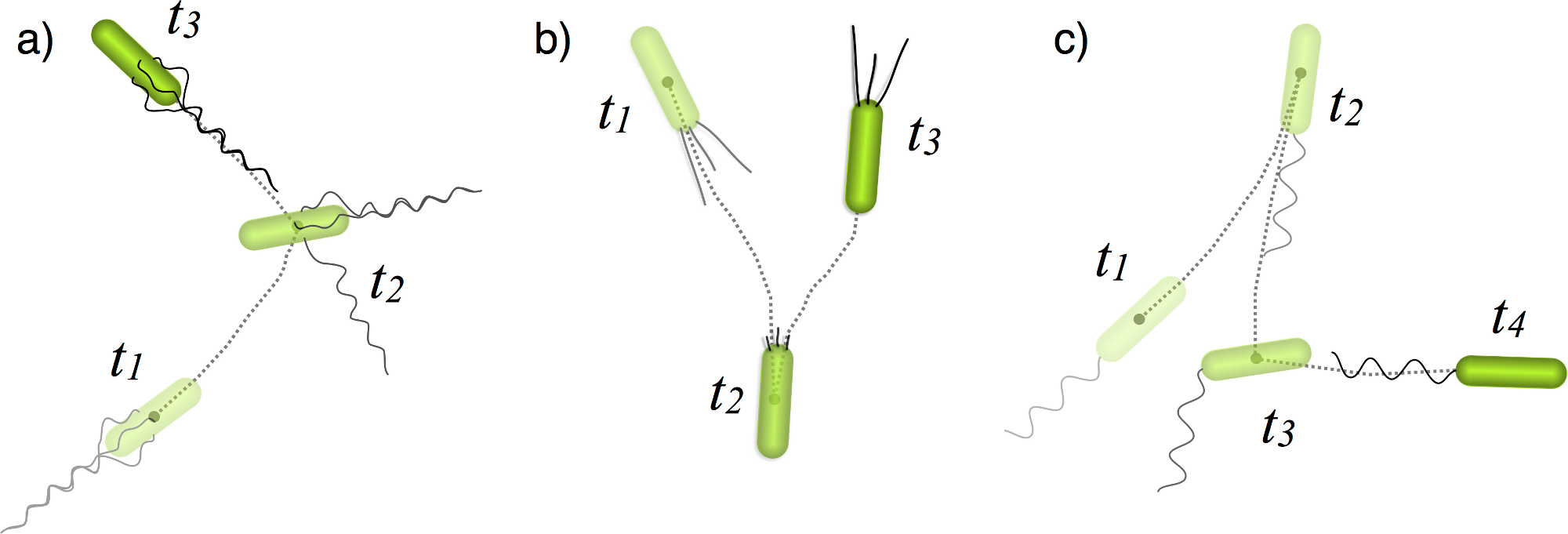}
\caption[Motility patterns of bacteria]
{(Color online) Different motility patterns of bacteria: 
(a) run and tumble motion, where straight runs are 
alternated by tumbling events; the angle between the
consecutive runs can have a certain preferred value (b) run reverse, where
 the next always has an opposite direction (c) run reverse and flick, where 
reversals strictly alternate with completely randomizing turn, similar 
to {\em E. coli}, in other words it is an alternation of (a) and (b) 
that makes up a more complex pattern of (c). From \textcite{taktikos2013}.}
\label{figb2}
\end{figure}

Interestingly, many bacteria (and some eukaryotic cells) swimming in fluid or moving by other means on surfaces  produce
similar patterns, reminiscent of the run-and-tumble motion. 
Those include, for example, run-and-reverse pattern, where the direction of the next run is  opposite to the previous one, 
or run-reverse-flick motion, where reversals are alternated with random turns \cite{Xie2011, taktikos2013}, 
see Fig.\ref{figb2}(c). 
Experimentally such trajectories are often observed within the single focal plane of the microscope or in a confined planar geometry. 
These biologically relevant motility patterns suggest an alternative description of a L\'{e}vy walk process in two dimensions, 
namely via the angle determining the orientation of the cell velocity, $\phi(t)$: $\mathbf{v}(t)=v(\cos\phi(t),\sin\phi(t))$. 
The time evolution of the angle can be split into two components, abrupt angular changes during the reorientation events 
and an independent component of the noise leading to the rotational diffusion, $\phi(t) = \phi_\text{rw}(t) + \phi_\text{rot}(t)$. 
$\phi_{\text{rw}}(t)$ characterizes a one dimensional CTRW in the angle space. 
The run-and-tumble motion is characterized by waiting times when the angle does not change (run) and random jumps 
according to the turning angle distribution (tumble). For example, in case of run-and-reverse, the jumps are of the size $\pm \pi$. 
Therefore the problem of the L\'{e}vy walk in two dimensions can be mapped onto a one-dimensional CTRW process for the angle. 
It has interesting consequences for the calculation of the standard quantities as, for example, the velocity autocorrelation function or the MSD. 
The velocity correlation function of bacteria's velocities at times $t_1$ and $t_2$, $C(t_1,t_2)$ is given by:
\begin{equation}  \label{corr1}
C(t_1,t_2)=\left<\mathbf{v}(t_1)\mathbf{v}(t_2)\right>=v^2  \left< e^{-i [\phi(t_2) - \phi(t_1)]} \right>.
\end{equation}
The contribution to correlations coming from the rotational diffusion is well known and appears as an exponential pre-factor,
a more non-trivial part is the random walking component of the angle. It can be shown that \begin{eqnarray}  \label{master1}
C_{\text{rw}}(t_1,t_2)\!
=\!v^2 \!\! \int_{-\infty}^{+ \infty}\!\!\!\! \!\! d\phi_1 \!\! \int_{-\infty}^{+\infty}\!\! \!\! \! d\delta\phi e^{-i \delta\phi} \!P(\phi_1,t_1 ; \delta\phi,t_2) .
\end{eqnarray}
Here, $P(\phi_1,t_1 ; \delta\phi, t_2)$ is the joint probability density 
to find a cell moving in 
direction $\phi_1$ at time $t_1$ and direction $\phi_1 + \triangle\phi$ at time $t_2$. It is easy to see that the above Eq.~(\ref{master1}) is the double Fourier transform with respect to $\phi_1$ and $\delta \phi$, where the corresponding coordinates in Fourier space are
set to $k_1=0$ and $k_2=1$, respectively. Therefore, to find the velocity autocorrelation function one needs to find the two-point PDF for the random walk of the angle. It is a non-trivial, especially for the case of power-law distributed waiting times, but exactly solvable problem \cite{baule2007,barkai2007,zaburdaev20082,Dechant2014}. The MSD can now be calculated by using the Kubo relation:
\begin{eqnarray}  \label{MSD_corr}
\left< [ \mathbf{r}(t) - \mathbf{r}(0)]^ 2  \right> = \int_0^t dt_1 \int_0^t dt_2 \; \langle \mathbf{v}(t_1) \cdot \mathbf{v}(t_2) \rangle.
\end{eqnarray}
There are two important particular cases of the above general formulas. For the exponentially distributed run times, $\psi_{\text{run}}(\tau)=\tau_{\text{run}}^{-1}\exp(-\tau/\tau_{\text{run}})$, many things simplify dramatically and yield the following answer for the MSD:
\begin{eqnarray}  \label{MSD_expo}
\left< [ \mathbf{r}(t) - \mathbf{r}(0)]^ 2  \right>_\text{rw} = 2 v^2 \widetilde{\tau}^2 \left( \frac{t}{\widetilde{\tau}} - 1 + e^{- t/\widetilde{\tau}} \right) ,
\end{eqnarray}
where the effective decorrelation time $\widetilde{\tau}$ depends on the average run time, $\tau_{\text{run}}$, and the average cosine of the turning angle, $\cos\phi_0$ \cite{Lovely1975}:
\begin{equation}\widetilde{\tau}=\frac{\tau_{\text{run}}}{1-\cos\phi_0}.\label{decor_time}\end{equation}
For {\em E.coli}, $\cos\phi_0\simeq 0.33$ whereas for reversing cells it is equal to $-1$. The Eq. (\ref{MSD_expo}) is a well known result for the Ornstein-Uhlenbeck process \cite{risken1996} which we already mentioned before, but here it was derived from the L\'{e}vy walk model and not from the Langevin equation. For short times $t\lesssim\widetilde{\tau}$ the MSD scales ballistically and then turns to the diffusive regime. 
In case of the power-law distributed run times (as in Eq. (\ref{psi})) with $1<\gamma<2$, the MSD scales as $t^{3-\gamma}$, a well known result. An interesting observation is that in the superdiffusive regime the turning angle distribution plays no role (unless the turning angle is not zero) in the asymptotic regime.  

To finalize this section we discuss two more modifications of the L\'{e}vy walk used to model motility  of  bacteria. 
Above we mentioned the run-reverse and flick motility pattern which was reported for {\em V.alginolyticus} 
bacteria by \textcite{Xie2011}. 
In this case, the reversals are alternating with completely randomizing turns with $\cos \phi_0=0$. 
In {\em V.alginolyticus} this happens because its single flagellum is unstable, 
when switching from CW to CCW rotation. 
The durations of runs after flick and reversals may also be governed by two different distributions. 
When translated into a CTRW model for the angle, 
that means that jumps with two distributions for the jump amplitude and waiting times are alternating. 
As a remarkable difference to the model with a single turning angle distribution where the velocity correlation function 
is always positive, run-reverse-flick model has an interval of negative velocity correlations \cite{taktikos2013}. 

For another type of swimming bacteria, {\em P. putida}, it was found that cells predominantly adopted 
the run-and-reverse pattern, but, in addition, the speed of a single cell changed roughly by a factor 
of two between forward and backward swimming directions \cite{Theves2013}. 
For the corresponding one dimensional L\'{e}vy walk model with two alternating speeds that would result 
in the back and forth motion, but with the ballistic scaling in the direction of the higher speed. 
The cells swimming in a fluid are subjected to fluctuations and therefore the rotational diffusion regularizes 
the ballistic scaling. 
As a result, bacteria undergoing run-and-reverse motion with alternating velocities, 
diffuse faster than bacteria showing run-and-reverse behavior but with a constant intermediate velocity. 

The above examples demonstrate that the class of L\'{e}vy walk models provides a perspective tool for the mesoscopic 
description of the bacterial motility. Whether the involved times are anomalously long or exponentially distributed, 
L\'{e}vy walk framework is flexible and can be adjusted to the needs of a particular problem -- 
rotational diffusion during runs, different turning angles and speeds, pausing during tumbles -- 
while remaining in the domain of analytically solvable models. 

\subsection{Short note on chemotaxis}\label{chemotaxis}
Bacteria, amoeba, sperms and many other cells and microorganisms are 
known to be able to perform chemotaxis: they can actively alternate motility in response to the gradients of 
certain chemicals, signaling molecules, nutrients, or waste products. 
Different organisms adopt different chemotactic strategies \cite{Eisenbach2004}. 
Larger cells, such as amoeba, can detect the gradients across their own cell body 
length via multiple chemoreceptors. 
Reacting to the occupancy of those receptors, 
amoeba can preferentially grow the pseudopodia in the corresponding direction and therefore continuously reorient during its motion. 
Bacteria are too small to do that and instead use a temporal integration of the chemical concentration 
which they experience along the trajectory. 
The chemical signal is passed on to the genetic pathway which regulates the flagella motor 
reversals (we are omitting a lot of interesting biological details, which, at least for {\em E.coli}, are well understood \cite{Berg2004}). 
If a cell swims in the direction of increasing concentration of the favorable signal it extends its run phase. 
The response of the cell to the pulses of certain chemicals was measured experimentally 
by observing the frequency of motor reversals; it revealed a non-trivial two lobed response function, 
showing the properties of adaptation \cite{Segal1986,Celani2010}. By assuming that tumbles follow after 
exponentially distributed run time, the rate of tumbling events in the presence of the signaling chemical 
with not too strong variations can be represented as:
\begin{equation}  \label{gennes_approach}
\lambda(t) = \lambda_0 \left( 1 - \int_{-\infty}^t d t' c(t') R(t-t') \right),
\end{equation}	
where $\lambda_0 = 1/\tau_{\text{run}}$ is the cell's tumbling rate in a homogeneous environment, $c(t)$ is the concentration of the chemical along the path, and $R(t)$ is the memory or response function of bacteria obtained from the experiments. \textcite{degennes2004} used this formula and the random walk model of run and tumble to calculate analytically the resulting average drift velocity along the small gradients of $c(x)$. This approach can be generalized to all above discussed motility patterns of bacteria and shows the importance of the theoretical modeling by means of simple random walks. One of the open questions in this field is how to generalize the de Genes' approach to a general distribution of tumbling events, going beyond the exponential function and including the power-laws.

\subsection{T-cells}\label{t-cells}
In a  recent experimental study by \textcite{Harris2012}, 
the migration of CD$8^{+}$ T-cells in the brain explant of mice was analyzed. 
CD$8^{+}$ T-cells are a special type of white blood cells which are responsible for killing cancer cells, 
those infected by viruses, or otherwise damaged or abnormal cells. 
Direct contact of the T-cell and the target cell is required for killing the abnormal cell. 
In this study, T-cells were targeting the cells infected by a parasite {\em T. gondii} 
which invades the cells of the central nervous system and causes the toxoplasmosis infection. 
T-cells which produce a fluorescent protein were imaged in 3D by using two-photon microscopy 
of the brain explant of mice with chronic toxoplasmic encephalitis in different experimental conditions. 
\begin{figure}[t]
\center
\includegraphics[width=0.45\textwidth]{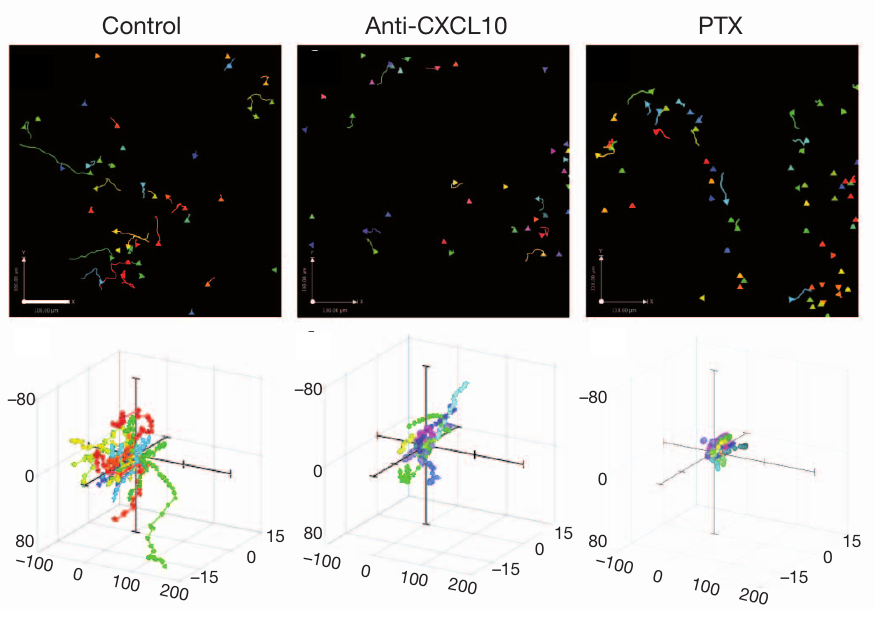}
\caption[3D trajectories of T-cells]
{(Color online) (top) Tracks of T-cells recorded by two-photon fluorescence 
microscopy in three different experimental conditions. (bottom) Reconstructed three-dimensional
trajectories of individual cells.
Adapted from  \textcite{Harris2012}.}
\label{figb3}
\end{figure}As one of the important factors involved in the regulation of T-cells motility, a small signaling protein, chemokine CXCL10, was noted. By varying the concentration of this chemokine the authors showed that T-cells were changing the average speed but not other statistical characteristics of their trajectories. Along with the standard MSD measurements, several additional properties of the acquired trajectories were analyzed: PDF of displacements at different times and its scaling properties, correlation function of cell displacements, and overall shape of the tracks. MSD showed a clear superdiffusive behaviour with the exponent 1.4: $\langle\mathbf{r}^2(t)\rangle\propto t^{1.4}$. Consistent with previous observations of runs and pauses in lymphocytes the authors suggested the model of L\'{e}vy walks with rests as the working hypothesis. Indeed by comparing this model with more than 10 other possible random walk models, it was shown to give the best representation of the experimental data (it is 
suggested to read the extensive, almost 20 pages, Supplementary material to the original paper). In Fig. \ref{figb3} we show the representation of 3D tracks and in Fig. \ref{figb4} the PDF of cell' displacements at different times (corresponds to the Control case in Fig. \ref{figb3}). 
\begin{figure}[t]
\center
\includegraphics[width=0.45\textwidth]{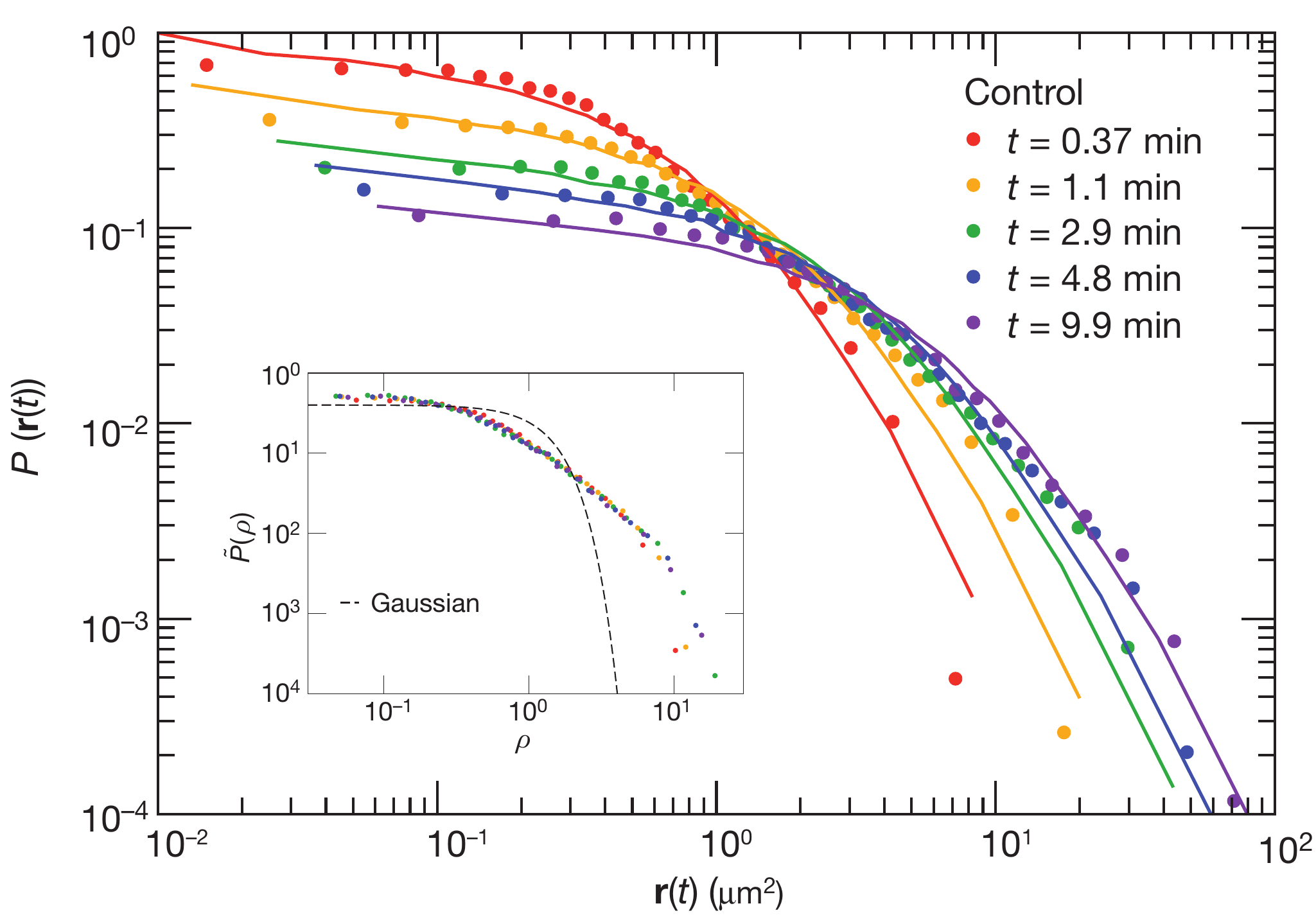}
\caption[PDF of T-cells' displacements]
{(Color online) PDF of T-cells' displacements for the control case at different time points. 
Symbols denote the experimental data, and lines are fits according to the generalized L\'{e}vy walk model. 
The inset shows that the profiles, rescaled according to $\tilde{P}=t^{\alpha}P$ and $\rho=r/t^{\alpha}$ with $\alpha=0.63$, collapse on top of each other. From \textcite{Harris2012}.}
\label{figb4}
\end{figure}

Rescaled profiles convincingly fall onto a single master curve. 
In the concluding remarks of the paper, it is mentioned that the L\'{e}vy walk model for the motility of 
T-cells is consistent with the idea of more effective search, 
as compared to Brownian motion in case of sparse targets. 
Overall it is one of the most thorough trajectory analyses to date which leads to the L\'{e}vy walk model. 
Probably because ten other models were shown to fail to reproduce the experimental data, 
it effectively exhausted the arsenal of arguments from the opponents of the L\'{e}vy walk foraging hypothesis [as a counterexample, 
see  a  trail of publications on the L\'{e}vy walk of mussels \cite{deJager2011}]. 

It is instructive to look at the scales involved in this study. 
A typical duration of the recorded trajectories was $15-30$ min with average moving speeds of 3 to 6 $\upmu \text{m}/\text{min}$, 
depending on the levels of the chemokine. 
With a size of the T-cell about 10 microns, 
similarly to the case of amoeba, cells travelled a couple of tens of their sizes. 
For bacteria that would correspond to a distance of a single run. 
However, for the case of T-cells that might be the relevant scale for finding the infected cells, 
and undoubtedly it is very intriguing that L\'{e}vy walks can be evoked in this context. 

\subsection{Humans}\label{humans}
\begin{figure}[t]
\center
\includegraphics[width=0.45\textwidth]{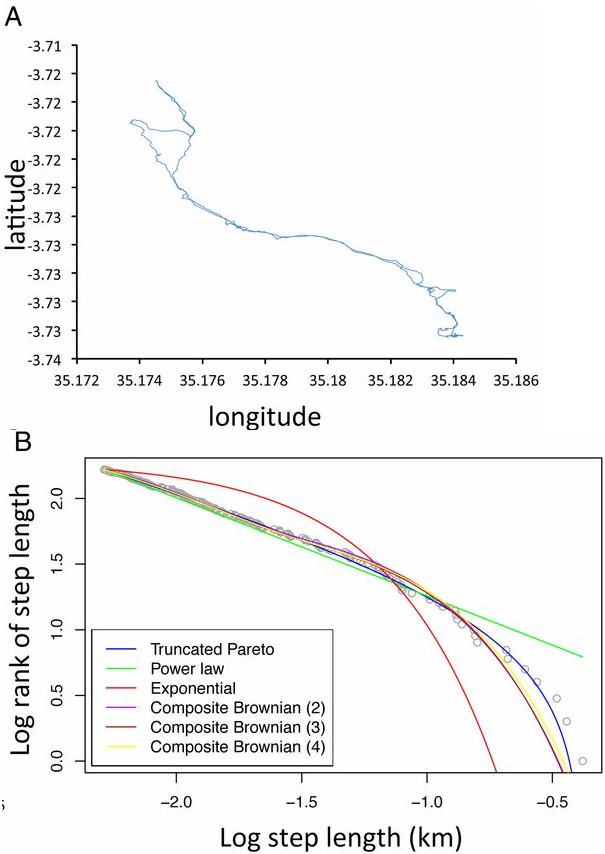}
\caption[Hadza gatherers trajectories]
{(Color online) L\'{e}vy walks of Hadza hunter-gatherers. 
(a) A representative trajectory of a Hadza hunter-gatherer 
bout obtained by GPS tracking. (b) 
The PDF of displacements during outbound parts of bouts, 
showing L\'{e}vy statistics in about a half of all cases. 
Symbols represent the experimental data and lines correspond 
to different theoretical approximations to this distribution. Adapted from \textcite{raichlen2014}.}
\label{figb5}
\end{figure}
Humans are most sophisticated organisms 
whose motility is governed by complex environmental, sociological, technological, 
and urban factors. 
The field of human mobility is an active  
domain of research because of its evident connection to real-life applications. 
Development of transportation systems, 
design of mobile networks, 
prevention of contagious disease spreading, 
all these issues are linked to the human mobility. 
Starting from dollar bill tracking by \textcite{Brockmann2006} and
to mobile phone tracking  by the group of Barab\'{a}si \cite{Gonzalez2008}, 
and to a recent study of influenza virus spreading by \textcite{Brockmann2013}, 
works on this topic gained a lot of attention, also in the public domain and media. 
Here we will review three empirical studies with very different settings, 
in which the L\'{e}vy walk patterns were found. 

An interesting experiment is described by \textcite{mendez2014} on the page 275 of their book. Nineteen 
blindfolded volunteers were ordered to search for targets randomly distributed over a soccer field. 
Each searcher was followed by a person who was recording the moving times 
of the searcher between the reorientation events. 
As an observed process it is a good example of the standard random walk with constant velocity. 
Searchers were not priorly informed of the purpose of the study, 
and about what was going to be measured. 
Each searcher was given ten minutes of time, 
and a prize was awarded to a person finding most of the targets. 
After a certain target was found on the field, it was returned to the field but displaced by a $1.5$ m distance in a random direction. 
In total there were $200$ targets with a characteristic size of 1 meter, distributed on a filed of the size $100 \times 50$ meters. 
Interestingly, after the data was analyzed and pooled according to the number of collected targets, $(0-1, 2-4, 5-8)$, 
and the distribution of run times was plotted, 
it appeared that the first two groups had an exponentially distributed run times, 
whereas the third group had a distribution reminiscent of the power law with 
an exponent of the tail $\mu=2.3$ 
[that corresponds to $\gamma=1.3$ in our notations for the flight time distribution, Eq. (\ref{psi})]. 
Certainly, the span of run times was only about one order of magnitude 
(there could be no runs longer than $\sim 2$ min because of the size of the field) 
and statistical tests could not give a clear preference to the power-law fit. 
Nevertheless, it is still very remarkable how the deviation from the exponential distribution 
arose and how this deviations correlated with the number of targets found. 
\begin{figure}[t]
\center
\includegraphics[width=0.5\textwidth]{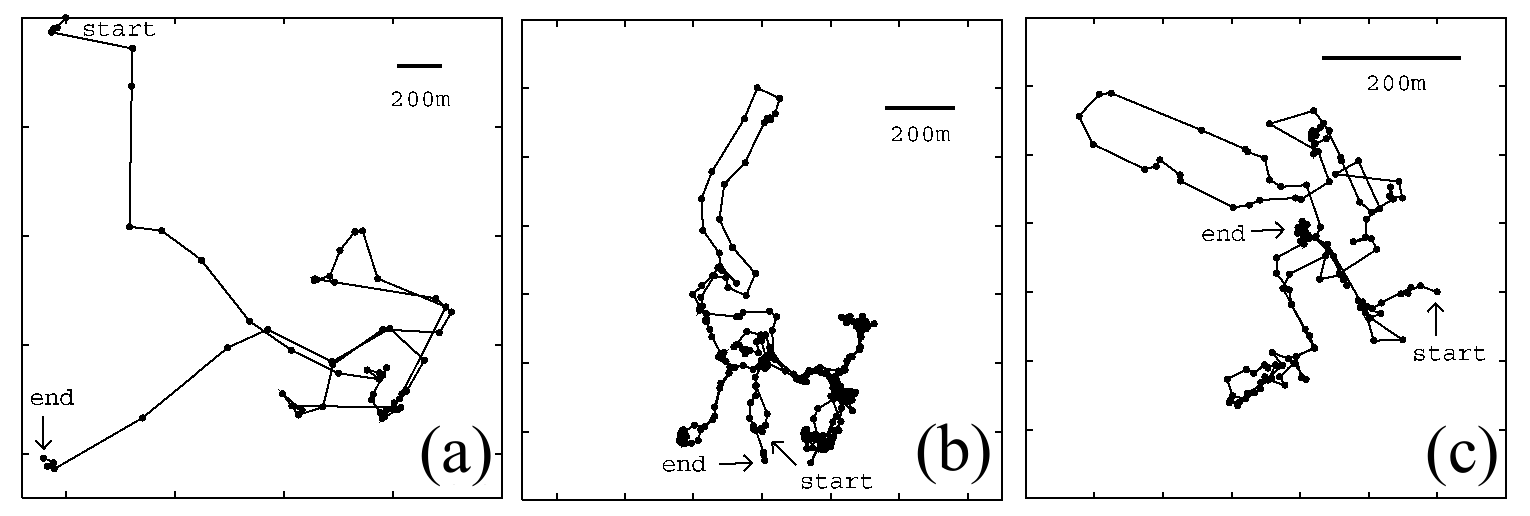}
\caption[Human trajectories]
{(Color online) GPS data on human movements
 during their daily activities in three different locations:
 (a) a University campus, (b) Disney World theme park, and (c) a state fair. 
Adapted from \textcite{rhee2011}.}
\label{figb6}
\end{figure}

\begin{figure*}[t]
\center
\includegraphics[width=0.999\textwidth]{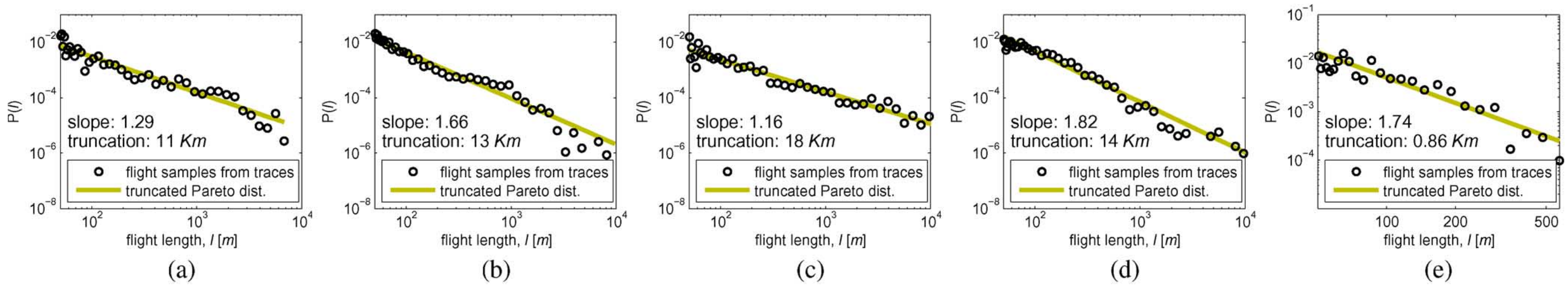}
\caption[Human displacements pdfs]
{(Color online) Distribution of human displacements.  
Step length distribution fitted with truncated Pareto 
distributions using the pause-based model to define the step lengths. (a) Campus I. (b) Campus II. (c) New York City. (d) Disney World. (e) State Fair. Reproduced from \textcite{rhee2011}.}
\label{figb7}
\end{figure*}

While the previous example might look like a 
fun experiment, some people rely on search for their survival. 
In another recent study by \textcite{raichlen2014}, human hunters-gatherers 
Hadza in northern Tanzania 
were shown to use L\'{e}vy walks in about a half of their foraging bouts, see Fig. \ref{figb5}. 
The Hadza hunter-gatherers have no modern tools or developed agriculture, 
they hunt with bow and arrow, and collect wild plant food. $44$
 subjects were monitored with the help of GPS devices during their foraging bouts for several days and at different seasons. 
The authors analyzed the step length distribution for outbound bouts (defined as travel between the camp and the furtherest 
away from the starting point). 
The steps were defined either by pauses or by turning angles, which in turn were analyzed with different threshold 
values from 0$^{\circ}$ to 180$^{\circ}$ with a step of 10$^{\circ}$. 
The obtained data were tested against L\'{e}vy walks, Brownian motion, or composite Brownian motion 
combining up to 4 exponential distributions. In around 50\% of all bouts the distribution 
of step lengths was best described either by a power-law or a truncated power-law with tail 
exponents ~1.9 and 1.5, respectively. The L\'{e}vy walk behavior appeared in both male and female 
subgroups despite the fact that they often had different goals of their bouts: hunting and searching 
for wild honey, or collecting berries and plant foods. The MSD of the corresponding tracks also 
showed an anomalous superdiffusive behavior. Inclusion of round bounds did  not change the results significantly. 
The authors argued that the human foragers, despite their higher cognitive complexity, still follow 
the same search pattern as used by other animals. Furthermore, the similar motion pattern of humans arises 
in much more complex urban environments, as we discuss next.

In a comprehensive study of \cite{rhee2011}, $226$ daily GPS traces were collected from 
101 volunteers in five different outdoor sites: two university campuses, 
state fair, theme park Disney World, and New York metropolitan area (see sample tracks in Fig. \ref{figb6}).

They acquired data with high space resolution of 3 meters and time step of 10 seconds, 
one of the most precise tracking to date. The following quantities were extracted from the traces: 
flight length, pause time, direction, and velocity. The authors used three different methods to 
define the flights on the smoothed data: rectangular (when a piece of trajectory between the two end points does not 
leave the boundary of a certain width from the line connecting those two points), based on the turning angle, 
and marked by pausing events. For all locations it was found that (truncated) power-law distribution fitted the data better than other 
model distributions. In comparison to previously discussed examples here the span of flight lengths covers four orders of magnitude. 
The tail exponents of those distributions were found to be in the range 1.2-1.9 based on pausing definitions of flights, see Fig. \ref{figb7}.  

Only in the case of the state fair, the exponential distribution was not so different from the power-law. 
Authors explain this by the truncation of the step length, as the state fair was 
indeed the smallest location of all five. The pausing events were also power-law distributed 
with a heavy tail exponents in the range 2.3 -- 3.5. The velocity of displacements was 
close to constant for short displacements, but increased steeply for larger travels. 
The reason behind this was that longer excursions could be made by using ground transportation, 
which was faster than walking. In terms of developing the appropriate L\'{e}vy walk type model 
that would require to introduce an additional coupling between the distance and velocity, which can be 
read out from the experimental data. The MSD for all five locations always had two regimes: superdiffusive 
at short times, less than 30 min to 1 hour, and subdiffusive afterwards. The superdiffusion is explained by 
long excursions, whereas the subdiffusive scaling was caused by the bounded travel domain and also due to the fact that 
humans do not do a completely random walk, but rather travel to certain destinations and often return to the same points, 
like home, office, or class. Because of similar factors, mobility of humans is certainly more complex than just a L\'{e}vy walk, 
but still this model appears to be one of the bests to describe human relocations as if they were a truly random process. 

To finalize this subsection we return to one of the first and influential studies of human travel data approximated 
by the dispersal of dollar banknotes. Some fraction of dollar bills in the US carry a stamp encouraging a person 
who gets a hold of it to visit a dedicated web-page, enter the bill number, current date, and location, and see 
its past trace. \textcite{Brockmann2006} used the databank of banknote traces and proposed a L\'{e}vy flight model 
combined with anomalously long traps, leading to a fractional diffusion equation, Eq. (\ref{FDE}). 
Although one could argue that instantaneous jumps
 might be not the most adequate representation of human travel, which on the vast scales of 
North America could happen by car, bus, train or air fair, each having its typical speed, for the 
data acquired it was practically impossible to take into account the finite velocity of travelers. 
Therefore, formally this study is outside our focus, but certainly deserves mentioning 
as one of the first works in this field. As the techniques of following individuals continue to progress, 
it is  to be expected that in the near future we will learn more about the human mobility.

\subsection{Bumblebees, seabirds, monkeys, and others}\label{others}
As we mentioned in the beginning of this section, the amount of data 
on  animal motions was constantly growing during the last decade. 
Not every new paper reports a L\'{e}vy walk motion pattern as a result, 
but at least tries to relate the observed motion patterns 
to the L\'{e}vy walk model. 
Current research trends in ecology were greatly influenced by the idea that under some circumstances a superdiffusive L\'{e}vy walk can be an advantageous search strategy when compared to classical Brownian-like diffusion pattern. We will review the search problem in the next section, 
and here we briefly list very diverse and interesting examples of data
on animal tracking. 

{\em Insects.} Insects can be traced by using different methods, 
such as traps, video cameras, entomological radars, 
or scanning harmonic radars combined with miniature transponders attached to individual insects. 
To the date, there is an impressive list of insects which were studied with respect to their motility 
patterns: ants \cite{Schultheiss2013}, bumblebees \cite{lenz2013}, honeybees \cite{Reynolds20072}, 
moth \cite{carde2012}, beetles \cite{Reynolds2013}, stoneflies \cite{Knighton2014}, and fruit flies \cite{Cole1995,Reynolds2007}.

{\em Sea animals.} Underwater creatures are much more difficult to follow and in general are traced by
 small, pressure-sensitive data-logging tags giving the depth information, 
or by high-frequency acoustic transmitter in combination with a directional hydrophone, 
or with satellite relayed data loggers. The list of tracked species is also quite long: 
various sharks, penguins, tuna \cite{Sims2008}, turtles \cite{Hays2006,dodge2014}, dolphins \cite{bailey2006}, 
mussels \cite{deJager2011}, cuttlefish, octopus, various rays, sole and anglerfish \cite{Wearmouth2014}, 
jelly fish \cite{hays2012}, grey seals \cite{austin2004}, and, finally, fishermen \cite{Bertrand2007}. 

{\em Birds.} Birds are usually tracked with the help of small GPS loggers attached to their bodies. 
One of the first studies in the field was done on wandering and black-browed 
albatrosses \cite{Viswanathan1996,Edwards2007,humphries2012} with several more to 
follow on pelagic seabird Cory�s shearwaters \cite{focardi2014}, frigatebirds \cite{demonte2012}, and Egyptian vultures \cite{lopez-lopez2013}.

{\em Mammals.} Most of observations of mammals foraging on terrain is done via visual contact and 
approximate GPS location determined by an observer using range finders. 
Several kinds of animals were tracked by this method: baboons \cite{schreier2014}, 
spider monkeys \cite{ramos2004}, fallow deer \cite{focardi2009}, jackals \cite{atkinson2002}, 
reindeer \cite{marell2002}, langurs \cite{vandercone2013}, and bearded sakis \cite{schaffer20142}.  

\section{L\'{e}vy walks and search strategies}\label{search}
Searching and foraging is enormously important in the ecological context and it is not surprising 
that more and more physicists and mathematicians contribute to this field. 
A growing database allows to propose and test various  models with increasing 
level of detail and complexity. 
The first mentioning of L\'{e}vy walks being advantageous 
in search as compared to classical random walks belongs to Shlesinger and Klafter \cite{shlesinger1986}.
Further on, L\'{e}vy-walk hunting strategy in the context of feeding behavior in grazing 
microzooplankton was discussed by \textcite{Levandowsky1988}.
It is widely recognized now that two papers by the group of Stanley, first on the L\'{e}vy flights of albatrosses \cite{Viswanathan1996}, 
and three years latter on optimality of the L\'{e}vy search \cite{Viswanathan1999}, 
lead to the birth of the new interdisciplinary field dealing with quantitative analysis of animal motility patterns and optimality of search. 
The maturity of the field is marked by several comprehensive monographs on the topic \cite{viswanathan2011, mendez2014, Benichou2011}, 
and the field itself spreads beyond animals and humans to robotics, see Section~\ref{robotics}. 

The problem of animal search is complex, as well characterized by \textcite{shlesinger2009}: 
``\textit{Actual search patterns of animals will depend on many factors: amount of energy expended in different modes of travel; 
the probability of finding food during various locomotions (flying, running, walking, hopping, etc); 
whether a single animal or a group is executing the search; day or night conditions; topography; weather; 
fixed food sources (water and vegetation) or moving targets (prey); homogeneous or scarce food sources; 
whether the animal randomly searches for food or has knowledge of food locations.}'' 
As an idealization of these features, when there is no prior information about the location of targets and 
complex interactions of a searcher with the environment and its prey, a so-called random search approach is used, 
which assumes that the searcher adopts a certain random motion pattern. The superdiffusive L\'{e}vy walk was proposed as an optimal search strategy \cite{Viswanathan1999} in case of sparse non-destructive targets, see below. However, validity and the straightforward use of the L\'{e}vy walk concept for the analysis of animal search 
patterns was questioned both experimentally and theoretically \cite{Benichou2006,Benichou2007,Edwards2007,Simon2007,Plank2009,Jansen2012,reynolds2014}. For an opinion on ``Should foraging animals really adopt L\'{e}vy strategies?'', see a recent review in \textcite{Benichou2011}. The resolution of this issue is out of our scope. Yet we do believe that there is a balanced middle point between the two extremes, ``L\'{e}vy'' and ``no L\'{e}vy'',
which follows from the universal principle: mathematics and physics can not take place of Nature but they certainly
can help to understand the former. 
Indeed, wandering albatrosses ``do not care about math'' \cite{Travis2007} and it is naive to think that 
a bird utilizes a L\'{e}vy walk when preying, by independently drawing a length of the next flight from a PDF with power law tails. 
L\'{e}vy walk-like motion patterns are not necessarily produced by a L\'{e}vy walk process\footnote{
In a recent study of trace fossils, \textcite{Sims2014} shown that the artificial trails produced by following three simple rules, 
(i) ``do not cross your trail'', (ii) ``stay close to it'', and (iii) ``make $U$-turns'', appeared to be L\'{e}vy walk patterns 
when analyzed with the conventional methods used in the field.}.
Moreover, patterns themselves -- even when they look very similar to those obtained in theory -- could not identify complex mechanisms of
animal locomotion that produced them. This  does not contradict the fact that the L\'{e}vy walk concept represents a powerful tool
for \textit{quantification} and \textit{analysis} of statistical data and provides with more insights into 
animal foraging strategies than the conventional Brownian-based  approach \cite{Buchanan2008}.

In the next three subsections we overview the current state of the field.  A special emphasis is put on the 
original paper by \textcite{Viswanathan1999} which greatly promoted the L\'{e}vy walk model 
as an advantageous search strategy. 

\subsection{L\'{e}vy walk as an optimal search strategy}\label{optimal_search}
\textcite{Viswanathan1999} considered a  walker which performed a L\'{e}vy walk in two dimensions and searched 
for targets, randomly distributed in space with a density $\rho$. The searcher can detect 
targets at the sight radius $r$. If a walker sees a target it proceeds straight to it. 
If there is no target in sight it chooses a random direction and moves for a random time with a fixed speed. If no target is found during a flight a new flight starts in an another random direction. The distribution of flight distances is chosen in the power-law form $g(l)\propto l^{-\mu}$.  Due to a simple coupling $l=v\tau$ we can identify 
\begin{equation}
\mu=\gamma+1,\label{munu}
\end{equation}
where $\gamma$ denotes the tail exponent of the flight time distribution, Eq. (\ref{psi}). As we discussed in the first section, $\mu>3$ will result in the finite mean squared length of the jump and normal diffusive dispersal. A regime of $1<\mu<3$ corresponds to the superdiffusive L\'{e}vy walks. In this model, it is important that the searcher keeps looking for a target while moving and that the current flight is terminated if the target is found. 
One of the ways to define the efficiency of the search is by the ratio of the number of targets found to the time spent in search or, in case of constant speed, to the total distance traveled:
\begin{equation}
\eta=\frac{1}{\langle l\rangle N},
\label{efficiency}
\end{equation} 
where $\langle l\rangle$ is the mean flight distance and $N$ is the average number of flights between the two successive targets. 
The only characteristic scale of the problem is given by an average distance between two detected targets, $\lambda=(2r\rho)^{-1}$. 
With its help, the mean flight distance can be approximated as:
\begin{equation}
\langle l\rangle=\frac{\int_{r}^{\lambda}l\cdot l^{-\mu}dl+\lambda\int_{\lambda}^{\infty}l^{-\mu}dl}{\int_{r}^{\infty}l^{-\mu}dl}
\label{average_flight}
\end{equation}
The first term in the nominator arises from the usual definition of the average flight length, but it 
has an upper bound of the typical distance between the two found targets. 
These flights do not terminate at the target. The second term counts the flights which were chosen to 
be longer than $\lambda$ but do terminate after the target encounter. The denominator is a normalizing factor. 
Next, the mean number of steps between the two successive targets needs to be found. At this point 
it is important to distinguish between two possible scenarios: targets can be either destroyed after 
being found (destructive case), or they become temporally depleted but can be revisited at later times (non-destructive). In these two cases, the average number of steps between two successive destructive and non-destructive targets can be estimated as [for detailed explanation see original paper by \textcite{Viswanathan1999}]:
\begin{equation}
N_{\text{d}}\simeq \left(\lambda/r\right)^{\mu-1};\quad N_{\text{n}}\simeq \left(\lambda/r\right)^{(\mu-1)/2}.\label{number_of_steps}
\end{equation}
Now the question of optimality may be asked: Is there an optimal value of $\mu$ which leads 
to maximal number of found targets, but keeps the length of excursions sufficiently short? 
If targets are plentiful, $\lambda\lesssim r$, then $N_{\text{d}}\approx N_{\text{n}}\approx1$ and $\langle l\rangle\approx \lambda$. 
In that case the search efficiency does not depend on $\mu$ at all. 
In the case of sparse resources, $\lambda\gg r$, situation is different. 
For destructive foraging the efficiency is maximal for smallest $\mu$ meaning that 
moving along one straight line is the best strategy in that case. However, situation 
is more interesting in case of non-destructive search. By substituting the expressions 
for $\langle l \rangle$, Eq.(\ref{average_flight}), and for the number of steps $N_{\text{n}}$, Eq. (\ref{number_of_steps}),
into  equation  (\ref{efficiency}), and equating its derivative with respect to $\mu$ to zero, we obtain the optimal value of the 
power-law exponent:
\begin{equation}
\mu_{\text{opt}}=2-1/[\ln(\lambda/r)]^2.
\label{muoptimal}
\end{equation}
The second term is a small correction in case of sparse targets, so roughly the 
exponent of $\mu\approx 2$ ($\gamma\approx 1$) arises as a solution. This value of power-law tail of the 
traveled distances corresponds to the border regime between superdiffusive and ballistic L\'{e}vy walks. 
Qualitatively the advantage of L\'{e}vy walks with $\mu\approx 2$ is explained by a compromise between diffusive trajectories returning to the same target zone ($\mu>3$) and ballistic motion ($\mu\sim1$) which is the best strategy to explore space. This result greatly promoted the notion of L\'{e}vy walks as 
an optimal search strategy in the case of randomly distributed, non-destructive, sparse targets.

One could question whether the assumptions made  when formulating the above search model
are realistic. An animal, even a protozoan, is a much more intellectual being  than 
a point-like particle driven by a finite-length  algorithm. After all, why should 
a donkey leave a water pond in the oasis (a non-destructive target following the nomenclature) 
he has once found in a desert?
Well, another could answer, the donkey has other needs also and he will 
turn to satisfy them once he has quenched his thirst and appeased his hunger; 
for example, he might like to find a mating partner. It is a perfectly correct argument but 
it goes far beyond the premises of the model. Animal search is a multi-layered  
activity determined by a vast number of external and internal (instincts, etc.) factors 
and it is  impossible to catch even most essential of them  with
a simple stochastic model. The good point is that the model introduced by \textcite{Viswanathan1999} 
allows for a gradual complexification 
and can absorb new assumptions and conditions. 
Since the paper was published,  many 
modifications  were 
proposed, which include, for example,  moving or/and regenerating, 
patchy targets \cite{Benichou2011,palyulin2014}. 
It was also found that, in some situations like searching for a single target in confinement \cite{Tejedor}, 
or under the presence of a bias \cite{palyulin2014}, persistent random walks 
or Brownian strategies perform better than L\'{e}vy walks.

\subsection{Intermittent search strategies}\label{intermittent}
A simple assumption that animals or humans have lower search capabilities 
when they are moving fast lead to the idea of the so-called \textit{intermittent search}, when 
periods of localized diffusive-like search activity are altered with  ballistic relocation to 
a new spot (searching for a lost key in an apartment is a good example). 
The intermittence has been detected in motion patterns 
of biological species ranging from protists to primates \cite{bartumeus2007,Benichou2011}. 
Different research fields contributed with different theories,
as, for example, ecologists discussed phases of  ``tactical habitat utilization'' (local search events)
and ``strategic displacements'' (ballistic relocations) \cite{Gautestad2006},  
while experts on random walks served a spectrum
of phenomenological models \cite{Benichou2011}. 
For us, further extensions of the standard L\'{e}vy walk model which are motivated 
by these studies are of interest.

\textcite{Lomholt2008}  suggested that an intermittent search in which the relocation 
happens according to the L\'{e}vy walk could lead to a more efficient search than, 
for example, exponentially distributed displacements between the diffusive search phases. 
From the point of view of modeling, such process might be called a composite process. The
L\'{e}vy walk is not simply diluted with resting events, when a walker is
immobile, like the process shown on Fig.~\ref{figt1}(c), but it is alternated with 
periods of different activity, for example, diffusion. 
Such processes are not new in the field of random walks but they experienced a 
revival of interest because of the new context. 
\textcite{Bartumeus2003} claimed that precisely this type of search strategy is realized by \textit{Oxyrrhis marina},
a dinoflagellate living in the sea depth, when it preys on a microzooplankton. Namely, when the prey decreases in abundance,
a predator switches from a slow-rate Brownian motion, characterized by an exponential PDF of flight time, to a helical L\'evy
motion, characterized by an inverse square power-law PDF.

In addition to the analysis of search patterns of biological species, the formalism of 
composite random walks allows to find analytic solutions for the density of particles and  calculate the scaling 
of the corresponding MSD. In a recent paper by \textcite{thiel2012}, a composite random walk was used 
to describe the run-and-tumble dynamics where the durations of the tumbles were explicitly taken 
into account and the runs were assumed to have a heavy tailed flight-time PDF. It was also assumed 
that during tumbling events particles perform  normal diffusion. Depending on the interplay 
between the tail of the flight times and durations of the tumbling phases (which also could, in principle, be 
characterized by a tunable power-law distribution) the MSD was shown to span the regimes from the normal 
diffusion to ballistic superdiffusion.

\subsection{L\'{e}vy walks for intelligent robotics: following suit}
\label{robotics}


Biological systems are a constant source of inspiration for the robot designers. 
It is not a surprise then that the wave of studies on L\'{e}vy-walk
foraging and animal search strategies has attracted attention of the researchers working
in the field of robotics. The current aim of the \textit{L\'{e}vy robotics} is twofold.
First, it is a development of new nature-inspired search algorithms for autonomous mobile robots
\cite{Pasternak2009,Nurzaman2009,Lenagh2010,Sutantyo2010,Ketter2012,Fujisawa2013,Sutantyo2013}. 
A complementary research line aims at the understanding of how  
L\'{e}vy-walk motion patterns emerge from combinations of different external factors and theoretical assumptions 
on  animal strategies and behavior \cite{Fricke2013}.

An idea to combine L\'{e}vy walks with chemotaxis in order to produce ``L\'{e}vy-taxis'', a search 
algorithm for an autonomous agent to find a source of chemical contamination in a turbulent aquatic environment, was proposed by 
\textcite{Pasternak2009}. It is not a
typical search task because the searcher should scan a constantly changing 
chemical field and follow plumes in order to find their origin. In the computational studies, 
a virtual AUV (Autonomous Underwater Vehicle), floating in a virtual two-dimensional river-like
turbulent flow, contaminated from a  point-like source, was used. Events of unidirectional motion,
characterized by a power-law distribution of their lengths and a wrapped Cauchy distribution of 
their direction angles, were intermingled with short re-orientation events. During the latter
the vehicle was randomly choosing a new movement direction 
along the local concentration upstream flow. 
This strategy somehow corresponds to a L\'{e}vy walk in a flow-oriented reference frame. 
When compared to other strategies, based on 
Brownian walk, simple L\'{e}vy walk, correlated Brownian walk and a brute-force zig-zag scanning, 
L\'{e}vy-taxis outperformed all of them, both in terms of detection success rate and detection speed.  

Another searching strategy for a mobile robot, a sequence of L\'{e}vy walks alternated with taxis 
events, was proposed by \textcite{Nurzaman2009}. In computer simulations, the robot task was to locate a 
loudspeaker by using the information 
on the local sound intensity obtained from a robot-mounted microphone. The loudspeaker was stationary and the robot's 
speed $\upsilon$ was constant.
\begin{figure}[t]
\center
\includegraphics[width=0.45\textwidth]{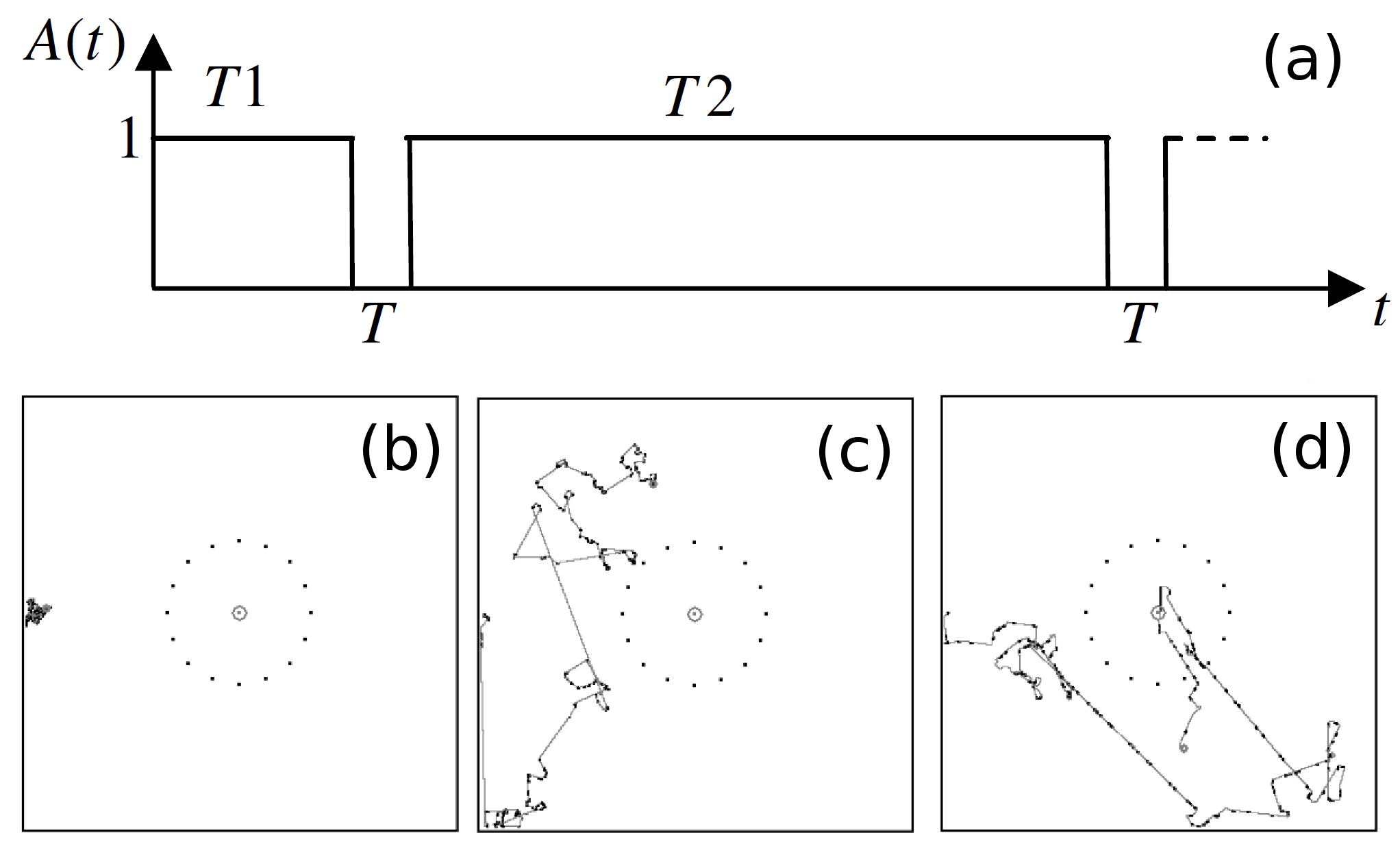}
\caption[robot1]
{Performance of a sonotactic robot. (a) Activity of the robot: Durations $T1,T2,...$ follow a power-law tail
distribution while the duration of  re-orientation events $T$ is constant; (b-d) Trajectories of the robot using the sonotaxis strategy (b),
the L\'{e}vy walk (c) and the combination of the two (d). Speaker (small solid circle) is located at the center of the squared 
test area (box) and dashed line encircles the area with sound gradient above a threshold. The starting point is located at 
the middle of the left box border.  Adapted from \textcite{Nurzaman2009}.}
\label{robot1}
\end{figure}
The robot orientation was defined by the angle $\theta$. The robot dynamics 
was governed by three stochastic equations,
\begin{eqnarray}
\left[\begin{array}{cc}
                  \dot{x}(t)\\
                \dot{y}(t)\\
                  \dot{\theta}(t)
             \end{array}
              \right] =
A(t)\left[\begin{array}{cc}
                  \upsilon\cos\theta(t)\\
                \upsilon\sin\theta(t)\\
                 0
             \end{array}
              \right]+
[1-A(t)]\left[\begin{array}{cc}
                  0\\
                0\\
                 \varepsilon_{\theta}(t)
             \end{array}
              \right] 
\label{msd_scaling}
\end{eqnarray}
where the Cartesian coordinates $x(t)$ and $y(t)$ specify the position of the robot at time $t$. 
Activity $A(t)$  is  a dichotomous function switching between $1$ and $0$ so that the robot
is either moving forward with velocity $\upsilon$ (activity is ``1'') or is choosing randomly a new direction of motion
(activity is ``0''). When the duration of a single  $1$-event is distributed according to a power-law, 
see Fig.~\ref{robot1}(a), the robot performs a two-dimensional version of the L\'{e}vy walk with rests shown on Fig.~\ref{figt1}(c).  
Alternatively,  a stochastic sonotaxis strategy by using which the robot tried to locate  and move towards the loudspeaker was probed. 
However, neither of the two strategies was able to accomplish the task when used alone.
The sonotaxis turned out to be effective in a close vicinity of the speaker only, 
and did not work when the sound gradient was small, see Fig.~\ref{robot1}(a). The L\'{e}vy walk did not care about the sound 
intensity by default and produced unbiased wandering only, Fig.~\ref{robot1}(b). 
The combination of the two solved the problem: 
the L\'{e}vy walk first brought the robot to the area where the sound-intensity gradient  was high enough and from there
the sonotaxis strategy was able to lead the robot to the loudspeaker, Fig.~\ref{robot1}(c).
A \textit{L\'{e}vy looped search} algorithm  to locate \textit{mobile} targets with a swarm of non-interacting  robots
was proposed by \textcite{Lenagh2010}. The idea was to replace straight ballistic segments with loops so that each searcher returns to its 
initial position. The length of each loop was sampled from a power-law distribution, whereas the starting angle 
was sampled from the uniform distribution in the interval $[0, 2\pi]$. The reported results showed that the looped search 
outperformed the standard L\'{e}vy search in tracking mobile targets.

\begin{figure}[t]
\center
\includegraphics[width=0.5\textwidth]{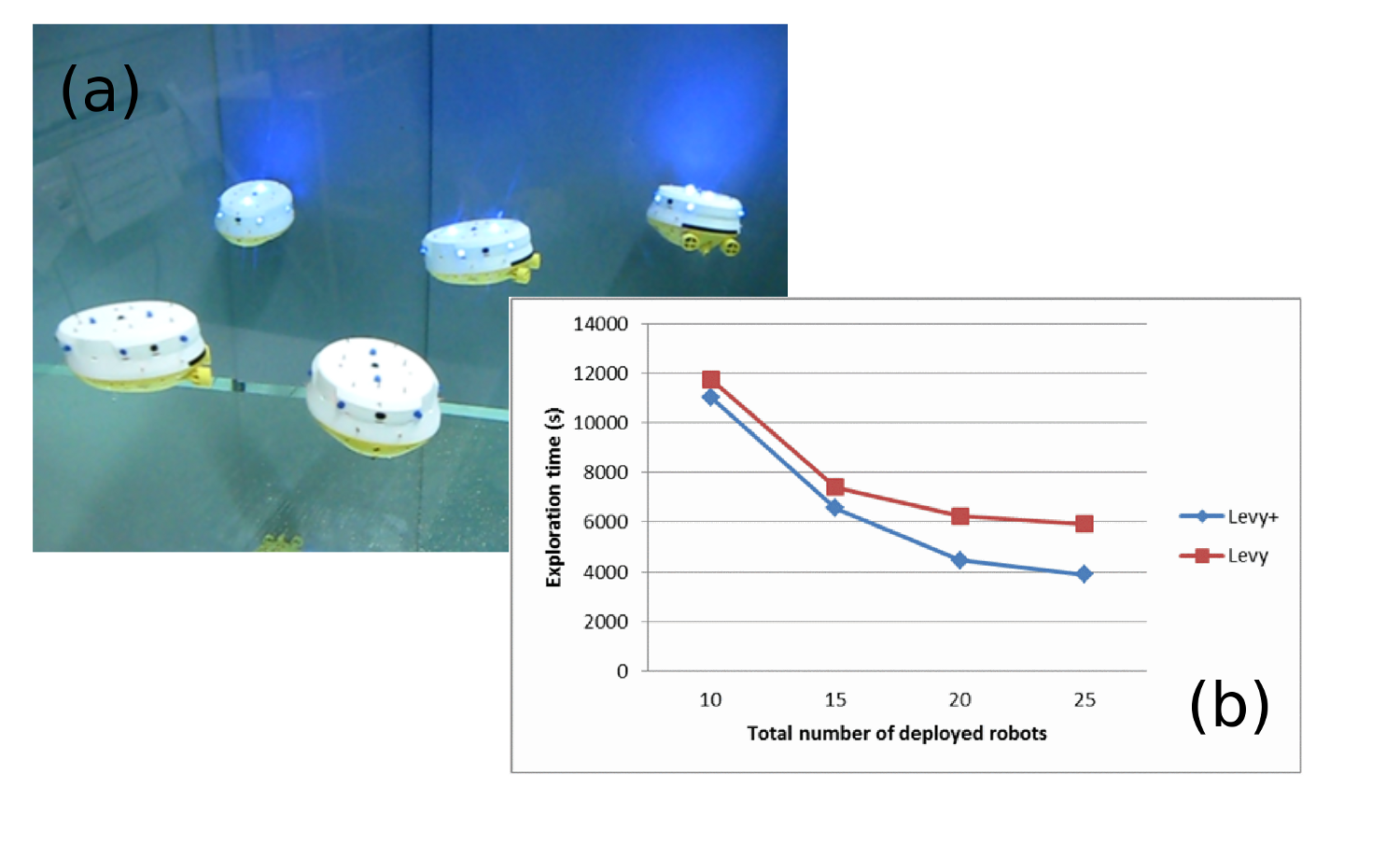}
\caption[robot2]
{Collective  multi-robot exploration. (a) Autonomous underwater vehicles used in the experiments; (b)
Targets searching experimental results: Exploration time vs number of robots for two strategies, 
with independent L\'{e}vy searchers (red line) and and interacting L\'{e}vy searchers (blue line). 
Adapted from \textcite{Sutantyo2013}.}
\label{robot2}
\end{figure}

The idea that a search efficiency can be increased by using a number of autonomous agents is natural and
relevant in many contexts. It is evident, for example, that the search time is inversely proportional to the number 
of independent searchers provided all  
other conditions remain the same. However, if an interaction or exchange of information between the searchers is allowed,
the search time can be decreased even further. Swarm communication is widely used among animals and insects, 
and it is known  among biologists and roboticists as ``stigmergy'' \cite{stigmergy}. A multi-robot searching algorithm based 
on a combination of a L\'{e}vy walk and an artificial potential field inducing repulsion among robots, was proposed and tested 
by \textcite{Sutantyo2010}. The obtained results for up to twenty robots showed that the repulsion increases search efficiency 
in terms of the search time. It is noteworthy that the effect diminishes with increase of the robot number,  because
crowding robots start to change their directions  earlier than expected from the governing power-law distribution. 
Experimental results obtained for 
two L\'{e}vy-swimming AUVs in a $3$-d  aquatic testbed \cite{Ketter2012} show that in this case the best performance corresponds 
to a simple divide-and-conquer strategy, when the tank is divided into two equal volumes and each submarine scouts its assigned region only.
However this situation may change when the number of AUVs is larger than two so that communication between searchers could be beneficial.
Group L\'{e}vy foraging with an artificial pheromone communication between  robots was studied recently by \textcite{Fujisawa2013}. 
Each robot had a tank filled with a ``pheromone'' (alcohol)  which was sprayed around by a micropump. 
Rovers also carried alcohol and touch sensors and  
their motion was controlled  by a program which 
took into account the local pheromone concentration. 
The swarm foraging efficiency  peaked when the robots were  programmed beforehand  to perform a L\'{e}vy walk  in the
absence of the communication. Multi-robot underwater exploration and target location were studied
with a swarm of L\'{e}vy-swimming AUVs by \textcite{Sutantyo2013}, see Fig.~\ref{robot2}(a). 
Interaction between the robots was introduced by using a modification of
the Firefly Optimization, an algorithm popular in the field of particle swarm optimization \cite{swarm}. The ``attractiveness'' of 
each AUV  was defined 
by the time since the robot last found a target; it increased every time a target was located and then slowly decayed. 
The task was for each searcher 
to find all  the targets. 
The results of the experiments showed that the interaction  decreases the averaged search time substantially,
see Fig.~\ref{robot2}(b).

Finally, an attempt to get insight into the machinery causing the emergence of L\'{e}vy walk-like  
patterns in the motion of different biological species was made recently by \textcite{Fricke2013}. Inspired by the results obtained for T-cells 
\cite{Harris2012}, see  Section \ref{t-cells}, researchers from the University of New Mexico and Santa Fe Institute 
used six small rovers,  equipped with ultrasound sensors, 
compasses, and cameras. This navigation set enabled  each robot to find patches of resources distributed over $2$-d area. 
Tunable adaptive algorithms  based on five different search strategies were tested. 
It turned out that the algorithm using correlated random walks, in which correlations between consequent step angles of a rover depend 
on the target last observed by the rover, produces L\'{e}vy-like motion patterns.

L\'{e}vy robotics is only one example that illustrates the practical  value of the LW-concept. 
We do believe that there are more to come and discuss potential candidates in the final section of the review.  

\section{Outlook}
L\'{e}vy walk concept is almost in its thirties and now 
possibly at the beginning of the most interesting phase of its life. 
The gradually developing theoretical framework was there 
in time to support the burst of applications across different fields. 
As can be seen from the previous sections, most of the empirical data obtained with 
cold atoms, nanostructured media, quantum dots, and ecology emerged only recently. 
L\'{e}vy walks remain in a stage of active development, and
we now see them being used in robotics and mobile communication technologies \cite{Lee2013}. 
In this concluding section  we  would like to discuss some open problems in the field, 
and to sketch what we think are the next perspectives and challenges.

For the physicists, probably one of the central questions is to understand how the L\'{e}vy walk, 
which is a mathematical model, emerges in diverse physical phenomena. 
There is a certain  progress in this respect in the fields of classical many-particle chaos \cite{Mendl2014} 
and cold atom dynamics \cite{barkai2014}. 
In the problem of light diffusion in hot atomic vapors, 
general principles of light emission/absorption were suggested to be relevant mechanisms \cite{Baudouin2014}. 
In experimental plasma physics, the anomalous nonlocal transport is regularly reported in various works, 
but its  origin remains a subject of ongoing debates. 
This can be partially explained by the high complexity of modern plasma experiments which are often performed 
in nonequilibrium regimes, involve nonlinear interactions, 
formation of large coherent structures, etc. 
We can see that even simple approximations of plasma ion dynamics by using the L\'{e}vy walk immediately
call for non-linear space-time couplings \cite{Zimbardo2000,Gustafson2012}. 

\begin{figure}[t]
\center
\includegraphics[width=0.45\textwidth]{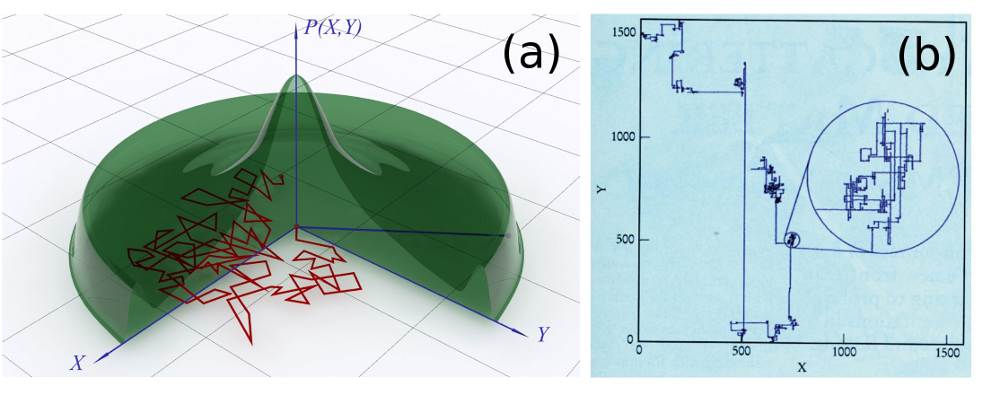}
\caption[Two-dimensional L\'{e}vy walks]
{(Color online) Two intuitive generalizations of the L\'{e}vy walk  to two dimensions.
Both models are characterized by power-law pdfs of the flight time, but in  
(a) an ``isotropic'' model the direction of a flight is given by a random 
angle uniformly distributed on the interval $[0, 2\pi]$, while in (b) a ``lattice'' model
ballistic flights happen only along one of the two basis vectors of a square lattice.
The latter process was observed when tracking trajectories of a Hamiltonian particle 
moving over an egg-grate potential \cite{klafter1994}. 
Figure (b) is adapted from \cite{Klafter1996}.}
\label{figOut}
\end{figure}

Most of the analytical results presented in this review are restricted to one-dimension. 
This reflects the current situation on the theory front. 
Although the analysis can be formally generalized to higher dimensions 
by replacing the Fourier coordinate with a Fourier vector, $k\rightarrow \mathbf{k}$, this technical step 
will immediately pose a number of questions. 
For example, a two-dimensional L\'{e}vy walk can be defined in two different intuitive ways, namely, as a process when 
(i) the  length of the upcoming flight and its random \textit{orientation} are 
both chosen from continuous PDFs (like in the case of run and tumble of bacteria, Section \ref{run_and_tumble})
or, alternatively, (ii) a random displacement is chosen independently but always \textit{along} one of the 
two basis vectors. How do the propagators of these processes look like? Evidently, because of the isotropy of 
the first process, the corresponding propagator will be circular symmetric, 
$P(\textbf{r},t) = P(r,t)$, $r = \sqrt{x^2 +y^2}$, 
see left panel of Fig.~\ref{figOut}(a). It is tempting to say that in this case the problem can be reduced to 
the one-dimensional setup by simply taking $|x|$ with $r$ in the propagator. 
However it has yet to be clarified which equation governs the evolution of $P(r,t)$.
A LW process of the type (ii) has  been observed in numerical studies of the
superdiffusion in two-dimensional chaotic Hamiltonian systems by \textcite{klafter1994}, see 
right panel of Fig.~\ref{figOut}(b). It was shown that at the asymptotic limit and far from the center $\textbf{r} = 0$, 
the corresponding propagator factorizes, $P(\textbf{r},t) \simeq P(x,t/2)P(y,t/2)$, where 
$\textbf{r} = x\cdot \mathbf{e}_x + y\cdot \mathbf{e}_y$ and $P(x,t)$, $P(y,t)$ are 
one-dimensional propagators. This type of two-dimensional L\'{e}vy walks with the exponent $\gamma = 2$ 
is relevant for the description of the diffusion in Sinai billiards with infinite horizon \cite{bouchaud1990},
as it has been shown recently by \textcite{Cristadoro2014}.

As a next step  one can consider L\'{e}vy walks on  lattices of different geometries and 
try to elucidate the effects of the 
underlying geometry on the corresponding propagators. 
This question is particularly motivated by the experimental studies of light propagation in regular foams, 
where, due to the effect of total internal reflection, the light gets trapped in the liquid phase 
of the foam \cite{gittings2004}. 
Theoretically it was shown that, in the case of a honeycomb foam lattice,  
the light propagation can be superdiffusive \cite{schmiedeberg2005,schmiedeberg2006}. 

An issue of correlated L\'{e}vy walks  
not only constitutes a theoretical challenge but is of relevance in the context 
of several recent experiments. 
One example is the diffusion of light in L\'{e}vy glasses, Section \ref{LWlight},
where the quenched disorder of scatterers may induce correlations between the flights. 
Independently, this question was posed  by theoreticians some time ago, 
see \textcite{kutner1998,kutner1997,levitz1997,barkai2000lor}, and still requires further analytical investigation. 
Correlated ballistic L\'{e}vy walk  could also serve as an advanced model to account for the correlations 
in blinking times of quantum dots \cite{stefani2009}.

A fundamental characteristic of any random walk process is the so-called first passage time, 
which defines how soon a random walker would visit a point located at a certain distance from the origin, 
see book by \textcite{Redner2001}. 
The first passage time problem for L\'{e}vy walks naturally occurs in the context of searching strategies, 
where it quantifies the time it takes to hit a target. 
Many of the results obtained for the first passage time and  related problems for the standard random 
walks \cite{Redner2001} 
were generalized to subdiffusion and L\'{e}vy flights. 
At the same time, the problem of the first passage time for L\'{e}vy walks remains largely 
unexplored \cite{korabel2011}. 

As already mentioned, the origins of L\'{e}vy walks in biology and ecological context is an unsettled issue. 
Although some examples exist that show how the power-law distributed run times emerge from the underlying genetic circuits 
of bacteria \cite{dobnikar2011}, 
it remains to be seen whether similar evidences can be found for more complex organisms that exhibit L\'{e}vy walk-like behavior. 
In meantime, L\'{e}vy walk strategies are implemented to construct robots that 
can assist humans in finding sources of contamination, and  
to develop efficient strategies to rescue people from disaster areas \cite{akpoyibo2014}. 
The concept of L\'{e}vy foraging has made its way into the field of criminology, 
potentially leading to implications in predictive policing. 
\textcite{levy_crime} used criminal records of more than a thousand offenders who committed series of 
crimes and found that the distribution of distances between the consequent events was consistent with L\'{e}vy walk dynamics. 
There is  also an interview with a burglar which corroborates the L\'{e}vy walk behavior as an optimal \textit{evading} strategy  and relates it to 
the perception of risk to be caught. In a very recent paper titled ``Voles don't take taxis'', \textcite{voles} comments on the 
work of \textcite{levy_crime} and puts it in the context of modern quantitative criminology research. 

To conclude, we provide an overview of the theoretical aspects of a simple but remarkably 
flexible model of L\'{e}vy walks. We illustrated theoretical considerations with a variety of examples 
where the model and its offspring served to quantify the stochastic transport phenomena and help  
elucidate underlying mechanisms. We would like to think that this review will stimulate  
researchers from even more distant fields to use the model in their studies and thus will help 
to advance L\'{e}vy walks into new unexplored territories.

\begin{acknowledgments}
We are grateful to our collaborators and colleagues who contributed their knowledge, 
time, enthusiasm, and support to this work: 
E. Barkai, A. Blumen, D. Brockmann, V. Belik, A. Chechkin, K. V. Chukbar, A. Dhar, S. Eule, R. Friedrich, 
D. Fr\"{o}mberg, P. H\"anggi, F. J\"ulicher, S. Lepri, R. Metzler,  K. Saito, M. Schmiedeberg, 
M. Shlesinger, I.M. Sokolov, G. Spohn, H. Stark, M. Timme, N. Watkins, and G. Zumofen.  
We thank O. Tkachenko and N. Denysov for the help with preparing the figures 
and L. Jawerth for proofreading of the manuscript. 

S.D. acknowledges support by  the German Excellence Initiative
``Nanosystems Initiative Munich'' and Grant No. N02.49.21.0003 (the agreement between the Ministry
of Education and Science of the Russian Federation and
Lobachevsky State University of Nizhni Novgorod).

\end{acknowledgments}

\end{document}